\documentclass[12pt, a4paper]{article}
\usepackage{styleBuding}
\usepackage{amsmath}
\usepackage{amssymb}
\usepackage{multirow}
\usepackage{subfigure}
\usepackage{siunitx} 

\newcommand{\rom}[1]{\uppercase\expandafter{\romannumeral #1\relax}}
\DeclareMathOperator{\sgn}{sgn}
\usepackage{csquotes}
\usepackage{multicol}
\title{\boldmath Electron and Muon Anomalous Magnetic Moments in the Inverse Seesaw Extended NMSSM}

\author[]{Junjie Cao,}
\author[]{Yangle He,}
\author[]{Jingwei Lian,}
\author[]{Di Zhang}
\author[]{and Pengxuan Zhu}

\affiliation{Department of Physics, Henan Normal University, Xinxiang 453007, China}
\emailAdd{junjiec@alumni.itp.ac.cn}
\emailAdd{heyangle@htu.edu.cn}
\emailAdd{ljwfly@hotmail.com}
\emailAdd{dz481655@gmail.com}
\emailAdd{zhupx99@icloud.com}

\abstract{The recently improved observation of the fine structure constant has led to a negative $2.4\sigma$ anomaly of electron $g-2$. Combined with the long-existing positive $4.2\sigma$ discrepancy of the muon anomalous magnetic moment, it is interesting and difficult to explain these two anomalies with a consistent model without introducing flavor violations. We show that they can be simultaneously explained in the inverse seesaw extended next-to-minimal supersymmetric standard model (ISS-NMSSM) by the Higgsino--sneutrino contributions to $(g-2)_e$ and $(g-2)_\mu$. The spectrum features prefer light $\mu$, which can predict $m_Z$ naturally, and it is not difficult to obtain a $\tau$-type sneutrino dark matter candidate that is compatible with the observed dark matter relic density and the bounds from dark matter direct detection experiments. Due to the compressed spectra and the undetectable decay mode of selectrons, they can evade the current Large Hadron Collider (LHC) constraints. 
}

\begin{document} 
\maketitle
\section{Introduction}
\label{sec:intro}
Since Schwinger showed that $a_\ell \equiv (g_\ell -2)/2 = \frac{\alpha}{ 2\pi}$~\cite{Schwinger:1948iu}, the anomalous magnetic moments of charged leptons have survived rigorous tests of the quantum electrodynamics and the later Standard Model (SM) of particle physics for more than half a century. Recently, an improvement of the measurement of the fine structure constant $\alpha$, via the recoil frequency of cesium-133 atoms, has yielded the most accurate measurement~\cite{Parker:2018vye}:

\begin{equation}
	\alpha^{-1}({\rm Cs}) = 137.035999046(27).
\end{equation}   
As a result, there is a negative $2.4\sigma$ discrepancy between the theoretical prediction $a^{\rm SM}_{e}$~\cite{Aoyama:2017uqe} and the existing experimental measurement $a^{\rm exp}_{e}$~\cite{Hanneke:2010au, Hanneke:2008tm} of the electron anomalous magnetic moment,

\begin{equation}\label{eq:amm_e}
	\Delta a_e \equiv a^{\rm exp}_{e} - a^{\rm SM}_{e} = (-87\pm 36)\times 10^{-14}. 
\end{equation}
Meanwhile, the long-standing discrepancy of the muon anomalous magnetic moment~\cite{Zyla:2020zbs} between the SM prediction $a^{\rm SM}_{\mu}$~\cite{Aoyama:2012wk,Aoyama:2019ryr,Czarnecki:2002nt,Gnendiger:2013pva,Davier:2017zfy,Keshavarzi:2018mgv,Colangelo:2018mtw,Hoferichter:2019gzf,Davier:2019can,Keshavarzi:2019abf,Kurz:2014wya,Melnikov:2003xd,Masjuan:2017tvw,Colangelo:2017fiz,Hoferichter:2018kwz,Gerardin:2019vio,Bijnens:2019ghy,Colangelo:2019uex,Blum:2019ugy,Colangelo:2014qya} and the combined results $a^{\rm exp}_{\mu}$ of Fermilab Muon g-2 experiment~\cite{Abi:2021gix} and the E821 experiment of Brookhaven National Laboratory~\cite{Bennett:2006fi,Aoyama:2020ynm} is 
\begin{equation}\label{eq:amm_mu}
	\Delta a_\mu \equiv a^{\rm exp}_{\mu} - a^{\rm SM}_{\mu} = (251 \pm 59) \times 10^{-11},
\end{equation}
corresponding to a $4.2\sigma$ discrepancy. 
\par There is insufficient evidence to show that these two anomalies 
 are indeed signs of new physics (NP). The discovery level confirmation of $\Delta a_\mu$, for example, requires efforts from the currently running E989 experiment at the Fermilab and the future J-PARC experiment and also progress in reducing the theoretical uncertainty. Providing a common explanation to these two anomalies in an NP model is very challenging. In general, in a complete renormalizable  model, $a_\ell$ can only be a quantum loop effect, because it comes from a dimension-5 operator. In a generic NP model without flavor violation, the new contribution to the anomalous magnetic moment $a_\ell^{\rm NP}$ is proportional to the mass square of the lepton times an NP factor $R^{\rm NP}_\ell$. Taking the central values of the two anomalies in Eqs.~(\ref{eq:amm_e}) and ~(\ref{eq:amm_mu}), one can easily find that there needs to be a difference of about $-15$ between $R^{\rm NP}_e$ and $R^{\rm NP}_\mu$, 
 
\begin{equation}
	\frac{R^{\rm NP}_e}{R^{\rm NP}_\mu} = \frac{m_\mu^2}{m_e^2} \frac{\Delta a_e}{\Delta a_\mu} \sim -15,
\end{equation} 
which is difficult to achieve from a common physical origin. 
\par At present, there have already been several discussions offering combined  explanations of the experimental results for electron and muon anomalous magnetic moments~\cite{Crivellin:2018qmi,Chun:2020uzw,Jana:2020joi,Arbelaez:2020rbq,Dorsner:2020aaz,Botella:2020xzf,Dutta:2020scq,Hati:2020fzp,Yang:2020bmh,Chen:2020jvl,Calibbi:2020emz,Jana:2020pxx,Haba:2020gkr,CarcamoHernandez:2019ydc,Badziak:2019gaf,Bauer:2019gfk,Abdullah:2019ofw,Endo:2019bcj,Crivellin:2019mvj,Dutta:2018fge,Davoudiasl:2018fbb,Han:2018znu,Hernandez:2021tii,Rose:2020nxm,Li:2020dbg,Chua:2020dya,Endo:2020mev,Hiller:2019mou,Gardner:2019mcl,Aebischer:2021uvt,Bodas:2021fsy,Liu:2018xkx,Bigaran:2020jil,Banerjee:2020zvi,Cornella:2019uxs}. Among these discussions, the supersymmetry (SUSY) framework includes a chiral enhancement factor $\tan{\beta}$, which has shown promising results~\cite{Crivellin:2019mvj}. Reference~\cite{Crivellin:2018qmi} argued that the combined explanation in the SUSY framework needs relatively large non-universal trilinear $A$ terms and also requires a flavor violation (for a more detailed discussion, see Refs.~\cite{Giudice:2012ms, Dutta:2018fge}). Due to the constraint from the lepton flavor violating process,  Ref.~\cite{Endo:2019bcj} examined the minimal flavor violation within the minimal supersymmetric standard model (MSSM) and found its compatibility with the Higgs mediation scenario. However, since the value of parameter $\mu$ needs to be at $\mathcal{O}(100~{\rm TeV})$, the parameter space of the explanation in Ref.~\cite{Endo:2019bcj} is unattractive.  
\par More recently, Ref.~\cite{Badziak:2019gaf} argued that in the MSSM without any flavor violation, a combined explanation can be achieved by setting the conditions that $\sgn(M_1 \mu) < 0$ and $\sgn(M_2 \mu) > 0$. The corresponding result features very light selectrons and wino-like charginos, which avoid the Large Hadron Collider (LHC) constraint due to  their degenerate spectra. The solution of Ref.~\cite{Badziak:2019gaf} is impressive, but it also has two unsatisfactory characteristics. One is that the solution prefers heavy Higgsinos with masses $\mu \sim \mathcal{O}(1~{\rm TeV})$. This leads to a relatively fine-tuned electroweak sector. In general, $\mu$ should be close to the $Z$ boson mass $m_Z$ to avoid large cancellation when predicting the observed value of $m_Z = 91.2~{\rm GeV}$~\cite{Baer:2020sgm,Cao:2018rix,Dimopoulos:1995mi,Giudice:2006sn,Baer:2012uy}. $\mu \sim 1~{\rm TeV}$ often induces tuning on the order of $1/10000$ to predict $m_Z$. The other is that wino-like particles are too light due to the current restrictions of the LHC direct SUSY searches. The wino exclusion planes reported by ATLAS and CMS within the simplified model framework are appropriate for the scenario of Ref.~\cite{Badziak:2019gaf}. According to Fig.~8 in the CMS report~\cite{Sirunyan:2018ubx}, for example, the benchmark points in Ref.~\cite{Badziak:2019gaf} with $M_2 \sim 200~{\rm GeV}$ are on the verge of being excluded by the multi-lepton plus $E_{\rm T}^{\rm miss}$ signal via the electroweakino channel $pp\to \tilde{\chi}^\pm_1 \tilde{\chi}^0_2$. 
\par In our previous work~\cite{Cao:2019evo}, we investigated the observation that in the inverse-seesaw mechanism extended next-to-minimal supersymmetric standard model (ISS-NMSSM), due to the $\mathcal{O}(0.1)$ level Yukawa coupling $Y_\nu$ of the Higgs field to the right-handed neutrino, the Higgsino--sneutrino (HS) loop can be a new source of $a_\mu$ to explain $\Delta a_\mu$. Unlike the MSSM, the newly introduced HS contribution $a_\ell^{\rm HS}$ in the ISS-NMSSM prefers a light $\mu$. The sign of $a_\ell^{\rm HS}$ is determined by the mass mixing effect of sneutrino fields $\tilde{\nu}_{L}^{\ell}$, $\tilde{\nu}_{R}^{\ell}$, and $\tilde{\nu}_{x}^{\ell}$ for a given flavor $\ell$, not by the mixing of charginos or neutralinos. In the ISS-NMSSM explanation, the masses of wino-like particles can be much heavier than the current LHC bounds. One can also assume one generation of sneutrinos to be the lightest supersymmetric particles (LSPs), which act as a dark matter (DM) candidate coannihilating with Higgsinos to achieve the observed relic density. Due to the singlet nature, the DM-nucleus scattering cross section is naturally suppressed below the current experimental detection limits~\cite{Cao:2019qng, Cao:2019aam, Cao:2019evo}. In this case, the neutral Higgsinos are the next-to-lightest supersymmetric particles (NLSPs) that decay into the invisible final states of the collider ($\tilde{H}^0 \to \nu \tilde{\nu}$). The charged Higgsino decays into a soft charged lepton and DM ($\tilde{H}^\pm \to \ell^{\pm} \tilde{\nu}$). Due to the lower production rate than that of winos and the degenerate mass spectrum, the current LHC data still allow a low Higgsino mass of around $100~\rm GeV$. Thus, compared with the MSSM framework, the ISS-NMSSM is more natural for providing common explanations for $\Delta a_e$ and $\Delta a_\mu$.
\par In this work, we investigate this issue by applying the ISS-NMSSM to explain $\Delta a_e$ and $\Delta a_\mu$. The remainder of this paper is organized as follows. First,  we briefly introduce the ISS-NMSSM and the properties of leptonic $g-2$ in Sec~\ref{sec:basic}. We then scan the parameter space that explains both the electron and muon $g-2$ discrepancies and analyze the characteristics of the input parameters and particle mass spectrum in Sec.~\ref{sec:scan}. In Sec.~\ref{sec:dm}, we find that our scenario can be embedded into a $\tau$-type sneutrino, which co-annihilates with a Higgsino to achieve the observed DM relic density and does not conflict with the current DM direct search observations. In Sec.~\ref{sec:collider}, we show the impact of the current LHC SUSY particle direct searches. Finally, we draw conclusions in Sec.~\ref{sec:summary}. 

\section{\label{sec:basic}Inverse seesaw mechanism extended next-to-minimal supersymmetric standard model and the lepton \texorpdfstring{$g-2$}{}}
\subsection{Brief introduction to inverse seesaw next-to-minimal supersymmetric standard model}
The complete definition of the ISS-NMSSM Lagrangian, such as quantum number setting, can be found in Ref.~\cite{Cao:2017cjf}. Here, we only briefly introduce the basic idea of the ``inverse-seesaw'' extension and the neutrino sector. 
\par The ``inverse-seesaw'' mechanism is added to the NMSSM framework by introducing two gauge singlet superfields $\hat{\nu}$ and $\hat{X}$ with opposite lepton numbers $L=-1$ and $L=1$, respectively \cite{Abada:2010ym}. With the assumptions of $R$-parity conservation, not introducing the $\Delta L = 1$ lepton number violation, the superpotential $W$ is given as follows:
\begin{equation}\label{eq:sp}
\begin{split}
	W &= Y_u\hat{Q} \cdot \hat{H}_u \hat{u} + Y_d \hat{H}_d \cdot \hat{Q} \hat{d} + Y_e \hat{H}_d \cdot \hat{L} \hat{e} + \lambda \hat{S} \hat{H}_u\cdot\hat{H}_d + \frac{\kappa}{3}\hat{S}^3\\
      &+ \frac{1}{2} \mu_X \hat{X}\hat{X} + \lambda_N \hat{S} \hat{\nu} \hat{X} + Y_{\nu} \hat{L}\cdot\hat{H}_u \hat{\nu}.      
\end{split}
\end{equation}
The first line of Eq.~(\ref{eq:sp}) is the standard NMSSM superpotential. The soft breaking terms of the ISS-NMSSM are given as follows:
\begin{equation}
\begin{split}
       V_{\rm soft} &= V_{\rm NMSSM} \\
       &+ M_{\nu}^2 \tilde{\nu}_R \tilde{\nu}_R^* + M_X^2 \tilde{x}\tilde{x}^* \\
       &+ \left(\frac{1}{2}B_{\mu_X} \tilde{x}\tilde{x} + \lambda_N A_{N} S \tilde{\nu}_R^* \tilde{x} + Y_{\nu}A_{\nu} \tilde{\nu}_R^* \tilde{L}\cdot H_u + {\rm H.c.}\right),
\end{split}
\end{equation} 
where $V_{\rm NMSSM}$ is the NMSSM soft breaking term, and $\tilde{\nu}_R$ and $\tilde{x}$ are the scalar parts of superfields $\hat{\nu}$ and $\hat{X}$, respectively. The dimensional parameter $\mu_X$ is a small $X$-type neutrino mass term, which often is treated as an effective mass parameter to obtain the tiny masses of active neutrinos. The introduction of $\mu_X$ violates the lepton number due to the $\Delta L=2$ term $\mu_X \hat{X}\hat{X}$ and the $\mathbb{Z}_3$ symmetry of the superpotential. 

	After the electroweak symmetry breaking, the $9\times 9$ complex and symmetric neutrino mass matrix $M_{\rm ISS}$  in the basis $(v_L, v_R^*, x)$ reads
	\begin{equation}
		M_{\rm ISS} = \begin{pmatrix}
			0 & m_D^T & 0 \\
			m_D & 0 & m_R \\
			0 & m_R^T & \mu_X 
		\end{pmatrix}
	\end{equation}	
	where $m_{D}=\frac{1}{\sqrt{2}} Y_{\nu} v_u$ and $m_{R}=\frac{1}{\sqrt{2}} \lambda_N v_s$ are the $3 \times 3$ neutrino Dirac mass matrices. The $M_{\rm ISS}$ can be diagonalized by a $9 \times 9$ unitary matrix $U_\nu$ according to 
	\begin{equation}
		U_\nu^* M_\nu U_\nu^\dag = {\rm diag}(m_{\nu_i}, m_{\nu_{hj}}), \quad i=1,2,3, \quad j=1,2,\cdots , 6. 
	\end{equation}
	This gives three active neutrino masses $m_{\nu_i}$ and six heavy neutrino masses $m_{\nu_{hj}}$. 
	$M_{\rm ISS}$ can be diagonalized by block to give the $3 \times 3$ light neutrino mass matrix $M_\nu$. Under the inverse seesaw limits $||\mu_X || \ll || m_D ||, || m_R||$, 
	\begin{equation}
		M_{\nu} \simeq m_D^T {m_R^T}^{-1} \mu_X m_R^{-1} m_D. 
	\end{equation}
	This $M_\nu$ is diagonalized by the unitary Pontecorvo-Maki-Nakagawa-Sakata (PMNS) matrix $U_{\rm PMNS}$: 
	\begin{equation}
		m_{\nu}^{\rm diag} \equiv {\rm diag}(m_{\nu_1}, m_{\nu_2}, m_{\nu_3}) = U^{T}_{\rm PMNS} M_{\nu} U_{\rm PMNS}
	\end{equation}
	 In this work, for our phenomenological purpose, we use the $\mu_X$-parametrization scheme that was introduced in \cite{Arganda:2014dta}\footnote{A detailed discussion of $\mu_X$-parametrization can also be found in \cite{Baer:2016ucr}. In principal, $Y_\nu$ in the $\mu_X$-parametrization can be arbitrarily large except for that the perturbativity of the theory requires  $|Y_\nu|^2/4\pi \leq 1.5$. }$^,$\footnote{The neutrino oscillation data can also be reconstruct by the Casas-Ibarra parametrization, in which the neutrino oscillations are generated by the off-diagonal term in $m_D$~\cite{Casas:2001sr}. Because of the special role of parameter $Y_\nu$ in this work, the $\mu_X$ parametrization scheme is more intuitive than the Cassas-Ibarra parametrization scheme. } to reproduce low-energy neutrino data 
	\begin{equation}\label{eq:muxpara}
	\mu_X = m_R^{T} {m_{D}^{T}}^{-1} U^{*}_{\rm PMNS} m_{\nu}^{\rm diag} U^{\dag}_{\rm PMNS} {m_{D}}^{-1} m_R. 
	\end{equation}
	$Y_\nu$ and $\lambda_N$ in this work are assumed to be flavor diagonal, so the neutrino oscillation data is attributed only to the flavor nondiagonal parameter $\mu_X$. In this case, the unitary constraint of neutrino sector can be translated into a constraint on the input parameters \cite{Cao:2019aam}
	\begin{equation}
	\label{eq:uc}
		\frac{\lambda_{N_{e}}}{Y_{\nu_e}} \frac{\mu}{\lambda v_u} > 14.1, \quad
		\frac{\lambda_{N_\mu}}{Y_{\nu_{\mu}}} \frac{\mu}{\lambda v_u} > 33.7, \quad
		\frac{\lambda_{N_\tau}}{Y_{\nu_\tau}} \frac{\mu}{\lambda v_u} > 9.4.
	\end{equation} 
	These inequalities indicate that, for given $\lambda_N$ and the Higgs sector parameters $\lambda$, $\tan{\beta}$, and $\mu$, this unitary constraint allows  $Y_{\nu_e}$ to be greater than $Y_{\nu_\mu}$, which is good for explaining $\Delta a_e$. 
	
\par The ISS mechanism preserve an approximate lepton number conservation. Equation~(\ref{eq:muxpara}) indicates that the light neutrino masses require the magnitude of lepton number breaking parameter $\mu_X$ to be highly suppressed. Numerically, both $\mu_X$ and the soft breaking mass term $B_{\mu_X}$ are extremely small ($|\mu_X| \lesssim \mathcal{O} ({\rm KeV})$ and $|B_{\mu_X}| \lesssim \mathcal{O}(100~{\rm GeV}^2)$), and they just slightly split the complex sneutrino field into the $CP$-even part and $CP$-odd part. When studying $a_\ell$, the influence of these two nonvanishing parameters can be ignored \cite{Cao:2019evo}. Because the dimension of the scan parameters is too high, for the sake of simplicity, we simply set the values of $\mu_X$ and $B_{\mu_X}$ to zero in the following discussion.

\par  The SUSY particles of particular importance to $a_\ell$ are sleptons $\tilde{\ell}$, $\ell$-type sneutrinos, neutralinos $\tilde{\chi}_i^0$, and charginos $\tilde{\chi}_i^\pm$.  The neutralino and chargino sector in the ISS-NMSSM were same as that of the NMSSM. On the basis of $\phi^0 = (\tilde{B}, \tilde{W}^3, \tilde{H}_d^0, \tilde{H}_u^0, \tilde{S} )^{T}$, the symmetric neutralino mixing matrix $\mathcal{M}_0$ is as follows:

\begin{equation}\label{eq:neutramass}
	\mathcal{M}^0	= \begin{pmatrix}
		M_1 & 0 & -\frac{g_1}{\sqrt{2}} v_d & \frac{g_1}{\sqrt{2}}v_u & 0\\
		& M_2 & \frac{g_2}{\sqrt{2}} v_d & -\frac{g_2}{\sqrt{2}} v_u & 0 \\
		& & 0 & -\mu & -\lambda v_u \\ 
		& & & 0 & -\lambda v_d \\
		& & & & 2\kappa v_s 
	\end{pmatrix}, 
\end{equation}
where the Higgsino mass $\mu = \lambda v_s$ is an effective $\mu$ term after the electroweak symmetry breaking, and $v_u$, $v_d$, and $v_s$ represent the vacuum expectation values (VEVs) of the Higgs field $H_u$, $H_d$, and S, respectively. The mass eigenstates $\tilde{\chi}_i^0 = N_{ij}\phi_j^0$ are arranged in ascending order of mass. With the basis $\phi^+ = (\tilde{W}^+, \tilde{H}_u^+)$ and $\phi^- = (\tilde{W}^-, \tilde{H}_d^-)$, the chargino mass term is given by $\phi^- \mathcal{M}^{\pm} \phi^+ + {\rm H.c.}$ with the mass matrix
 
\begin{equation}
	\mathcal{M}^{\pm} = \begin{pmatrix}
		M_2 & g_2 v_u \\
		g_2 v_d  & \mu 
	\end{pmatrix}.  
\end{equation}
     The corresponding mass eigenstates are defined by 
     
\begin{equation}
	\tilde{\chi}_i^+ = V_{ij} \phi_j^+ , \quad \tilde{\chi}_i^- = U_{ij} \phi_j^-. 
\end{equation} 
The symmetric mass matrix $\mathcal{M}_{\tilde{\ell}}^2$ for slepton $\tilde{\ell}$ for each flavor $\ell$ in the $(\tilde{\ell}_L , \tilde{\ell}_R)$ basis is 

\begin{equation}
	\mathcal{M}_{\tilde{\ell}}^2 = \begin{pmatrix}
		M_{L_\ell}^2 + m_\ell^2 + m_Z^2 \cos{2\beta}(\sin^2{\theta_W} -\frac{1}{2})  & m_\ell(A_{E_\ell} - \mu \tan{\beta})\\
		  & M_{E_\ell}^2 - \sin^2{\theta_W} m_Z^2 \cos{2\beta}   
	\end{pmatrix},
\end{equation}
and the corresponding rotation matrix is represented by $X^{\ell}$. 
\par In terms of the particle composition, the ISS-NMSSM differs from the NMSSM~\cite{Ellwanger:2009dp} only in the neutrino sector. For each generation $\ell = e,\mu, \tau$, the sneutrino fields are the mixtures of left-handed sneutrino $\tilde{\nu}_L^\ell$, right-handed sneutrino $\tilde{\nu}_R^\ell$, and $x$-type sneutrino $\tilde{\nu}_x^{\ell}$. On the basis of $\phi_\nu^\ell = (\tilde{\nu}_L^\ell, \tilde{\nu}_R^\ell, \tilde{\nu}_x^\ell)$, the symmetric mass matrix $\mathcal{M}_{\tilde{\nu}}^2$ is given by 

\begin{small}
\begin{equation}\label{eq:snmass}
	\mathcal{M}_{\tilde{\nu}}^2 = \frac{1}{2}\begin{pmatrix}
	2 M_{L_\ell}^2 +  Y_{\nu_\ell}^2 v_u^2 +  m_Z^2 \cos{2\beta}~	& Y_{\nu_\ell} (\sqrt{2} A_{\nu_\ell} v_u - \lambda v_d v_s)~	&  Y_{\nu_\ell} \lambda_{N_\ell} v_u v_s \\
		& 2 M_{\nu_\ell}^2 +  Y_{\nu_\ell}^2 v_u^2 +  \lambda_{N_\ell}^2 v_s^2 ~&  \lambda_{N_\ell} (\sqrt{2} A_{N_\ell} v_s + \kappa v_s^2 - \lambda v_u v_d) \\
	 	&  & 2 M_{X_\ell}^2 + \lambda_{N_\ell}^2 v_s^2
	\end{pmatrix}.
\end{equation}
\end{small}
The mass eigenstate of one generation sneutrino is $\tilde{\nu}_i^\ell = \sum_j Z_{ij}^\ell \phi_{\nu,j}^\ell$, with $Z^\ell$ denoting the unitary matrix to diagonalize $\mathcal{M}_{\tilde{\nu}}^2$. From Eq.~(\ref{eq:snmass}), one can easily find that the diagonal elements can be adjusted by the soft breaking parameters $M_{L_\ell}^2$, $M_{\nu_\ell}^2$, and $M_{X_\ell}^2$. For the off-diagonal elements, Higgs VEV terms, such as $Y_{\nu_\ell} v_u$ and $\lambda_{N_\ell} v_s$, provide a scale for the mixing of the three fields of left-handed, right-handed, and $x$-type sneutrinos. The relative signs and the magnitude of different sneutrino components can be adjusted by two $A$-term soft breaking parameters $A_{\nu_\ell}$ and $A_{N_\ell}$.  

\subsection{Lepton anomalous magnetic moment in ISS-NMSSM}
The lepton anomalous magnetic moment $a_\ell$ always corresponds to lepton chirality-flipping interactions. In the ISS-NMSSM, the chirality of the $\ell$-lepton number can be flipped by $Y_{e_\ell}$ or $Y_{\nu_\ell}$. All the SM-like diagrams (the $\ell$-lepton number is carried only by lepton $\ell$ and/or neutrino $\nu_\ell$) involve only SM particles, so their contribution to $a_\ell$ is identical to the SM prediction $a_\ell^{\rm SM}$. Therefore, the SUSY contribution $a_\ell^{\rm SUSY}$, in which the $\ell$-lepton number is carried also by a scalar lepton $\tilde{\ell}$ and/or $\ell$-type sneutrino $\tilde{\nu}_\ell$,  provides the source of the observed anomaly $\Delta a_\ell$. 

\par The general one-loop SUSY contribution to $a_\ell$ in the ISS-NMSSM is given as follows~\cite{Cao:2019evo}: 
\begin{equation}\label{eq:damufull}
\begin{split}
    &a_{\ell}^{\rm SUSY} = a_{\ell}^{\tilde{\chi}^0 \tilde{\ell}} + a_{\mu}^{\tilde{\chi}^{\pm} \tilde{\nu}},\\
    a_{\ell}^{\tilde{\chi}^0 \tilde{\ell}} &= \frac{m_{\ell}}{16 \pi^2}\sum_{i,l}\left\{
    -\frac{m_{\ell}}{12 m_{\tilde{\ell}_l}^2} \left( |n_{il}^{\rm L}|^2 + |n_{il}^{\rm R}|^2 \right) F_1^{\rm N}(x_{il}) + \frac{m_{\tilde{\chi}_i^0}}{3 m_{\tilde{\ell}_l}^2} {\rm Re}(n_{il}^{\rm L} n_{il}^{\rm R}) F_2^{\rm N}(x_{il})
    \right\},\\
    a_{\ell}^{\widetilde{\chi}^\pm \tilde{\nu}} &= \frac{m_{\ell}}{16 \pi^2}\sum_{j,n}\left\{
    \frac{m_{\ell}}{12 m_{\tilde{\nu}_{\ell,n}}^2} \left( |c_{jn}^{\rm L}|^2 + |c_{jn}^{\rm R}|^2 \right) F_1^{\rm C}(x_{jn}) + \frac{2 m_{\widetilde{\chi}_j^\pm}}{3 m_{\tilde{\nu}_{\ell,n}}^2} {\rm Re}(c_{jn}^{\rm L}c_{jn}^{\rm R}) F_2^{\rm C}(x_{jn})
    \right\}.
\end{split}
\end{equation}
Here, $i = 1, \cdots, 5$ and $j=1,2$, respectively, denote the neutralino and chargino indices, $l=1,2$ denotes the slepton index, $n = 1,2,3$ denotes the sneutrino index, and 

\begin{equation}\label{eq:coupl}
	   \begin{split}
        n_{il}^{\rm L} = \frac{1}{\sqrt{2}}\left( g_2 N_{i2} + g_1 N_{i1} \right)X^{\ell,*}_{l1} -Y_{e_\ell} N_{i3}X^{\ell,*}_{l2}, \quad
        &n_{il}^{\rm R} = \sqrt{2} g_1 N_{i1} X^\ell_{l2} + Y_{e_\ell} N_{i3} X^\ell_{l1},\\
        c_{jn}^{\rm L} =   -g_2 V_{j1} Z_{n1}^{\ell, *} + Y_{\nu_\ell} V_{j2} Z_{n2}^{\ell, *}, \quad
        &c_{jn}^{\rm R} =  Y_{e_\ell} U_{j2} Z_{n1}^\ell.
    \end{split}
\end{equation}
 
The kinematic loop function $F(x)$ is normalized with condition $F(1)=1$, which is given as follows:

\begin{equation}\label{eq:loopfunc}
    \begin{split}
        F_1^{\rm N}(x) &= \frac{2}{(1-x)^4} \left( 1 - 6x + 3x^2 + 2x^3 - 6x^2 \ln{x} \right),\\
        F_2^{\rm N}(x) &= \frac{3}{(1-x)^3} \left( 1 - x^2 + 2x \ln{x} \right),\\
        F_1^{\rm C}(x) &= \frac{2}{(1-x)^4} \left( 2 + 3x - 6x^2 + x^3 + 6x \ln{x} \right), \\
        F_2^{\rm C}(x) &= -\frac{3}{2(1-x)^3} \left( 3 - 4x + x^2 + 2\ln{x} \right),
    \end{split}
\end{equation}
with definitions $x_{il} \equiv m_{\tilde{\chi}^0_i}^2 / m_{\tilde{\ell}_l}^2$ and $x_{jn} \equiv m_{\tilde{\chi}^\pm_j}^2 / m_{\tilde{\nu}_n}^2$.

\begin{figure}[ht]
\centering
\includegraphics{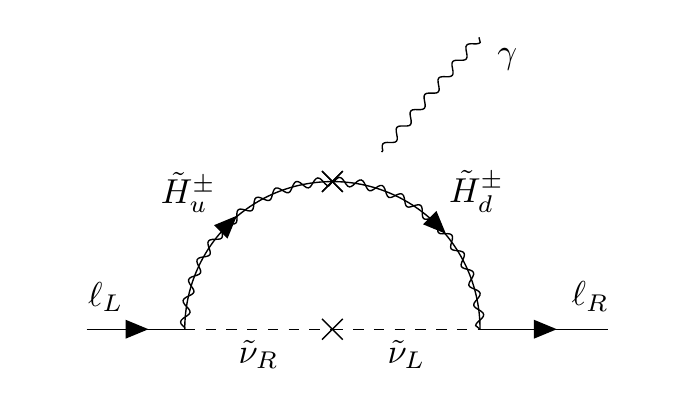}
\caption{\label{fig:feynloop}One-loop diagram of Higgsino--sneutrino contribution, the additional contribution to $a_{\ell}^{\rm SUSY}$ in the ISS-NMSSM compared to the MSSM.}
\end{figure}

\par Comparing with the MSSM contribution to $a_\ell$ given in \cite{Martin:2001st}, the ISS-NMSSM contribution contains an extra HS term. As depicted in Fig.~1, it is easy to understand the leading behaviour with the help of diagram that are written in terms of interaction eigenstates, where the insertions of mass and mixing terms and lepton number chirality flips. Accordingly, this HS contribution $a_\ell^{\rm HS}$ can be approximately expressed as  
	\begin{equation}\label{eq:ahs}
	a_{\ell}^{\rm HS} \approx m_\ell^2 \times \frac{1}{48\pi^2 v}\frac{Y_{\nu_\ell}}{\mu \cos{\beta}}
	\left\{  \sum_n Z^{\ell}_{n1}Z^{\ell}_{n2} \times x_n F_2^C (x_n) \right\}, \quad x_n \equiv \frac{\mu^2}{m_{ \tilde{\nu}_{\ell, n} }^2},
	\end{equation}
	where $v=\sqrt{v_u^2 + v_d^2} = 173~{\rm GeV}$, and the function $xF_2^C(x)$ increases monotonically with $x$, and its value ranging from 0 to 1.5. Equation~(\ref{eq:ahs}) shows that $a_{\ell}^{\rm HS}$ is enhanced by the factor $\cos{\beta}\approx1/\tan{\beta}$ for large $\tan{\beta}$. A relatively large $Y_{\nu_\ell}$ is expected to achieve a larger $a_\ell^{\rm HS}$. The lepton chirality flipping in the HS contribution is reflected in the left--right mixing term $Z_{n1} Z_{n2}$ for each sneutrino $\tilde{\nu}_n$. If the $x$-field-dominated sneutrino is too heavy, then the contribution from the left-handed dominated sneutrino and the right-handed dominated state to $a_\ell$ cancel each other because $Z^\ell_{11}Z^\ell_{12} \approx - Z^\ell_{21}Z^\ell_{22}$. This cancellation can be alleviated by rendering a sizable mixing between $x$-field and the right-handed sneutrino field, which, as shown in Eq.~(\ref{eq:snmass}), can be obtained by choosing a large mixing term $\lambda_N v_s$. Furthermore, A light $\mu$ is favored for large $a_\ell^{\rm HS}$, so the small $\lambda$ at $\mathcal{O}(0.01)$ often has a larger $a_\ell^{\rm HS}$. This small $\lambda$ is also preferred by the leptonic unitary condition in Eq.~(\ref{eq:uc}).
		
\par Looking back to the MSSM explanation to both anomalies, the first difficulty is that the most of the relevant parameters (except the lepton soft trilinear breaking term $A_e$) are lepton flavor independent. Specifically, the full expression of $a_\ell^{\rm SUSY}$ in MSSM is very similar to that in ISS-NMSSM, and the difference is that $c_L$ in Eq.~(\ref{eq:coupl}) of MSSM does not contain the $Y_{\nu_\ell} V_{j2}$ term. The $\mu$ parameter governs the Higgsino $\tilde{H}_u - \tilde{H_d}$ transition and the dominant part of scalar lepton mixing term, so that the MSSM contribution is proportional to $\mu \tan{\beta}$. Compared with $\mu\tan{\beta}$ , $A_e$ is usually ignored in the left-handed and right-handed scalar lepton mass mixing. The gaugino mass parameters $M_1$ and $M_2$ affect the sign of MSSM contribution to $a_\ell$. Thus, in general, $a_\mu^{\rm SUSY}$ and $a_e^{\rm SUSY}$ are highly correlated in MSSM. The second difficulty comes from the correlation between $a_\ell^{\tilde{\chi}^0 \tilde{\ell}}$ and $a_\ell^{\tilde{\chi}^\pm \tilde{\nu}}$. In MSSM, the mass of the left-handed slepton $\tilde{\ell}_L$ is slightly larger than the mass of sneutrino $\tilde{\nu}^\ell$ by about $15~{\rm GeV}$ for large $\tan{\beta}$. As a result, the wino-Higgsino loop in $a_\ell^{\tilde{\chi}^0 \tilde{\ell}}$ is about $-1/2$ times that in $a_\ell^{\tilde{\chi}^\pm \tilde{\nu}}$. 

\par The most attractive property of the HS term in explaining both anomalies in ISS-NMSSM is that the left-right mixing $Z_{n1}Z_{n2}$ is positively related to $A_\nu$, which is lepton flavor relevant. The magnitude and sign of the HS contribution can be adjusted by $A_\nu$. Numerically, an $|A_\nu|$ at $\mathcal{O}(100~{\rm GeV})$ to $\mathcal{O}(1~{\rm TeV})$ is sufficient for $a_\ell^{\rm HS}$ to explain both anomalies. Furthermore, the weakening of mass relevance between left-handed dominated slepton $\tilde{\ell}_L$ and left-handed dominated sneutrino $\tilde{\nu}^\ell_L$ gives ISS-NMSSM parameter space more room for interpreting both anomalies.

\section{\label{sec:scan}Features of combined explanation to \texorpdfstring{$\Delta a_e$}{} and \texorpdfstring{$\Delta a_\mu$}{} in ISS-NMSSM}
The relevant SUSY particle masses have different contributions to $a_{\ell}^{\rm SUSY}$. To reveal the detailed features of $a_{\ell}^{\rm SUSY}$ in the ISS-NMSSM, which is covered up by the complexity of the loop functions $F(x)$ in Eq.~(\ref{eq:loopfunc}), we use the \textsc{MultiNest} technique~\cite{Feroz:2008xx, Feroz:2013hea} to scan the ISS-NMSSM parameter space with the following parameter values and ranges: 

\begin{equation}\label{eq:paras}\begin{split}
	&0< \lambda < 0.7, \quad |\kappa|<0.7, \quad 100~{\rm GeV} < \mu < 500~{\rm GeV}, \quad \tan{\beta} = 60,\\
	&A_{\lambda} = 2~{\rm TeV}, \quad -1~{\rm TeV}< A_{\kappa} < 1~{\rm TeV},\quad -5~{\rm TeV} < A_t = A_b < 5~{\rm TeV},\\  
	&|M_1| < 1~{\rm TeV}, \quad |M_2| < 1.5~{\rm TeV},\\
	&100~{\rm GeV} < M_{L_e} < 1~{\rm TeV}, \quad 100~{\rm GeV} < M_{E_e} < 1~{\rm TeV}, \quad A_{E_e} = 0\\
	&0< Y_{\nu_e} < 1.0,\quad |A_{\nu_e}| < 2~{\rm TeV},  \quad 0 < \lambda_{N_e} < 1.0, \quad |A_{N_e}| < 2~{\rm TeV}, \\
	& |M_{\nu_e}^2| < 1~{\rm TeV}^2, \quad |M_{X_e}^2| < 1~{\rm TeV}^2, \\
	&100~{\rm GeV} < M_{L_\mu} < 1~{\rm TeV}, \quad 100~{\rm GeV} < M_{E_\mu} < 1~{\rm TeV}, \quad A_{E_\mu} = 0\\
	&0< Y_{\nu_\mu} < 1.0,\quad |A_{\nu_\mu}| < 2~{\rm TeV},  \quad 0 < \lambda_{N_\mu} < 1.0, \quad |A_{N_\mu}| < 2~{\rm TeV}, \\
	& |M_{\nu_\mu}^2| < 1~{\rm TeV}^2, \quad |M_{X_\mu}^2| < 1~{\rm TeV}^2, \\
	\end{split} 	
\end{equation} 
where all parameters are defined at the scale of $1~{\rm TeV}$. All the other soft breaking parameters, like those first two generation squark, are fixed at a common value of $3~{\rm TeV}$. The prior probability distribution function (PDF) of these inputs are set as uniform distributions, and the $n_{\rm live}$ parameter, which indicates the number of the active points to determine the iso-likelihood contour in the $\textsc{MultiNest}$ algorithm iteration, is set at 10000. The likelihood function $\chi^2$ in the scan is taken as 

\begin{equation}
	\chi^2 = \chi^2_{a_\ell} + \chi^2_{\rm Higgs} + \chi^2_{\rm veto}, 
\end{equation}
where $\chi^2_{a_\ell} = \frac{1}{2} \left(\frac{a_e^{\rm SUSY} + 87\times10^{-14}}{36\times 10^{-14} }\right)^2 + \frac{1}{2} \left( \frac{a_\mu^{\rm SUSY} - 267\times10^{-11}}{76\times10^{-11}} \right)^2$ is a standard Gaussian form of two anomalies. $\chi^2_{\rm Higgs}$ requires the sample in the parameter space given by~Eq.~(\ref{eq:paras}) to predict an SM-like Higgs boson compatible with current experimental observations using the \textsc{HiggsSignals-2.2.3} code~\cite{Bechtle:2013xfa, Stal:2013hwa, Bechtle:2014ewa} and to satisfy the constraints of a direct search of the Higgs boson using the \textsc{HiggsBounds-5.3.2} code~\cite{Bechtle:2013wla}. $\chi^2_{\rm veto}$ is introduced to ensure that the LSP is a Higgsino, the electroweak vacuum is stable, sneutrino fields have not developed nonzero VEVs, and the unitary constraint of Eq.~(\ref{eq:uc}) is satisfied. If the parameter point satisfies the above assumption, $\chi^2_{\rm veto}$ is equal to 0; otherwise, $\chi^2_{\rm veto} = 10000$.  Technically, the introduction of $\chi^2_{\rm veto}$ highlights the characteristics of the parameter space of ISS-NMSSM explaining the two anomalies, and minimizes the influence of other phenomenological constraints on the statistics of the scanning result. After several repeated scans, we checked that the statistical distribution of the parameters in the results is reproducible.

\begin{figure}[h]
\makebox[\textwidth][c]{
\includegraphics[width=0.28\paperwidth]{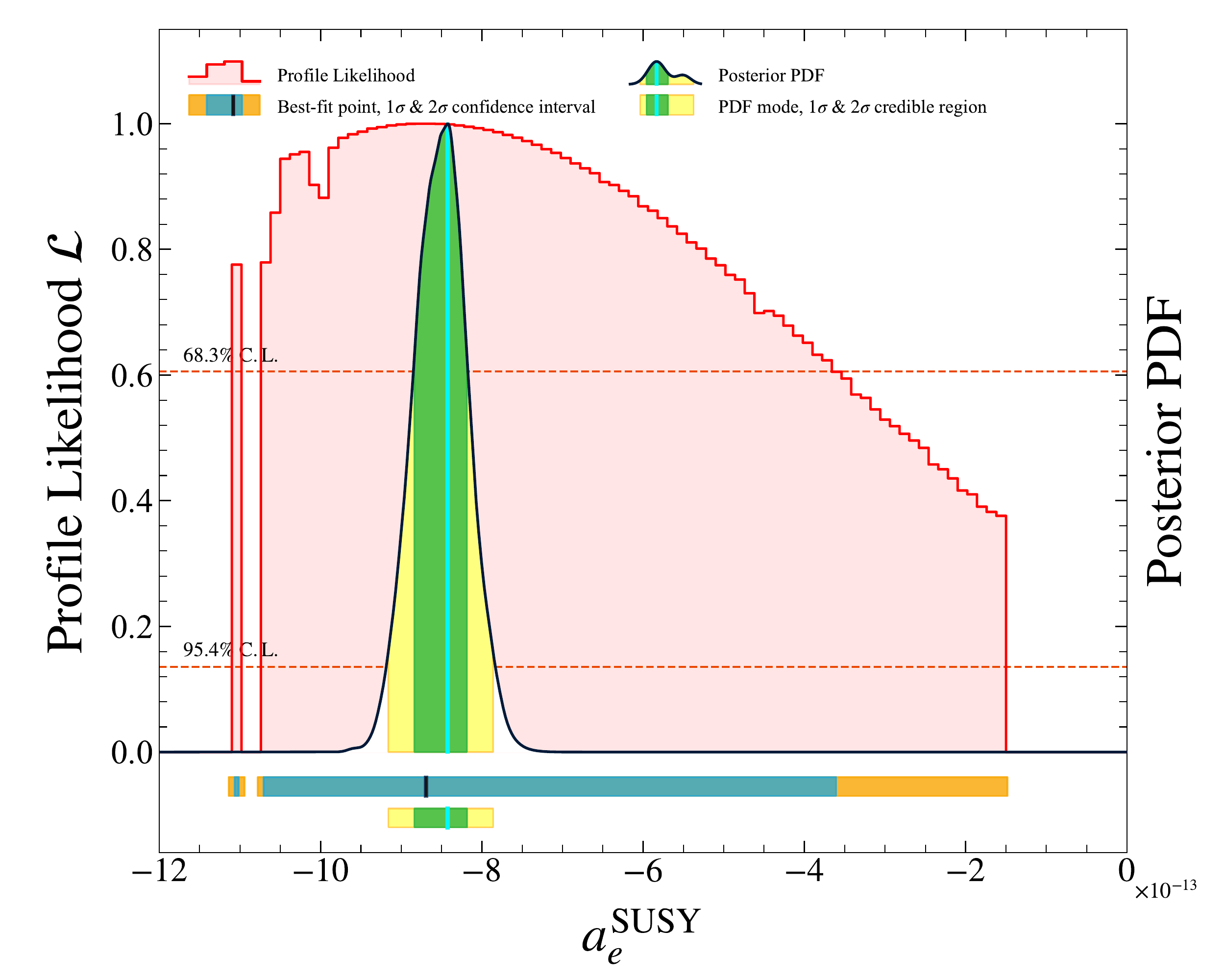}\hspace{0.5cm}
\includegraphics[width=0.28\paperwidth]{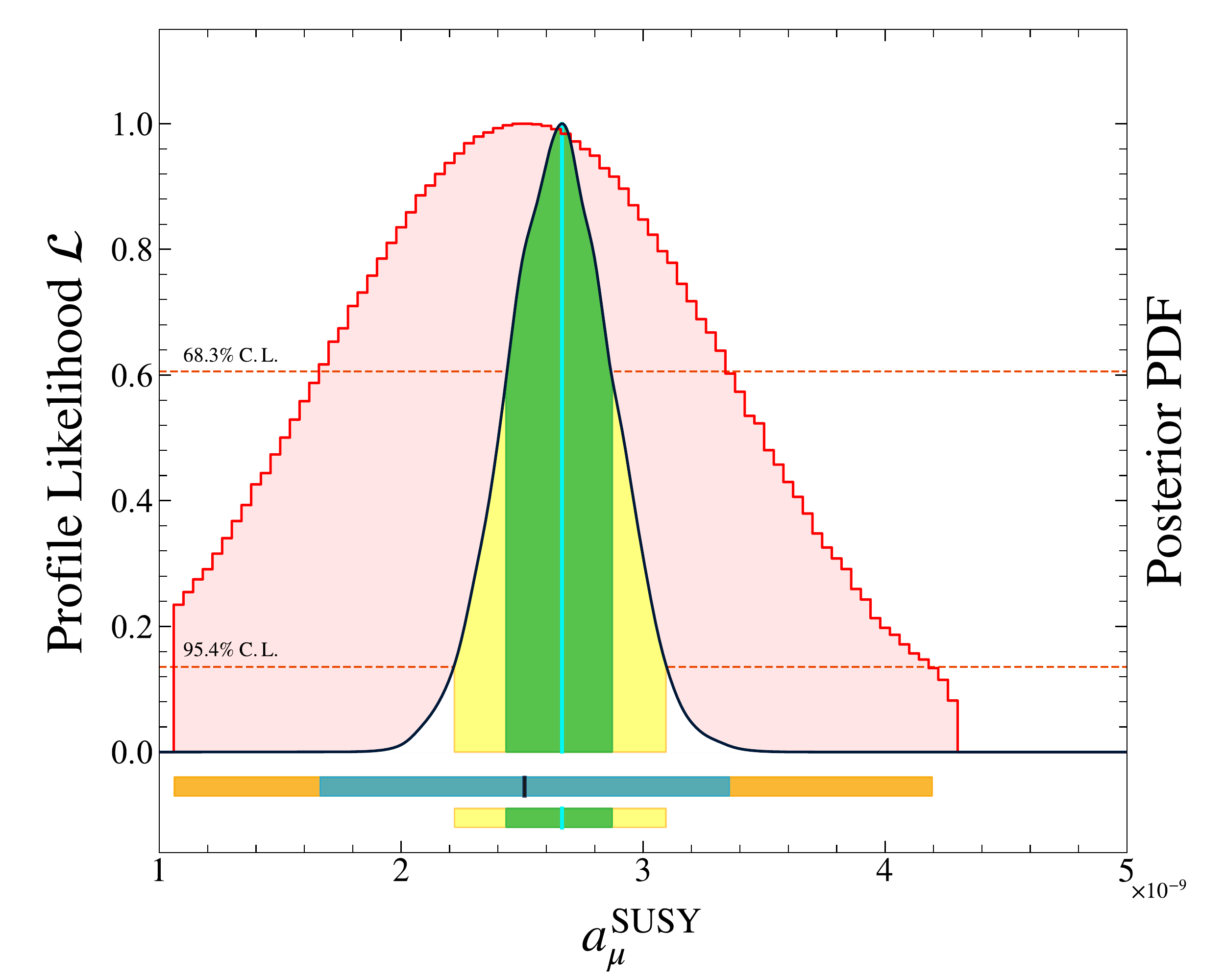}}
\vspace{-0.1cm}
\makebox[\textwidth][c]{
\includegraphics[width=0.28\paperwidth]{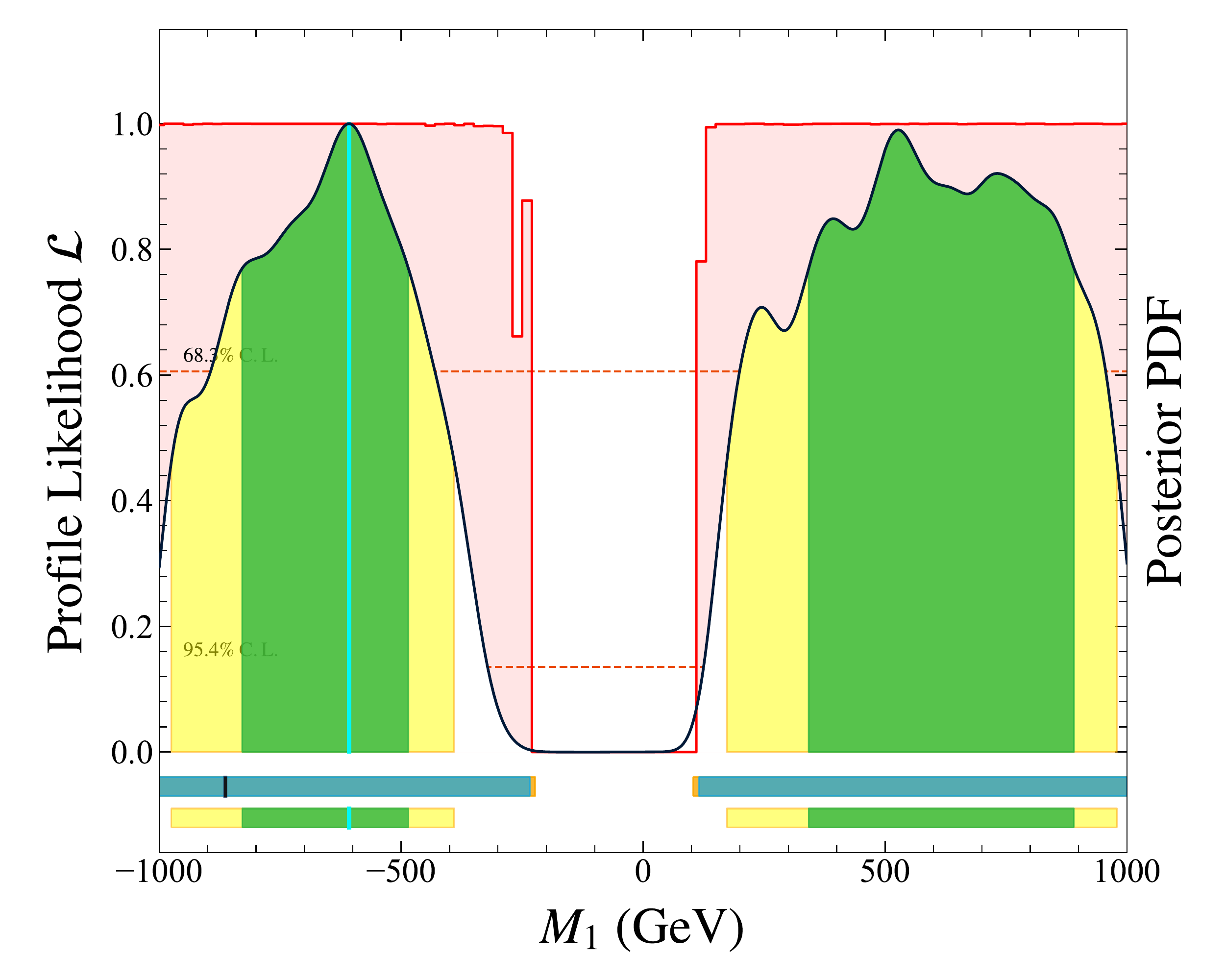}\hspace{0.5cm}
\includegraphics[width=0.28\paperwidth]{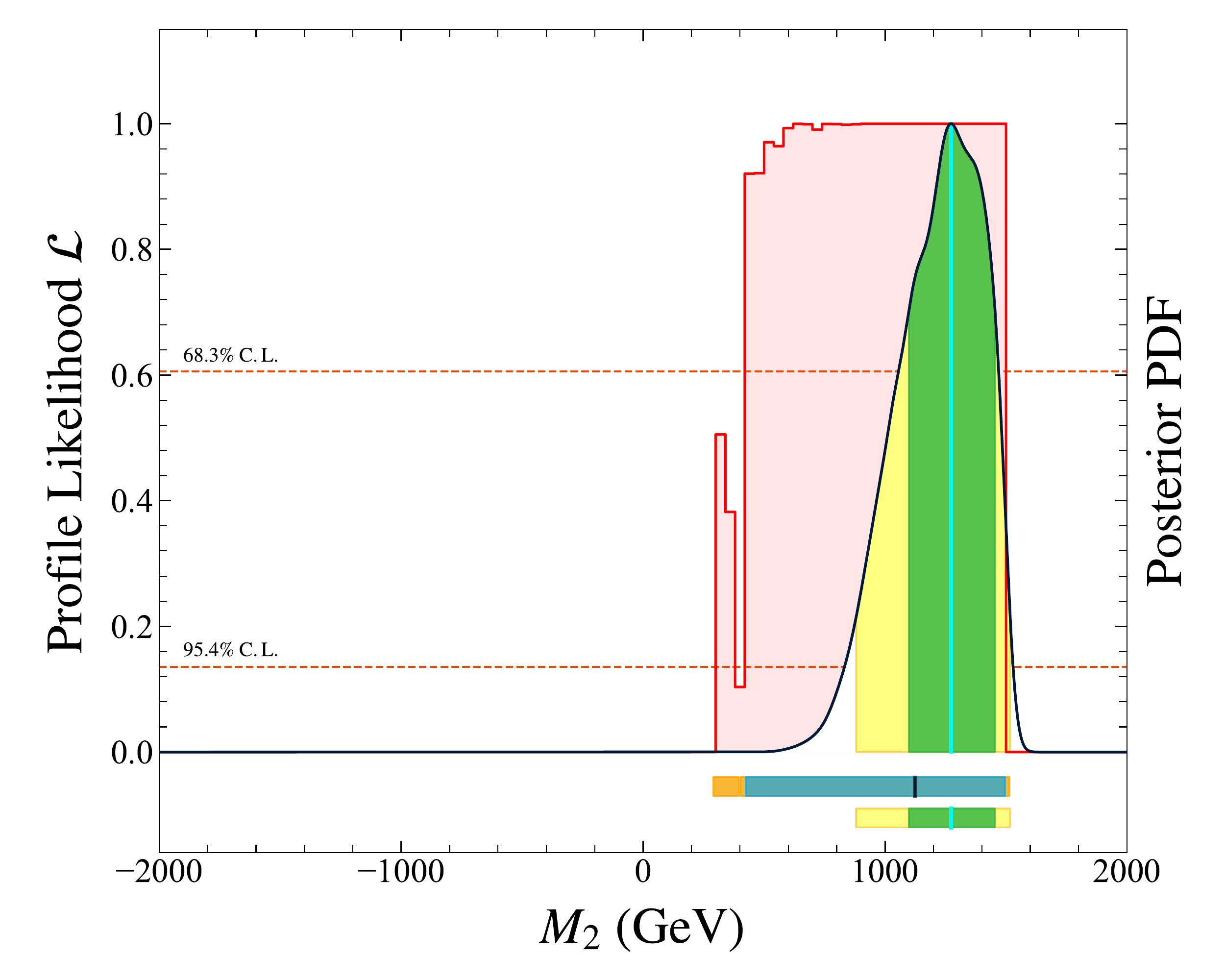}}
\vspace{-0.1cm}
\makebox[\textwidth][c]{
\includegraphics[width=0.28\paperwidth]{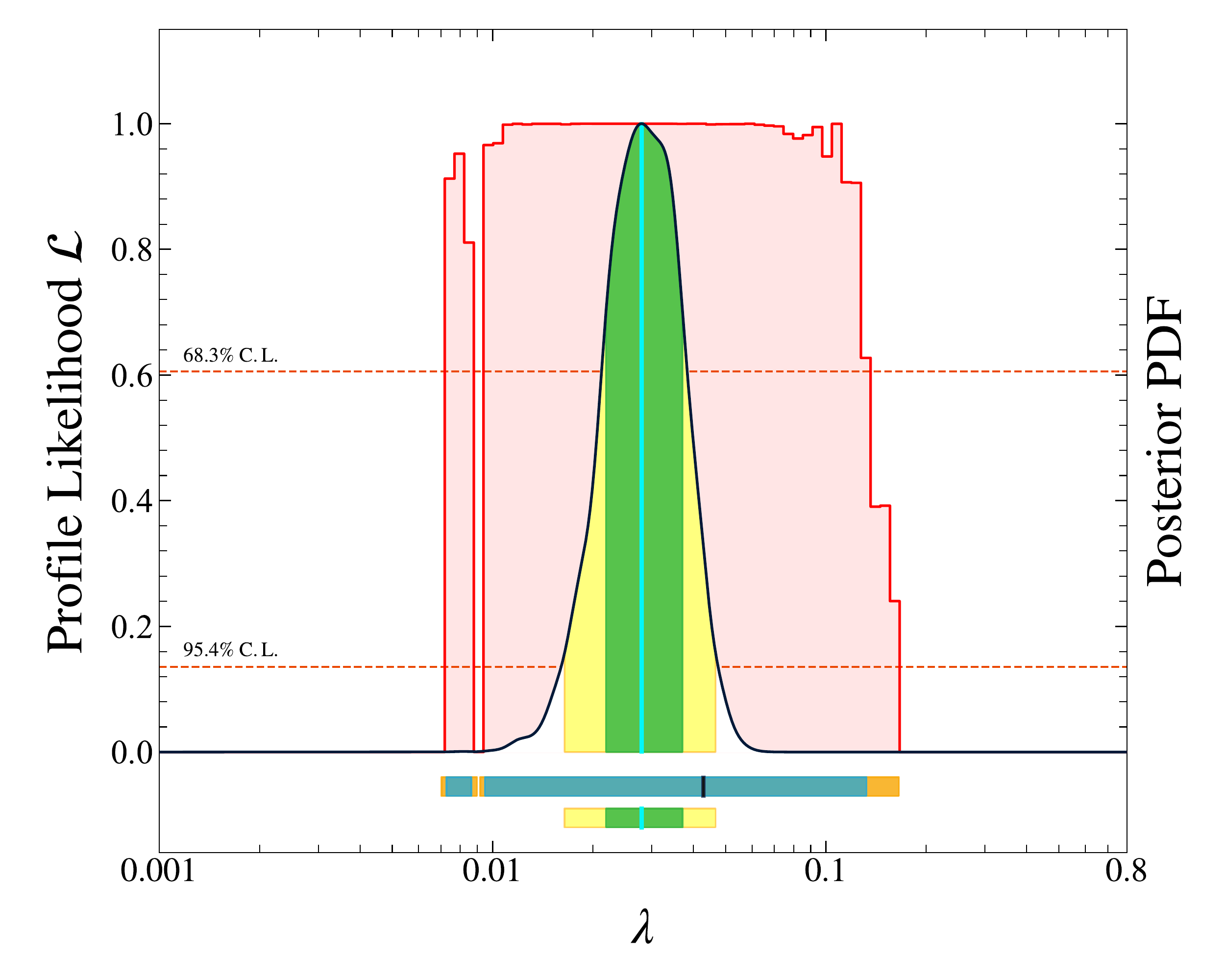}\hspace{0.5cm}
\includegraphics[width=0.28\paperwidth]{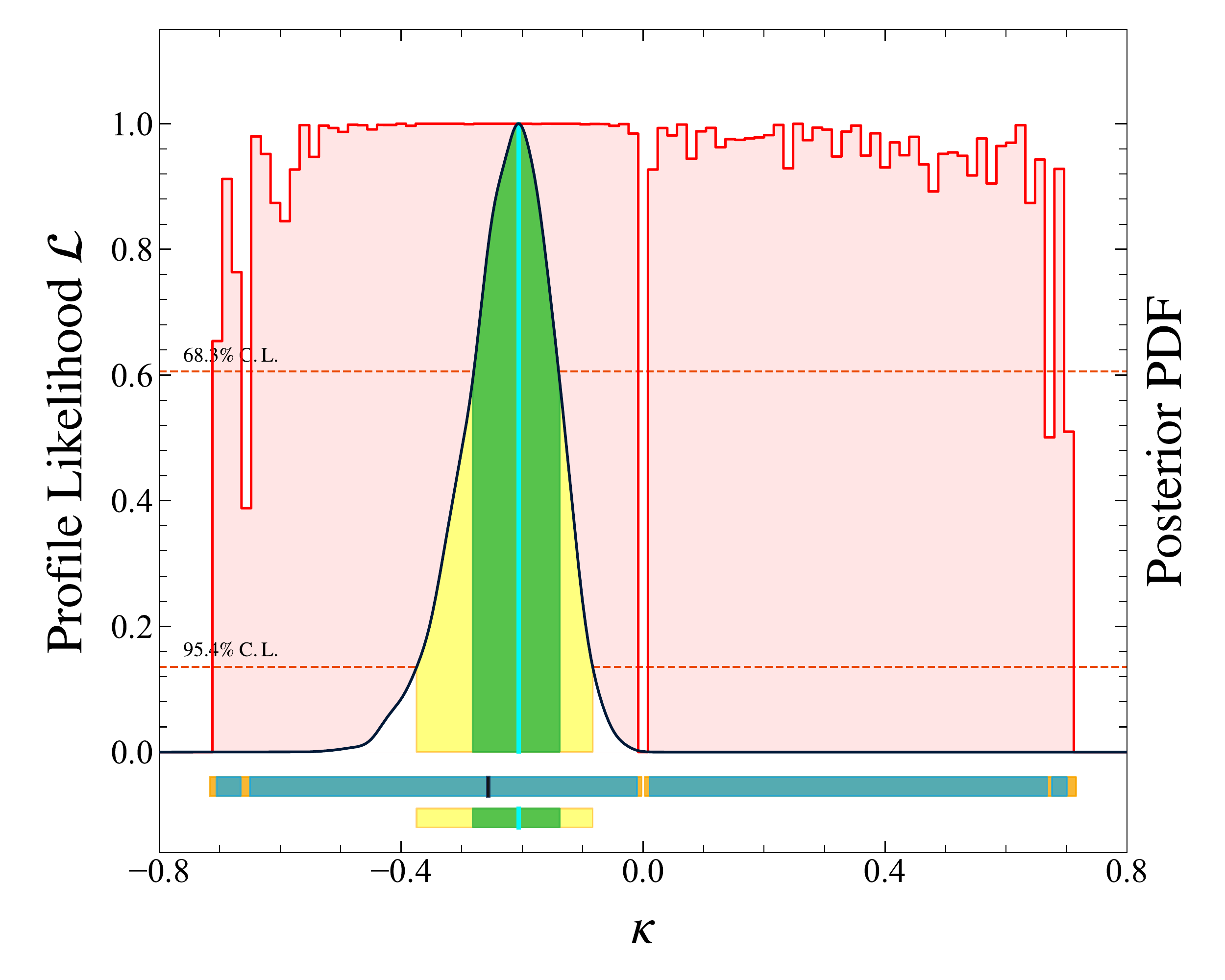}}
\vspace{-0.1cm}
\makebox[\textwidth][c]{
\includegraphics[width=0.28\paperwidth]{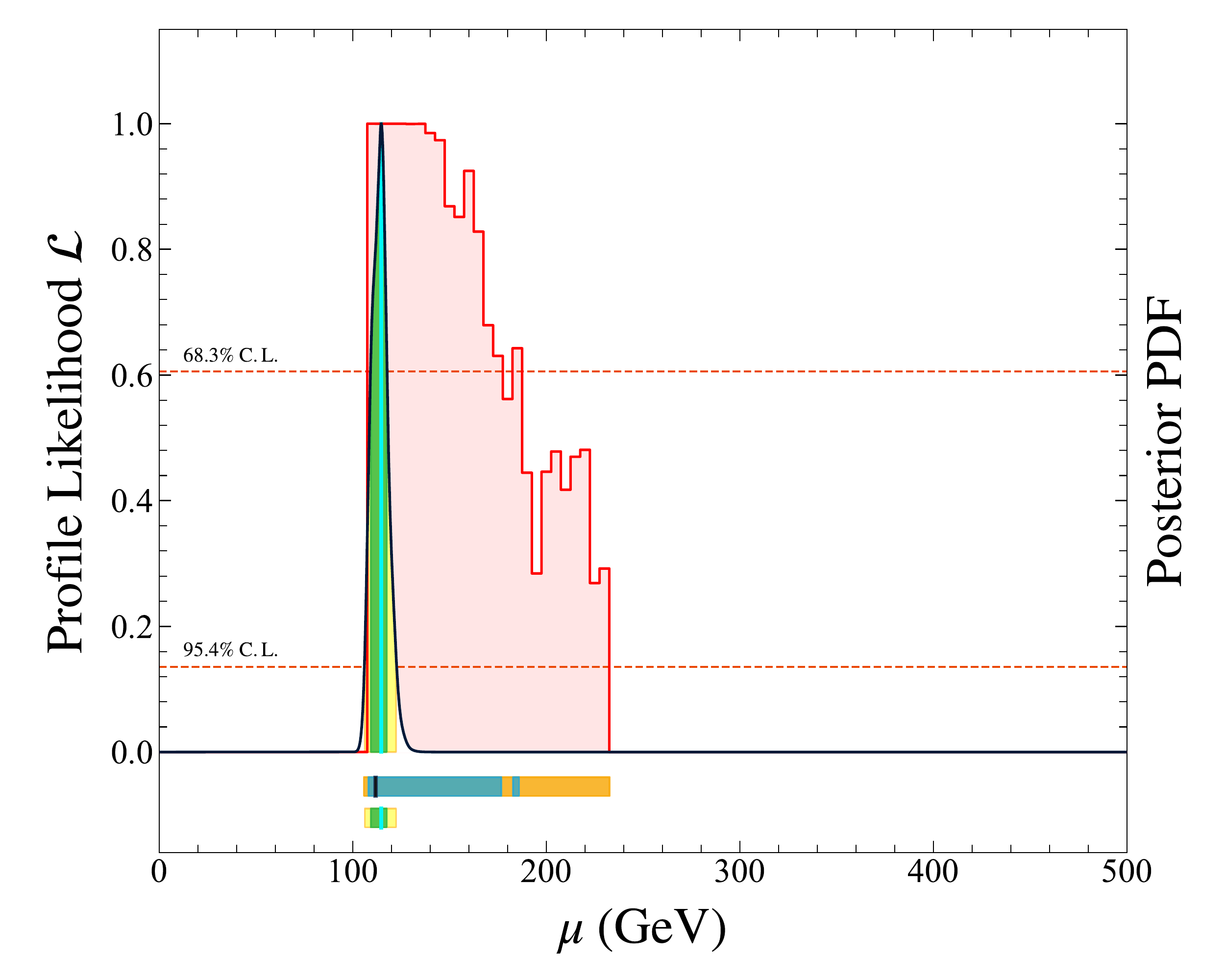}\hspace{0.5cm}
\includegraphics[width=0.28\paperwidth]{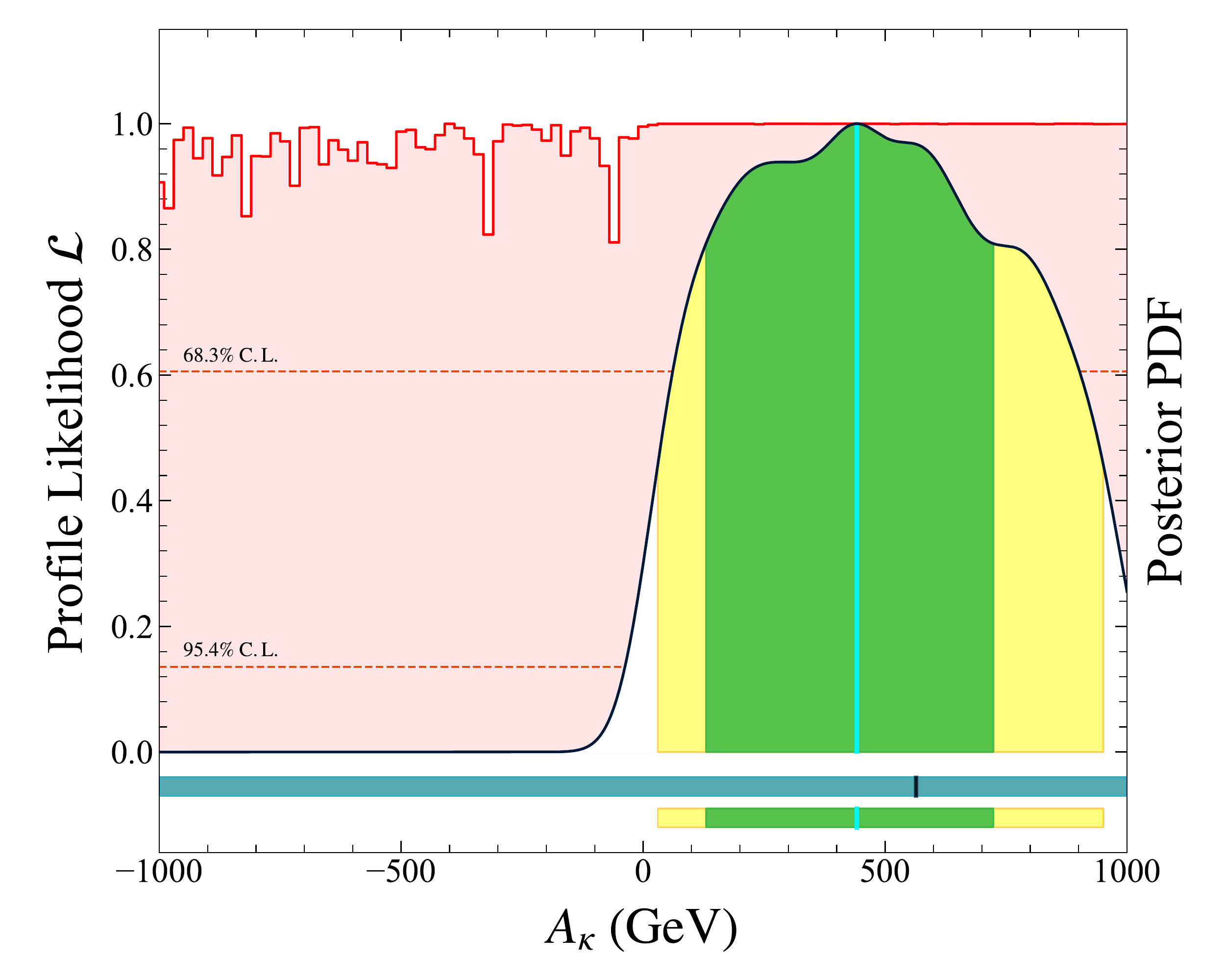}}
\vspace{-0.1cm}
\caption{\label{fig:1dcopara} One-dimensional profile likelihood $\mathcal{L}$ and posterior PDF distributions of $a_{\ell}^{\rm SUSY}$ and electroweakino input parameters. Regions of orange areas colored blue show the $1\sigma$ ($2\sigma$) confidence interval, and the best point is marked by a black vertical line. Regions of yellow areas colored green represent the $1\sigma$ ($2\sigma$) credible region.}
\end{figure}

\begin{figure}[th]
\makebox[\textwidth][c]{
\includegraphics[width=0.28\paperwidth]{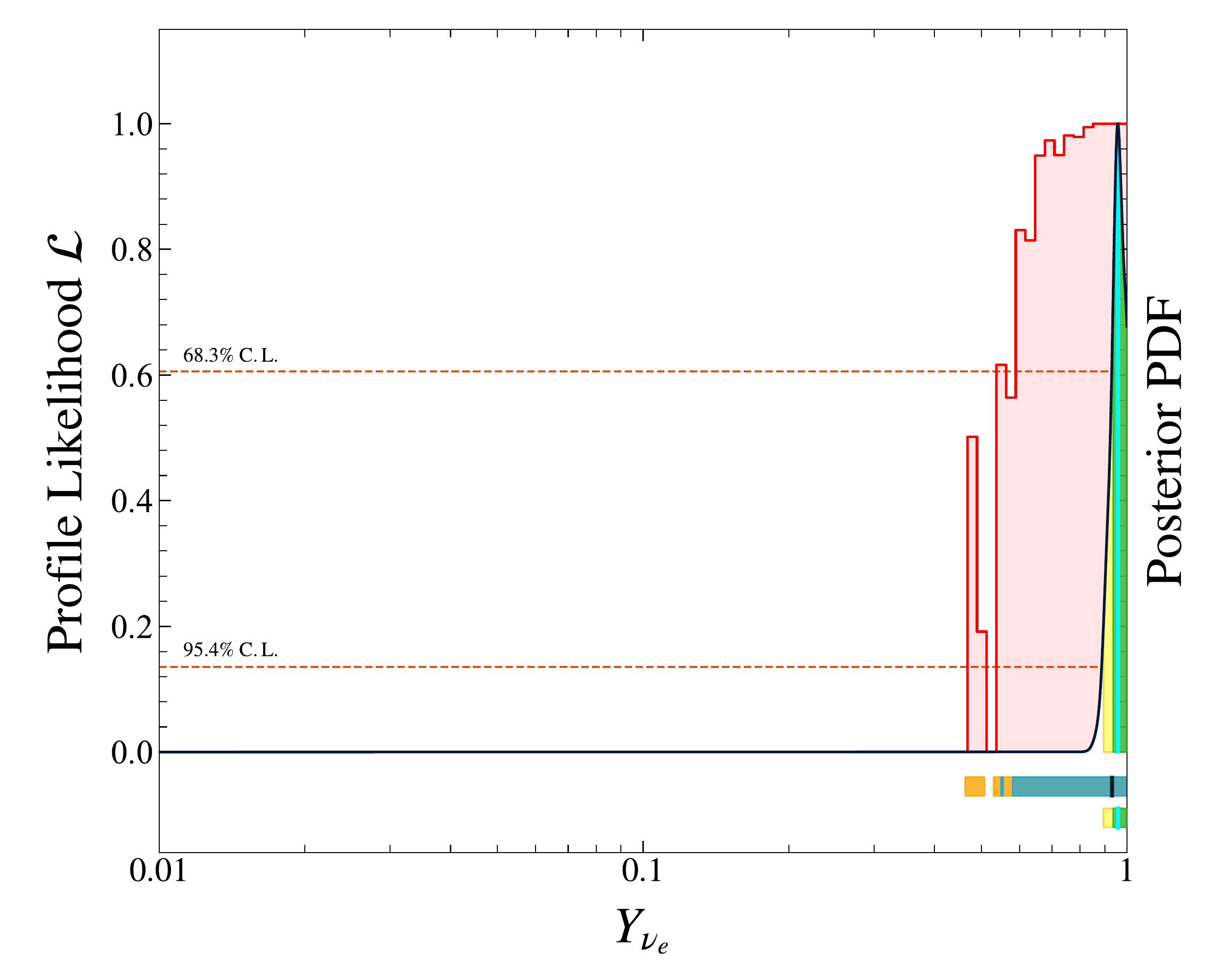}\hspace{0.5cm}
\includegraphics[width=0.28\paperwidth]{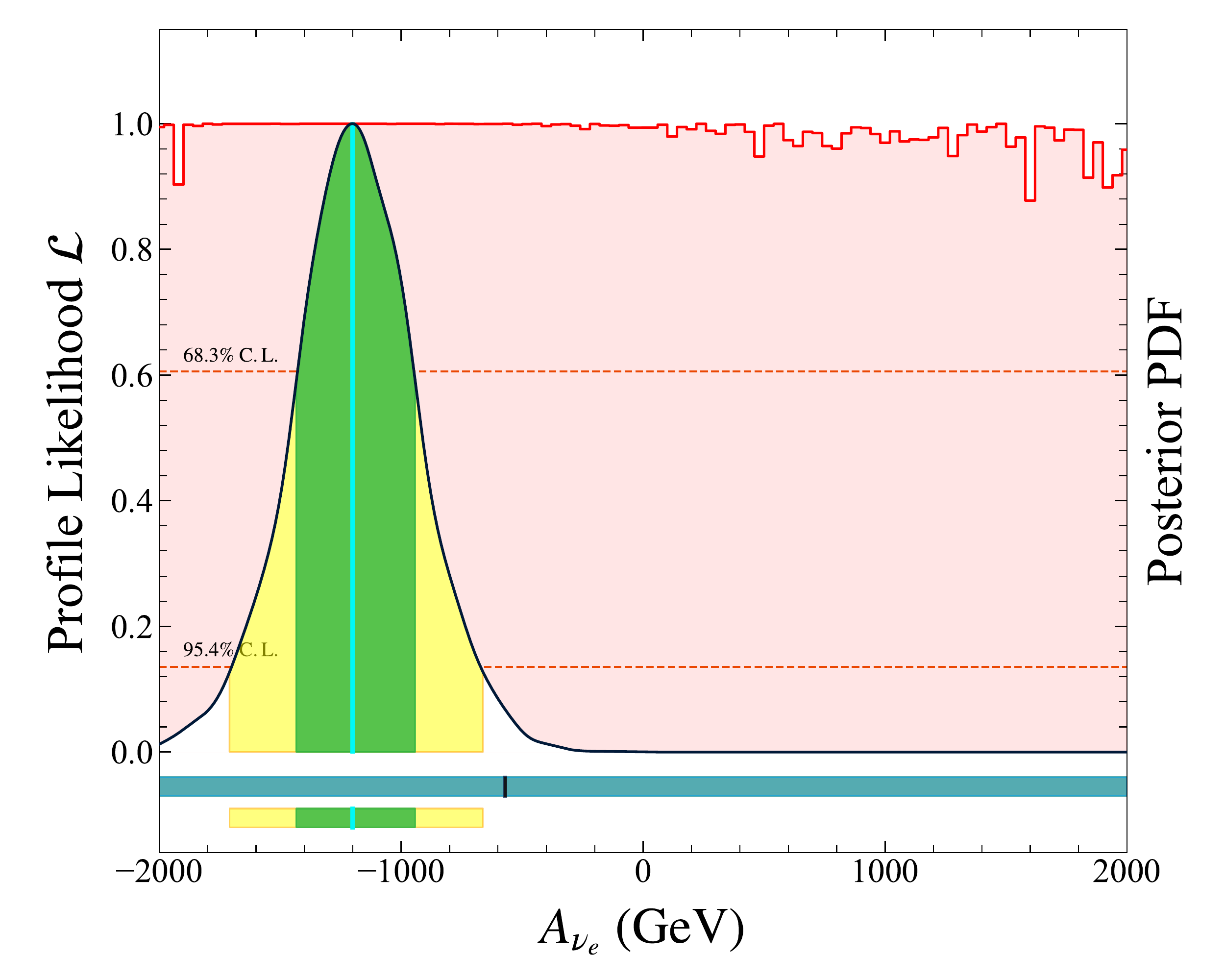}}
\vspace{-0.1cm}
\makebox[\textwidth][c]{
\includegraphics[width=0.28\paperwidth]{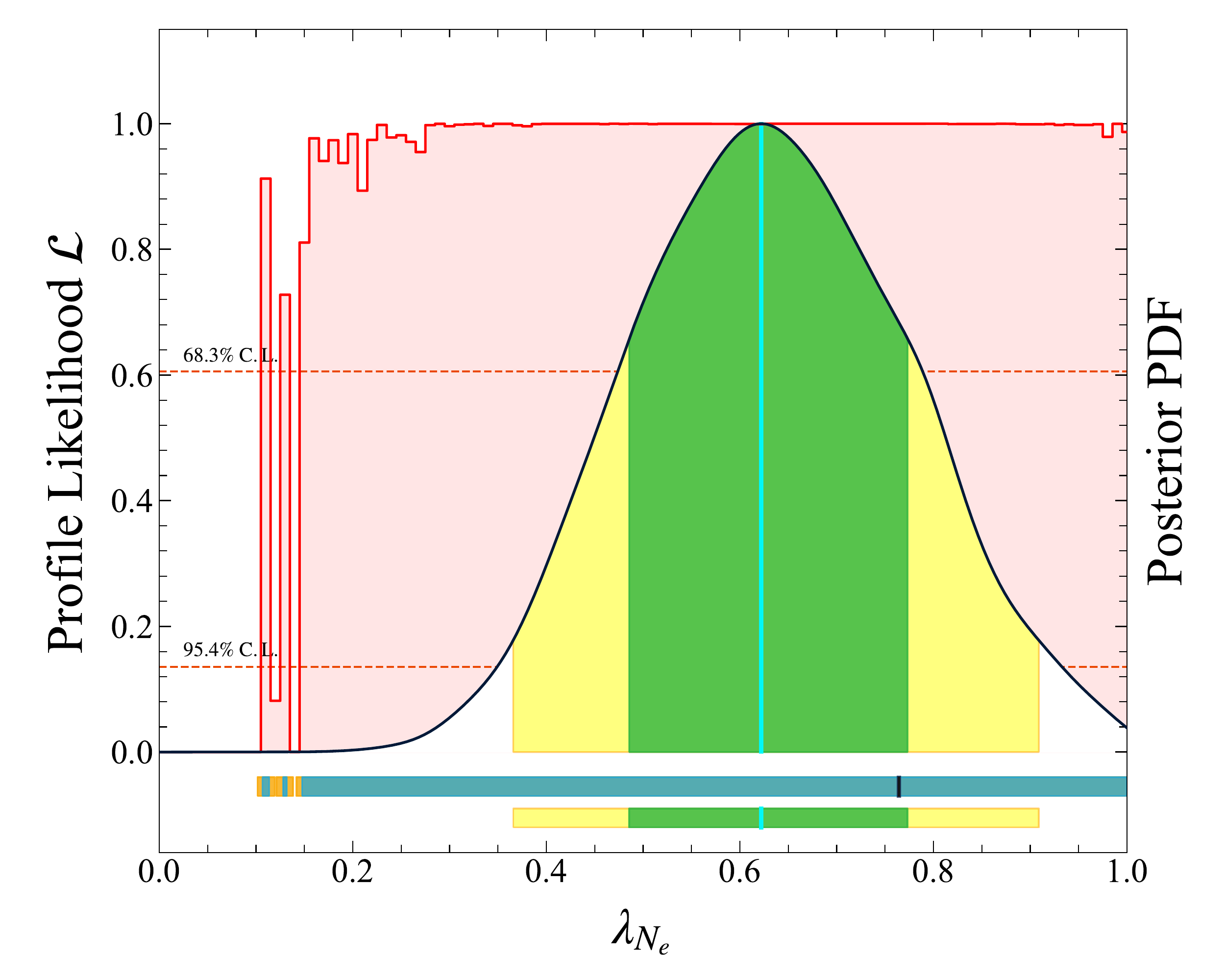}\hspace{0.5cm}
\includegraphics[width=0.28\paperwidth]{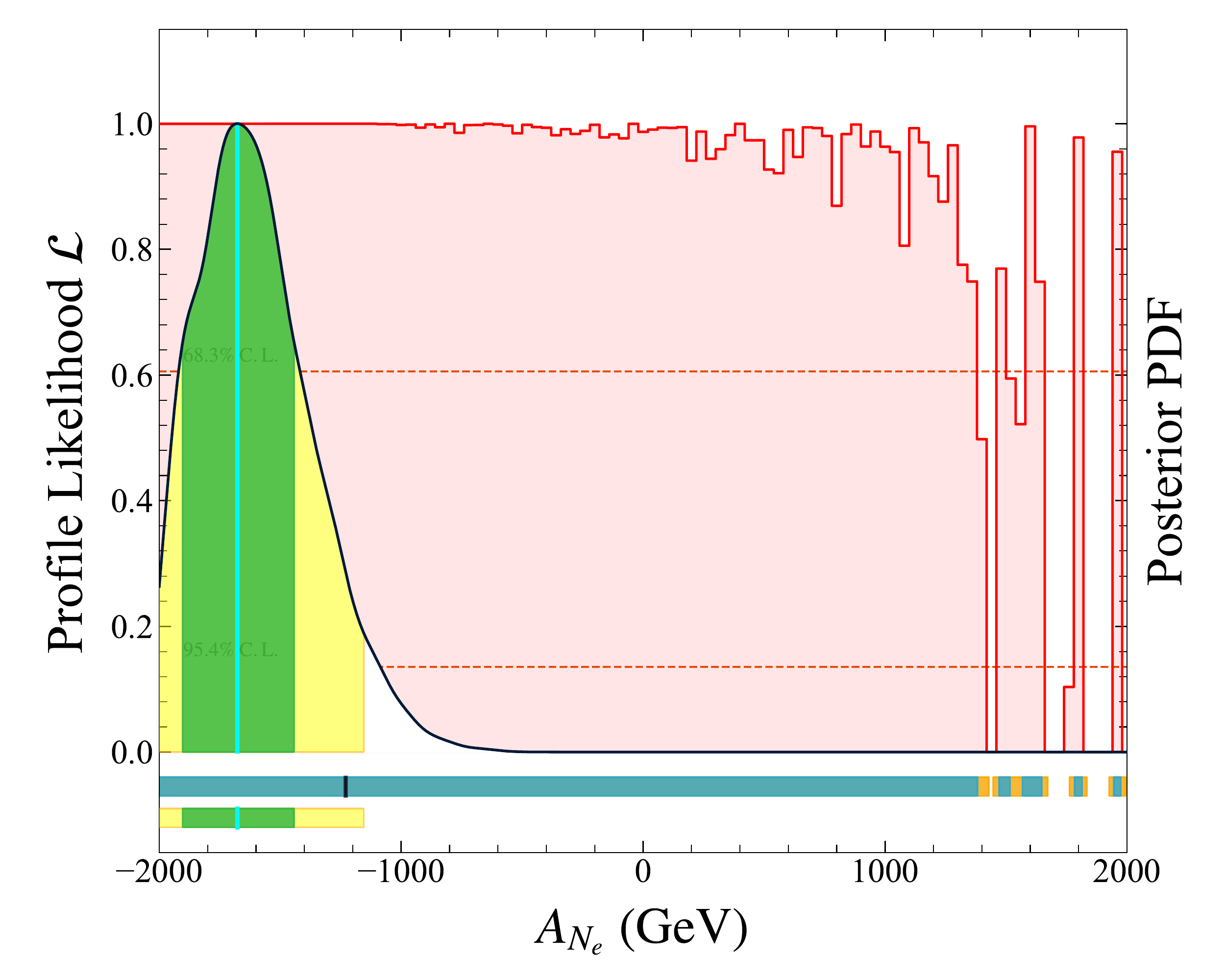}}
\vspace{-0.1cm}
\makebox[\textwidth][c]{
\includegraphics[width=0.28\paperwidth]{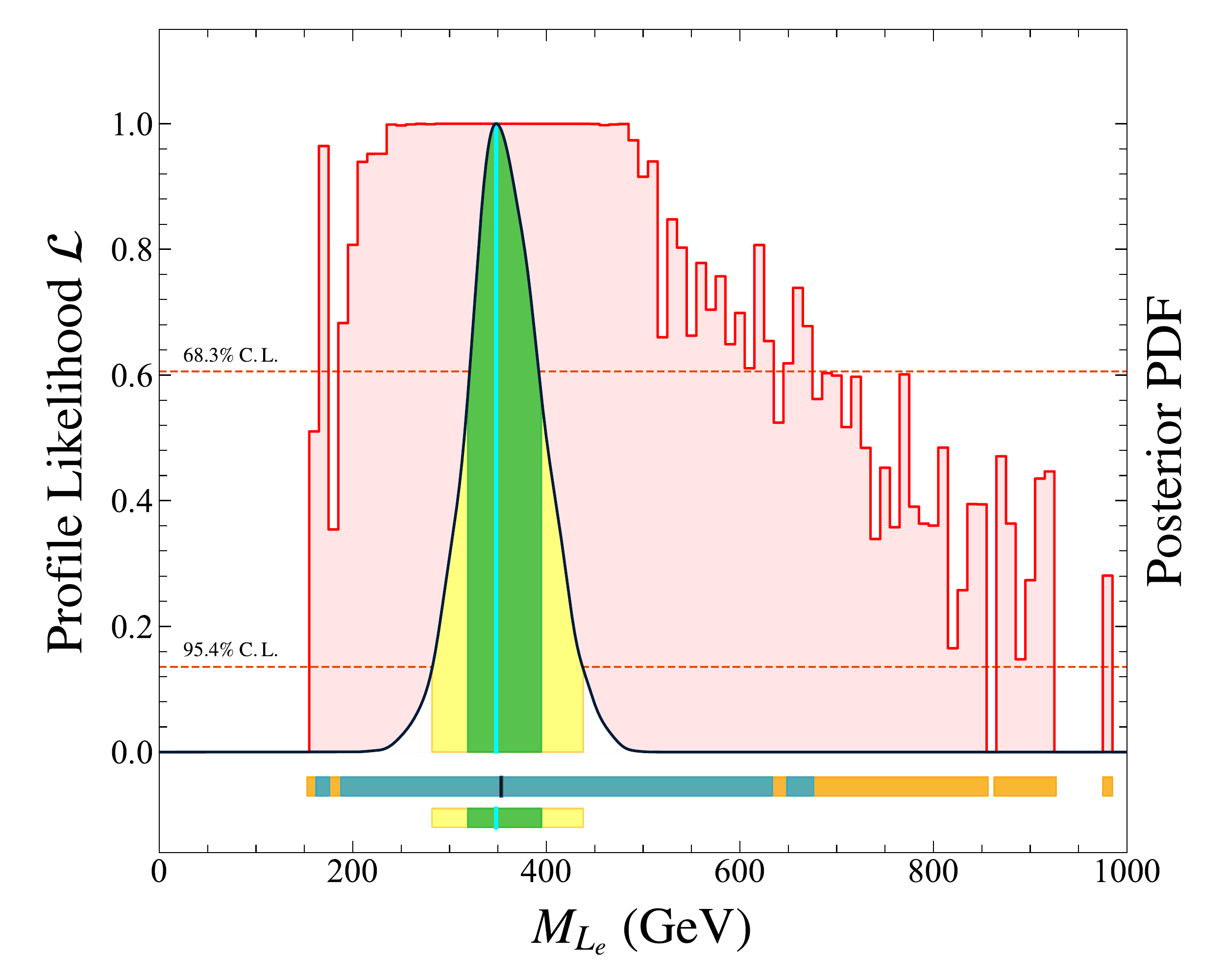}\hspace{0.5cm}
\includegraphics[width=0.28\paperwidth]{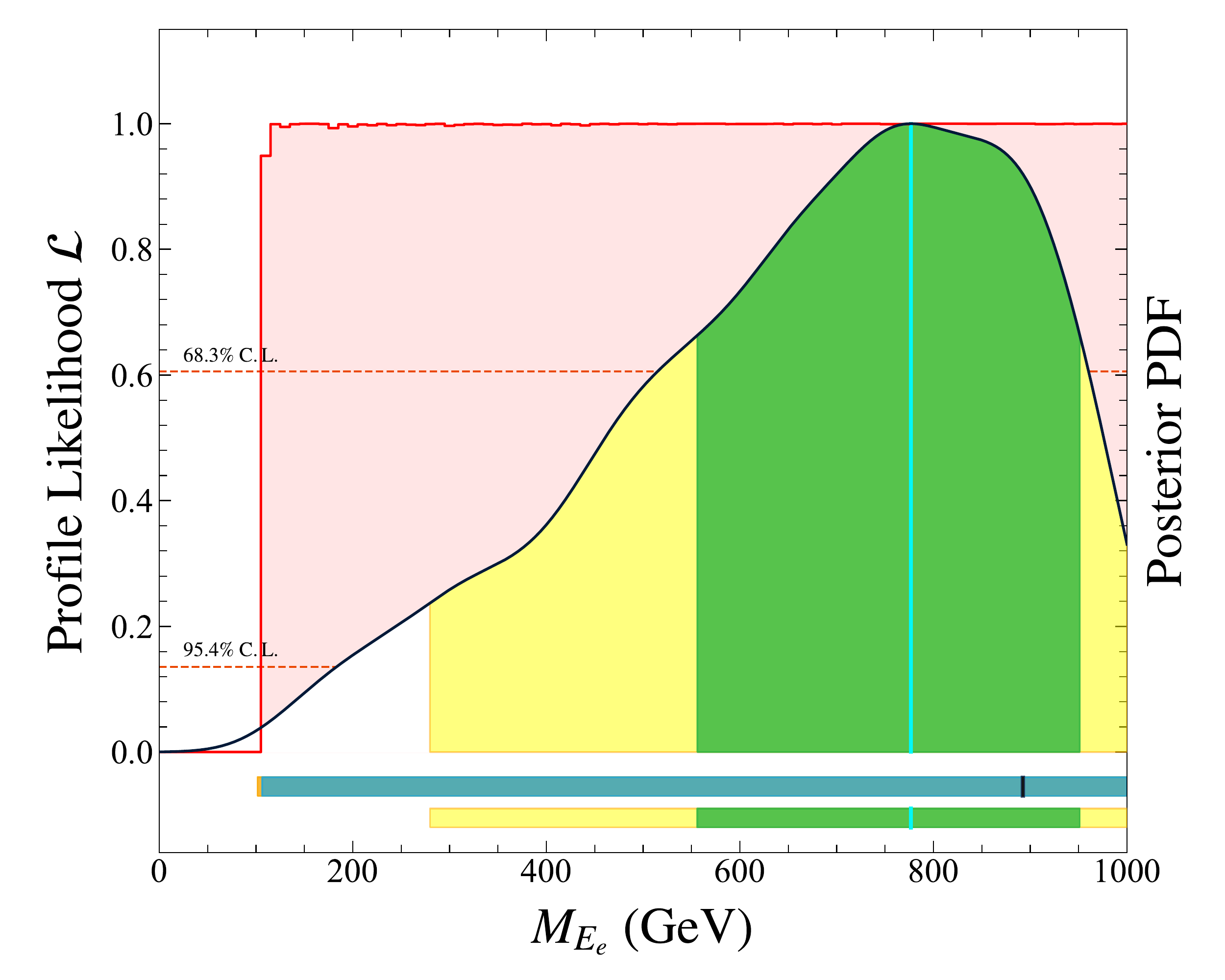}}
\vspace{-0.1cm}
\makebox[\textwidth][c]{
\includegraphics[width=0.28\paperwidth]{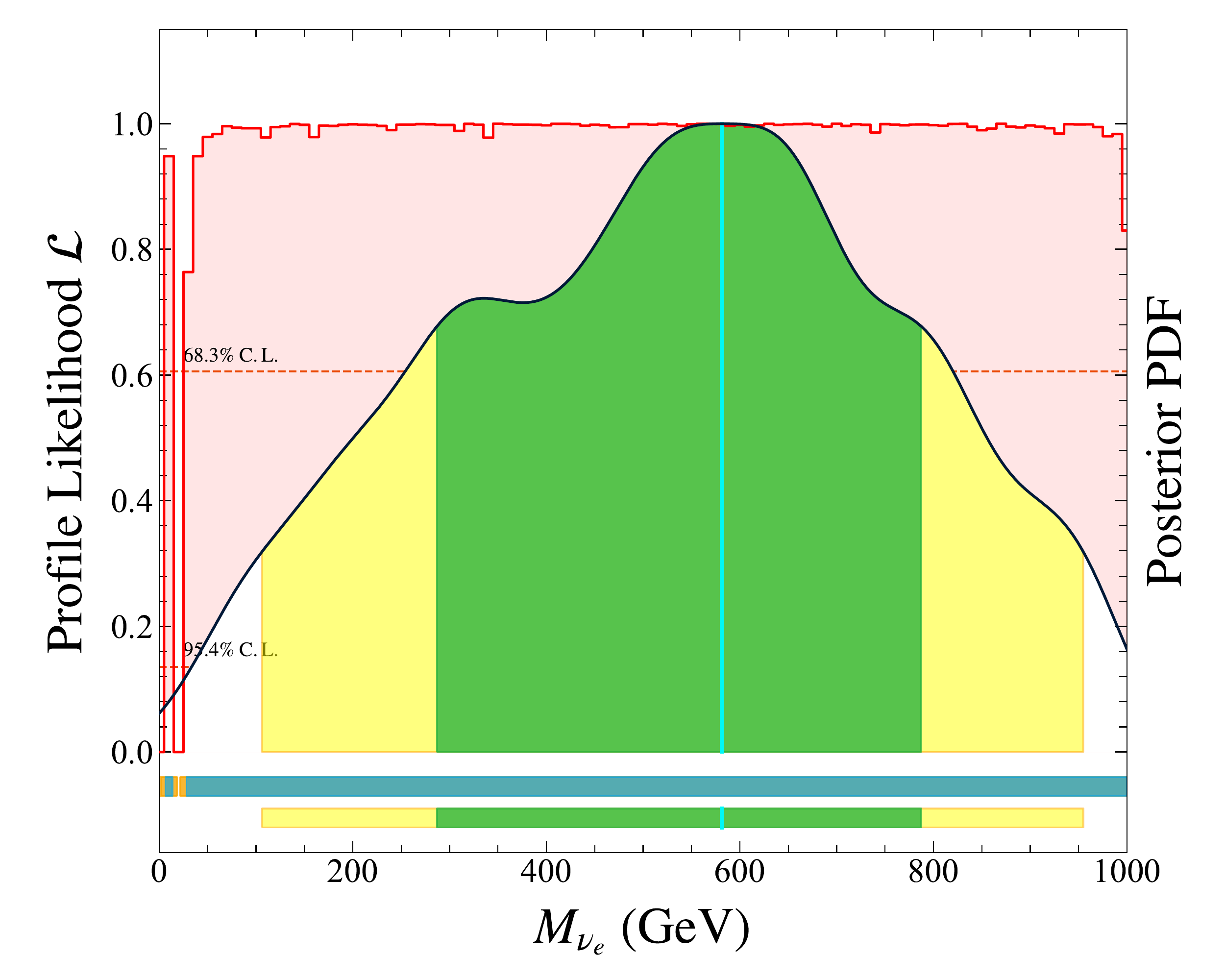}\hspace{0.5cm}
\includegraphics[width=0.28\paperwidth]{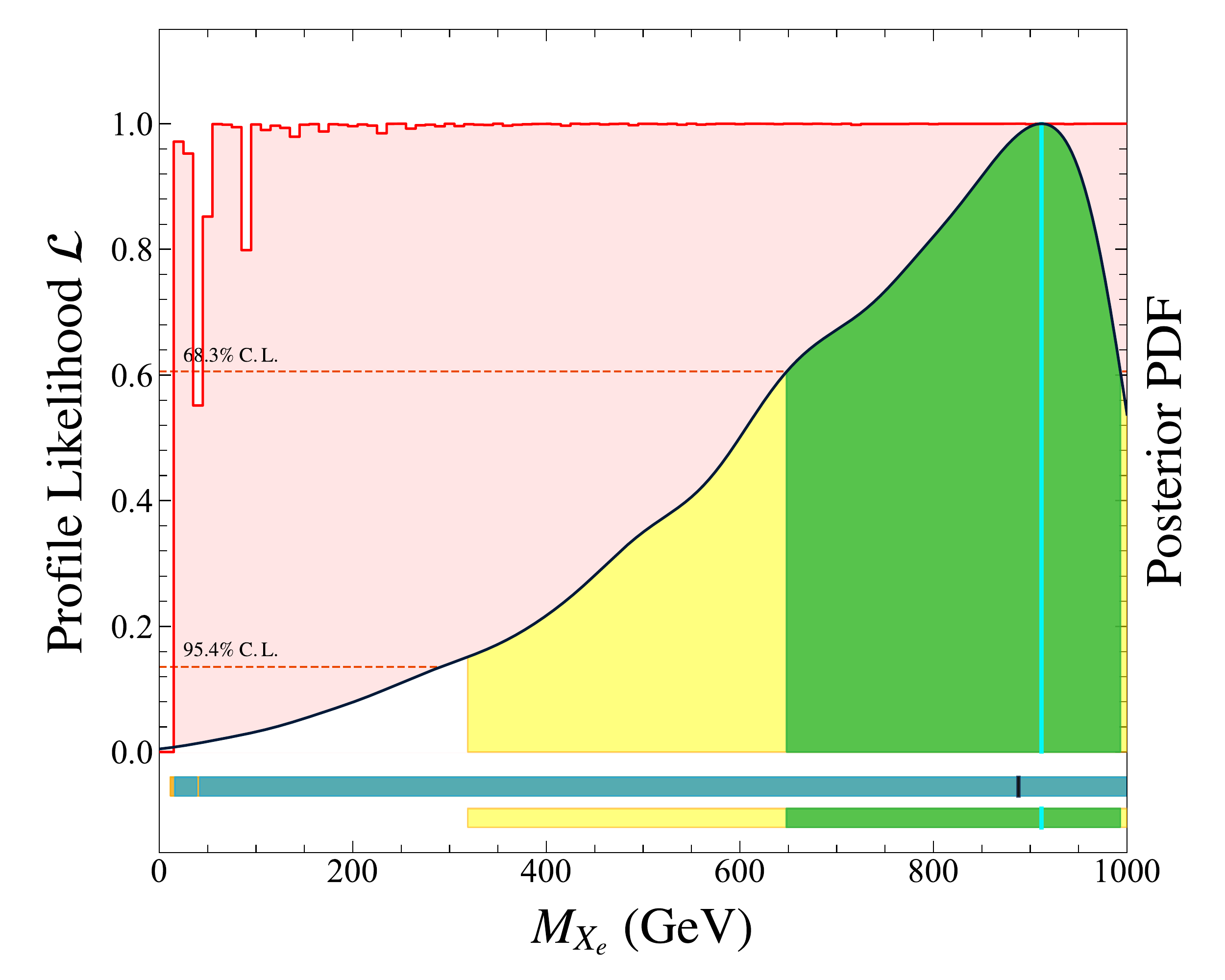}}
\vspace{-0.1cm}
\caption{\label{fig:1devpara} One-dimensional profile likelihood $\mathcal{L}$ and posterior PDF distributions of the input parameters of $e$-type slepton and sneutrino. Regions of orange areas colored blue show the $1\sigma$ ($2\sigma$) confidence interval, and the best point is marked by a black vertical line. Regions of yellow areas colored green represent the $1\sigma$ ($2\sigma$) credible region.}
\end{figure} 

\begin{figure}[th]
\makebox[\textwidth][c]{
\includegraphics[width=0.28\paperwidth]{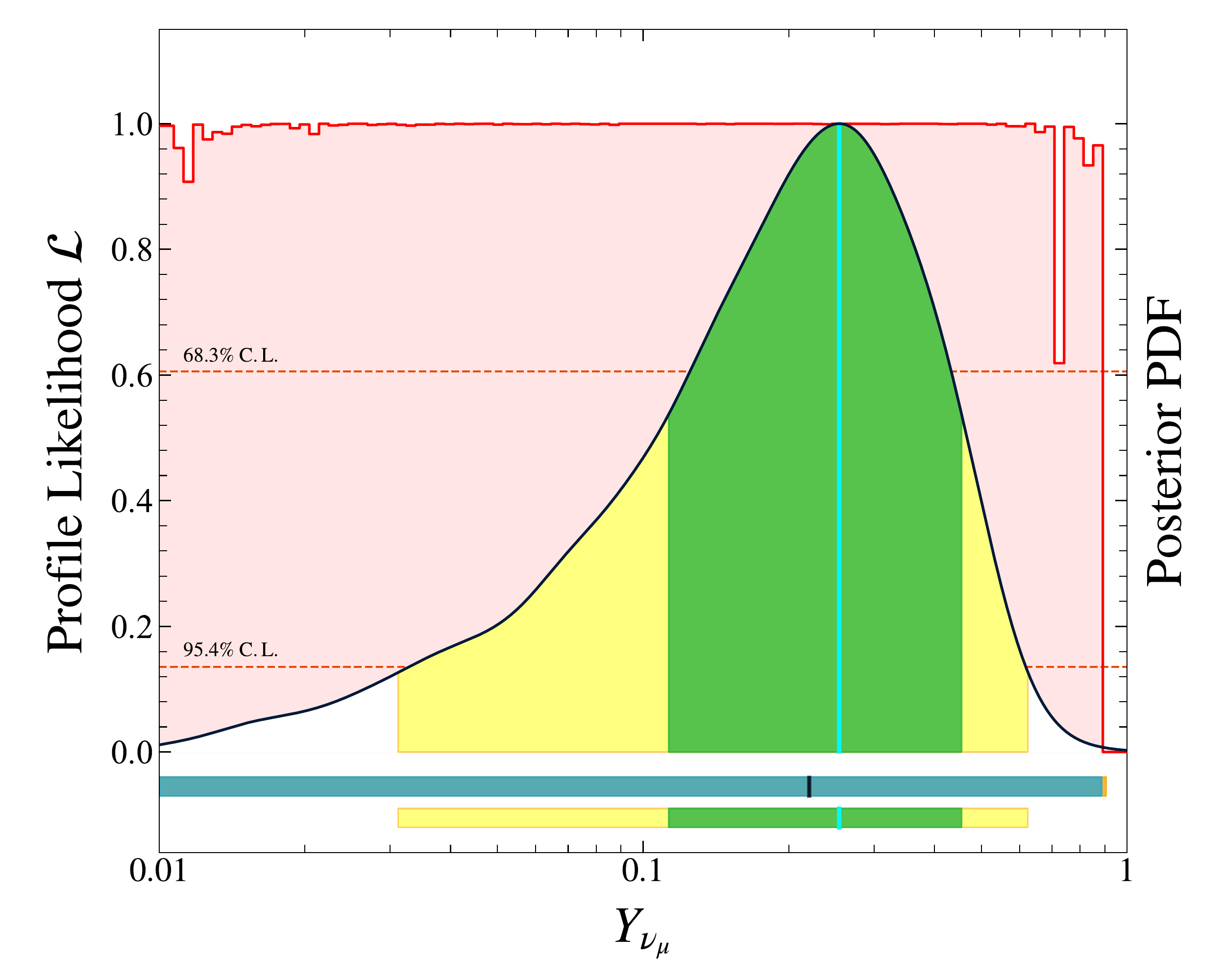}\hspace{0.5cm}
\includegraphics[width=0.28\paperwidth]{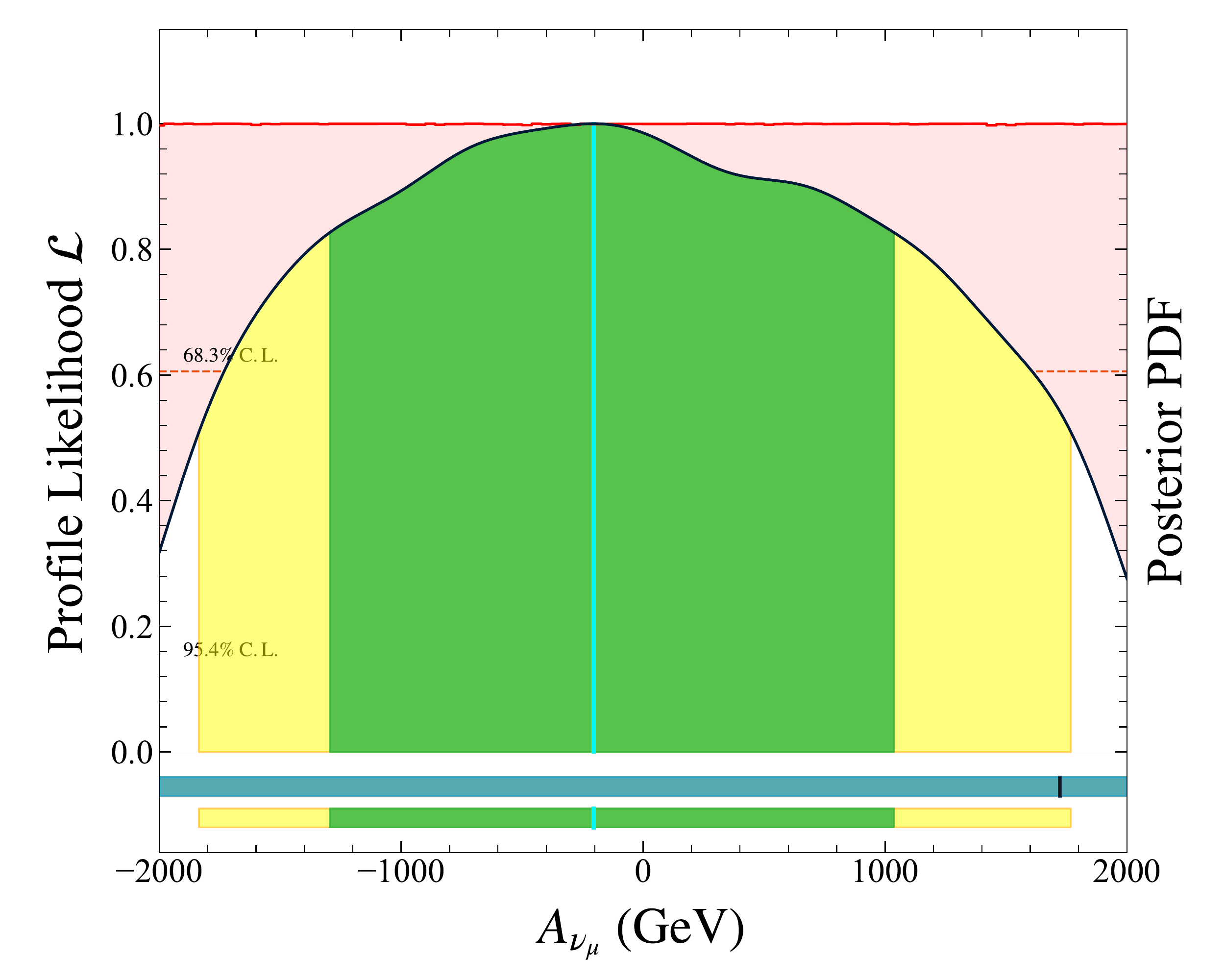}}
\vspace{-0.1cm}
\makebox[\textwidth][c]{
\includegraphics[width=0.28\paperwidth]{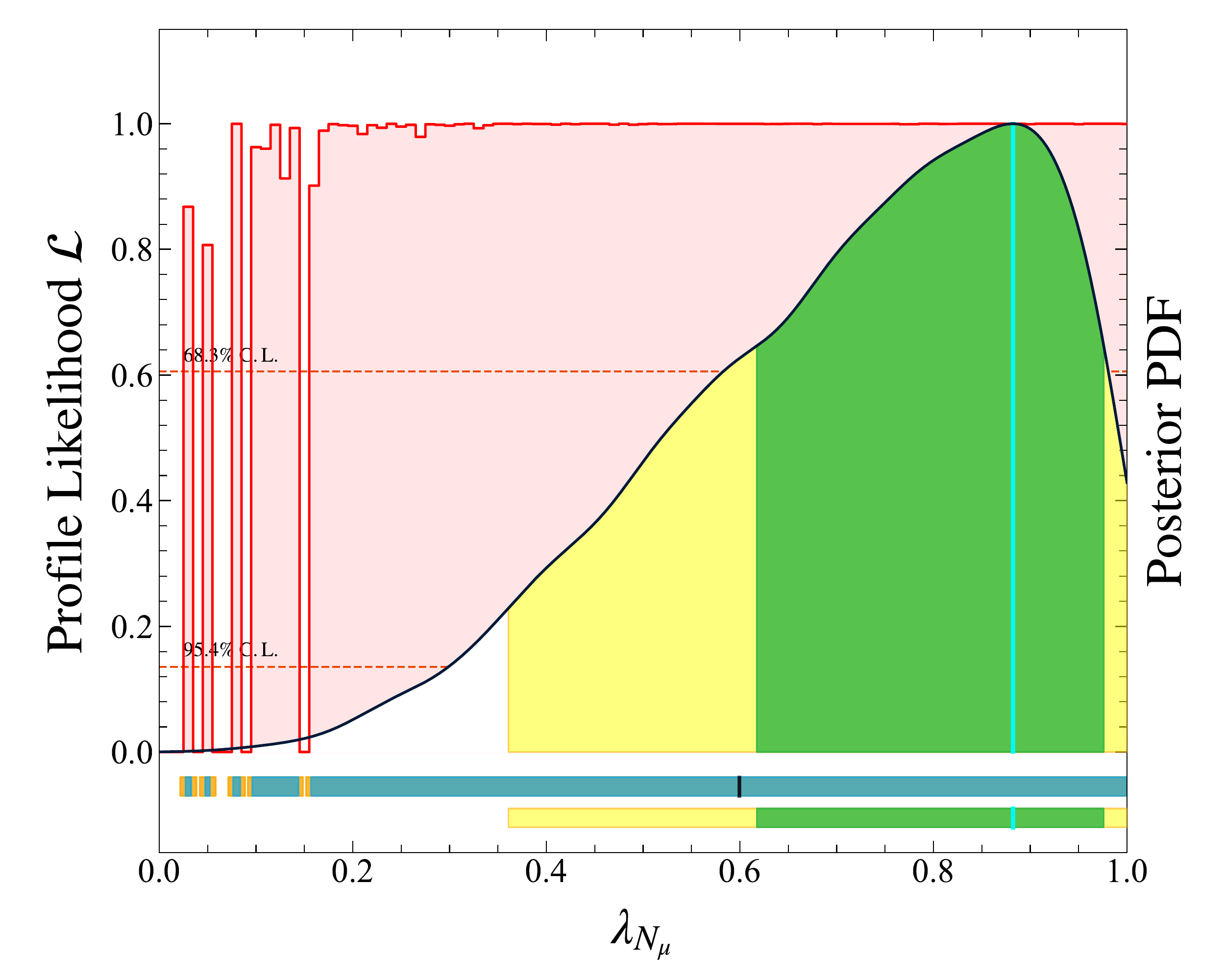}\hspace{0.5cm}
\includegraphics[width=0.28\paperwidth]{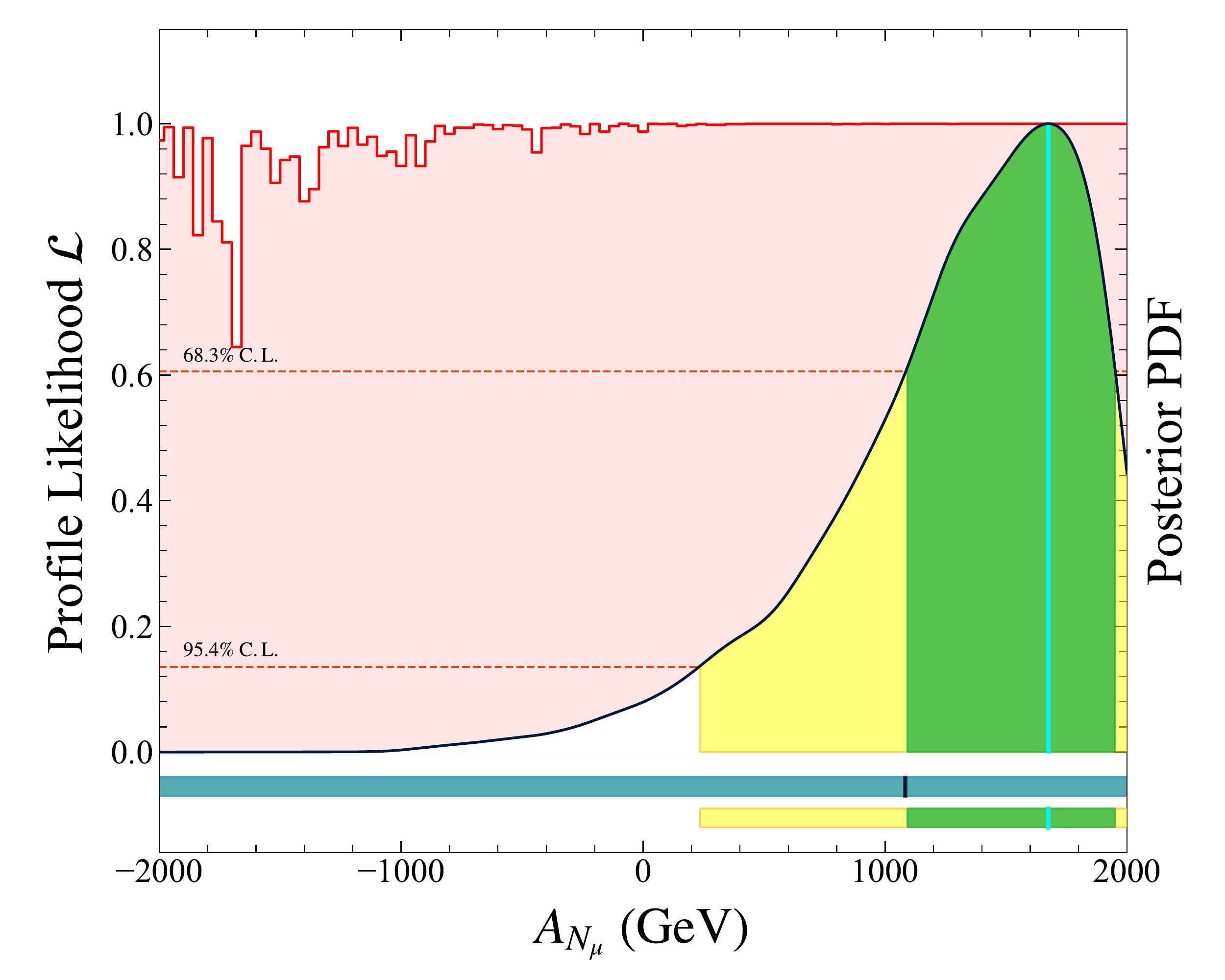}}
\vspace{-0.1cm}
\makebox[\textwidth][c]{
\includegraphics[width=0.28\paperwidth]{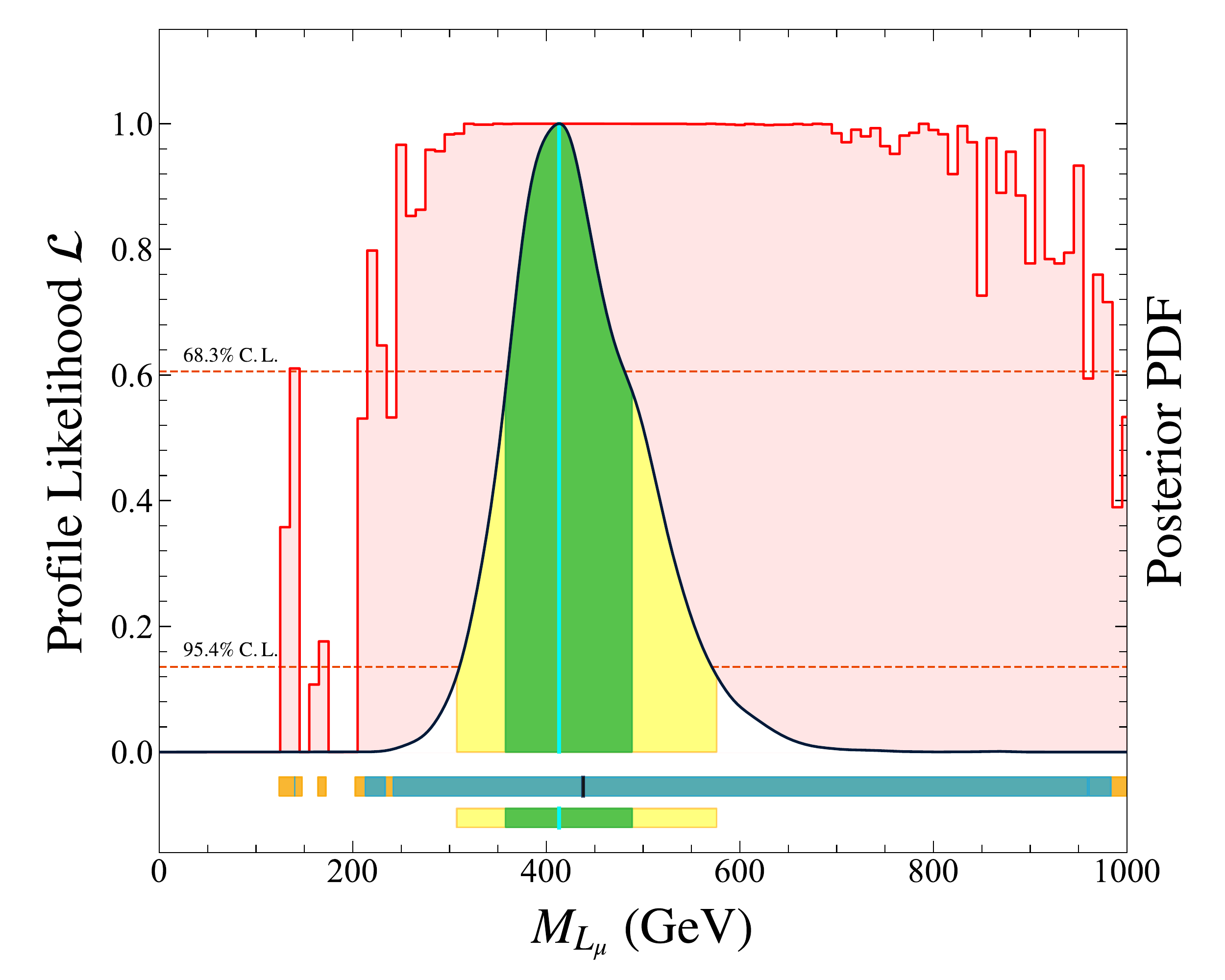}\hspace{0.5cm}
\includegraphics[width=0.28\paperwidth]{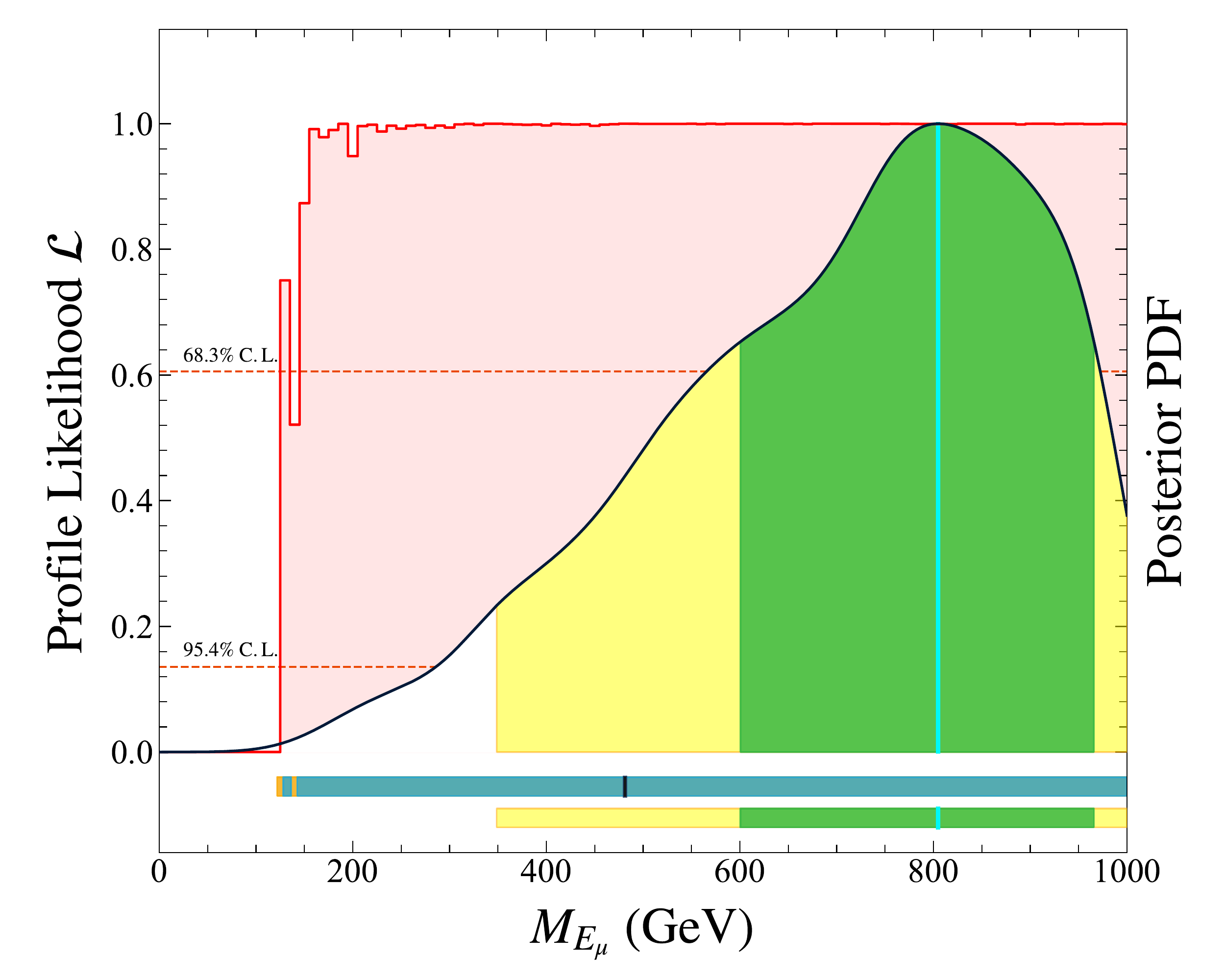}}
\vspace{-0.1cm}
\makebox[\textwidth][c]{
\includegraphics[width=0.28\paperwidth]{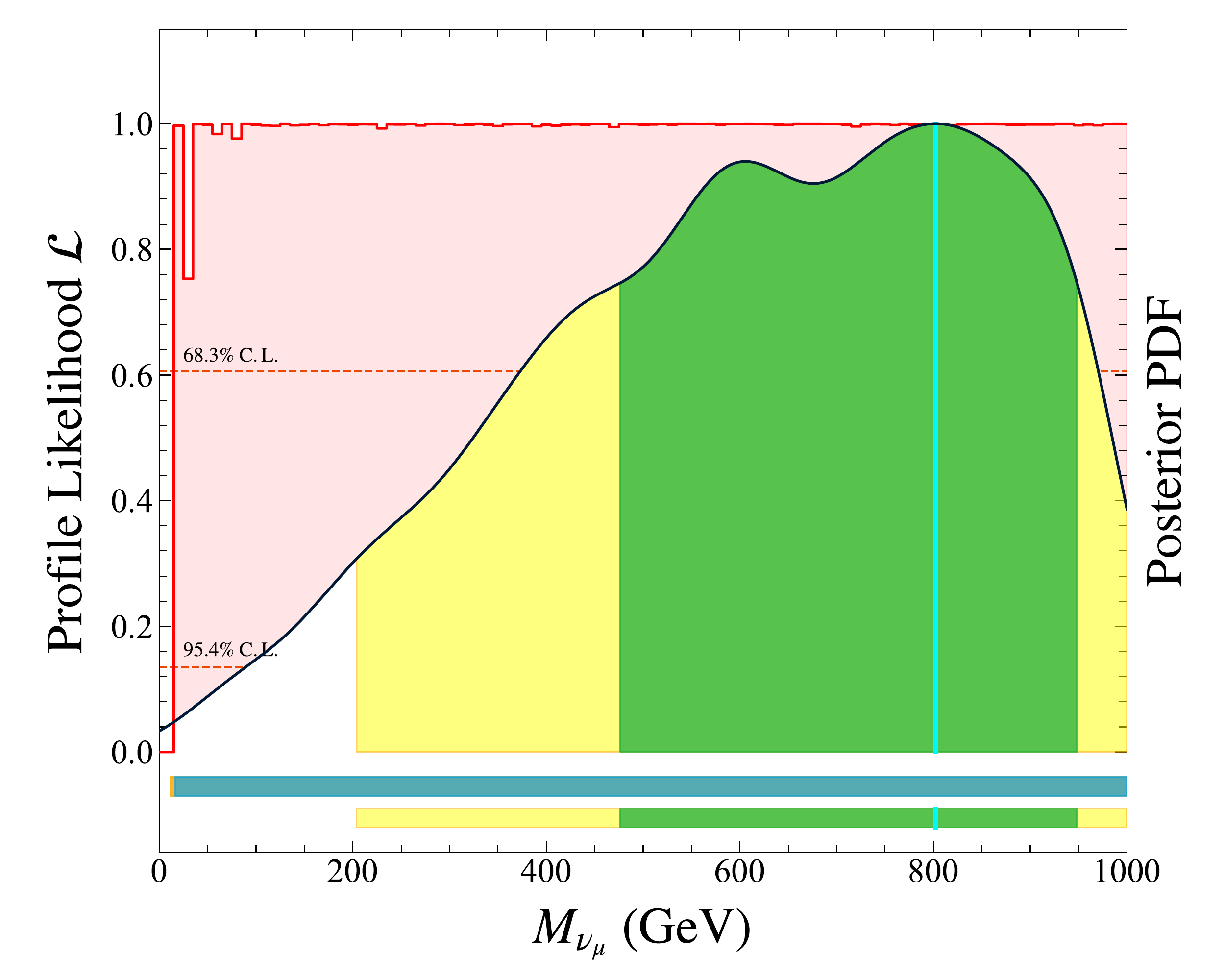}\hspace{0.5cm}
\includegraphics[width=0.28\paperwidth]{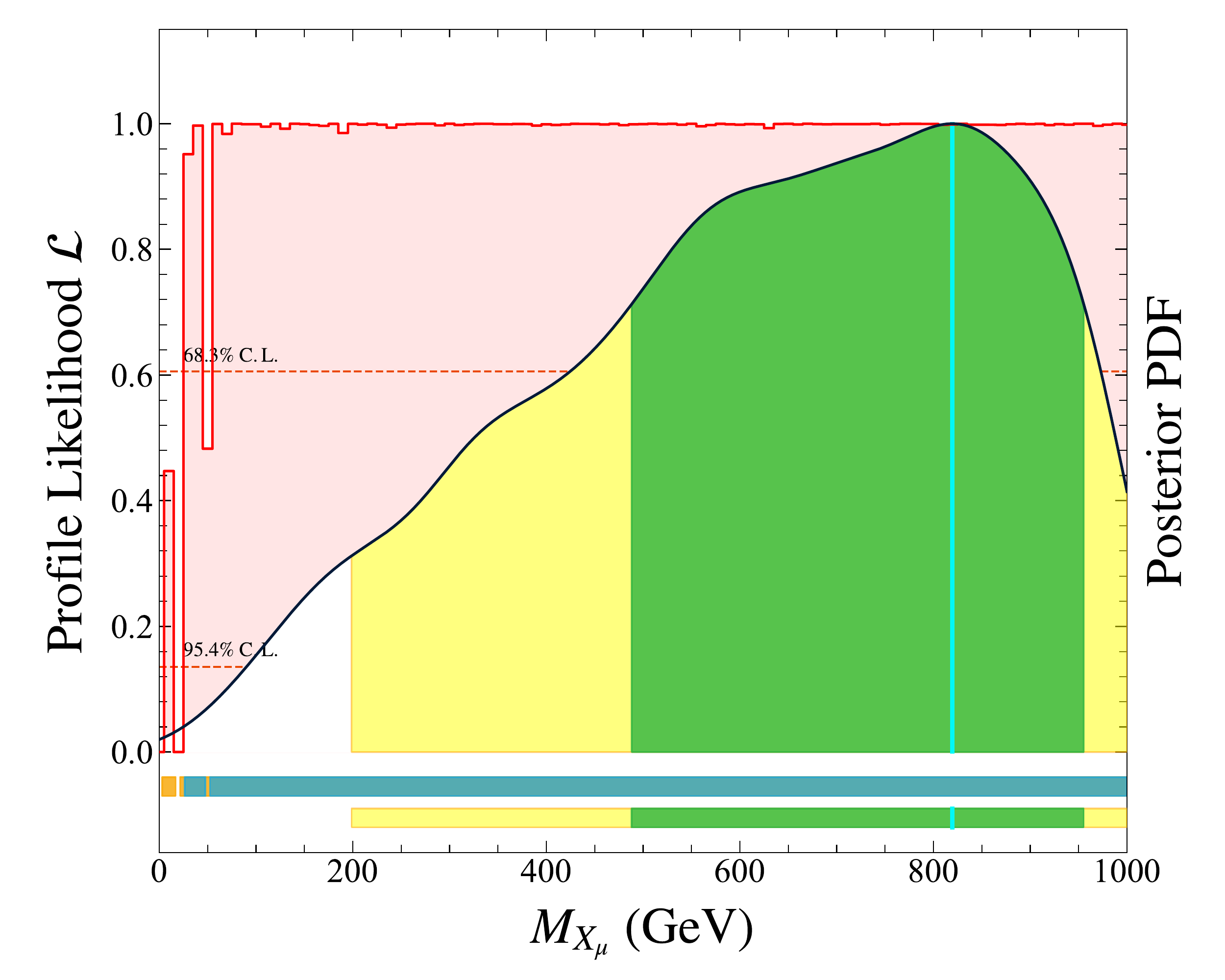}}
\vspace{-0.1cm}
\caption{\label{fig:1dmuvpara} One-dimensional profile likelihood $\mathcal{L}$ and posterior PDF distributions of the $\mu$-flavor related input parameters. Regions of orange areas colored blue show the $1\sigma$ ($2\sigma$) confidence interval, and the best point is marked by a black vertical line. Regions of yellow areas colored green represent the $1\sigma$ ($2\sigma$) credible region.}
\end{figure} 

\par In the numerical calculation, an ISS-NMSSM model file is generated by the \textsc{SARAH-4.14.3} package~\cite{Staub:2015kfa, Vicente:2015zba}, the particle spectra of the parameter samples are calculated by the \textsc{SPheno-4.0.3}~\cite{Porod:2003um, Porod:2011nf} and \textsc{FlavorKit}~\cite{Porod:2014xia} packages, and electroweak vacuum stability and sneutrino stability are tested by the \textsc{Vevacious}~\cite{Camargo-Molina:2013qva, Camargo-Molina:2014pwa} package, in which the tunneling time from the input electroweak potential minimum to the true vacuum is obtained via the \textsc{CosmoTransitions} package~\cite{Wainwright:2011kj}. 

\begin{figure}[ht]
\centering
\includegraphics[width=\textwidth]{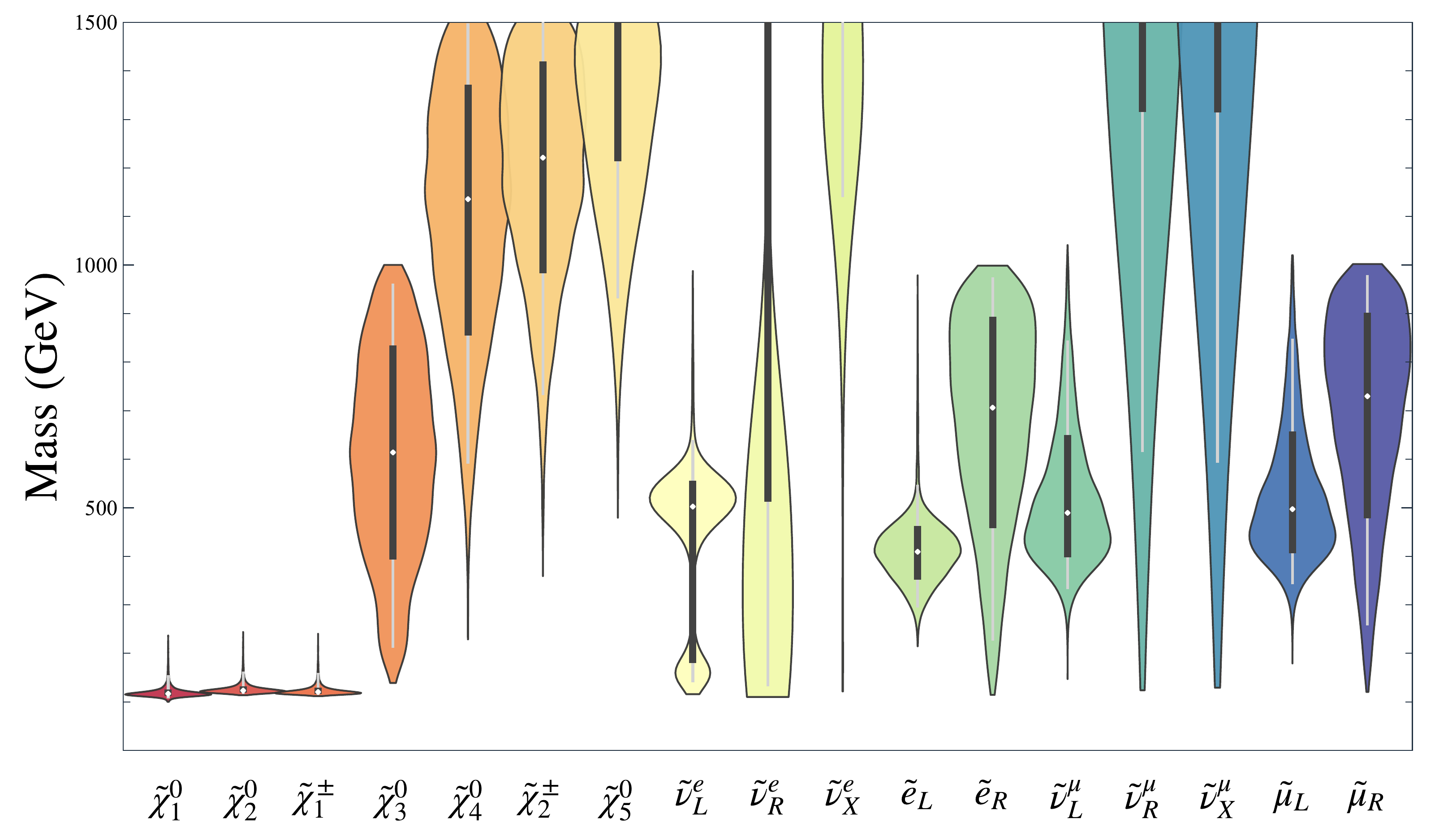}
\caption{\label{fig:spmass} Violin plots showing the mass distributions of supersymmetry (SUSY) particles. Sleptons and sneutrinos are labeled by their dominated components. The violins are scaled by count. The thick vertical bar in the center indicates the interquartile range with the white dot representing the median, and the long vertical line represents the 95\% confidence interval.}
\end{figure}
\par The one-dimensional profile likelihood (PL) plots of $\chi_{a_\ell}^2$ with the values normalized to the maximum and marginal posterior PDF of the input parameters are shown in Figs.~\ref{fig:1dcopara}, \ref{fig:1devpara}, and \ref{fig:1dmuvpara}. The value of PL for a fixed value of the parameter of interest, $\theta = \theta_0$, is the profiled value of nuisance parameters that maximized $\mathcal{L}$. Therefore, in a multidimensional model, the PL can be treated as an indicator of how well the theory can explain the experiments in the subspace of $\theta = \theta_0$. Complementarily, the one-dimensional marginal posterior PDF reflects the volume of subspace $\theta = \theta_0$. 
\par These figures show that ISS-NMSSM has a large parameter space to give a common explanation of $\Delta a_e$ and $\Delta a_\mu$. The results are summarized as follows:
\begin{itemize}
	\item  In the view of the PL, the smooth bell-shaped PL curves of $a_{e}^{\rm SUSY}$ and $a_{\mu}^{\rm SUSY}$ indicate the ability of the ISS-NMSSM to fit the two anomalies. The missing part of the PL curve where $a_{e}^{\rm SUSY} < -1 \times 10^{-12}$ indicates that a large $a_e^{\rm SUSY}$ is difficult to achieve in theory. 
	\item The distributions of the input parameters $M_2$, $\lambda$, $\mu$, $Y_{\nu_e}$, $Y_{\nu_\mu}$, and $A_{\nu_e}$ show some trends that verify the discussion in the previous section. The narrow PL picks of $100~{\rm GeV} \lesssim \mu \lesssim 150~{\rm GeV}$ and $Y_{\nu_e} \geq 0.7$ reflect the properties of Eq.~(\ref{eq:ahs}) and the difficulty of explaining $\Delta a_e$. The wider credible region of the $\mu$-type input parameter than that of $e$-type parameter also confirms this feature.   
	\item For most input parameters, the credible regions are smaller than the confidence interval. The PL values are very close to 1 almost in the entire space. This implies that the value of these parameters are not affected by the two anomalies.  
	\item  $\mu \sim 110~{\rm GeV}$ and $0.015 \lesssim \lambda \lesssim 0.05$ cause the masses of all the singlet Higgs particles to be on the order of several TeV. 
	\item $M_1$ and $M_2$ are often much greater than $\mu$ to reduce the correlations in the MSSM contributions to $a_{e}^{\rm SUSY}$ and $a_{\mu}^{\rm SUSY}$. The distributions of $M_1$ shows that the bino--slepton loop contribution is far from dominant. The distribution of $M_2$ shows that the wino-related contributions are suppressed in this explanation. 
\end{itemize}

\par There were 7989 samples obtained that explained two anomalies within the $2\sigma$ range in total. In Fig.~\ref{fig:spmass}, we plot the mass distributions of the SUSY particles via violin plots\footnote{A violin plot is similar to a box plot; it shows the probability density smoothed by kernel density estimation~\cite{hintze1998violin}.}, where sleptons and sneutrinos are labeled by their dominating components. The masses of the Higgsino-dominated triplet states were around $110~{\rm GeV}$, and the wino particles were often heavier than $700~{\rm GeV}$. The masses of left-handed selectron $\tilde{e}_L$ and $\tilde{\mu}_L$ were distributed around $400~{\rm GeV}$, and the right-handed sleptons were relatively heavy. In contrast to the MSSM spectrum, the mass of $\tilde{\nu}_L^e$ can be much lower than that of $\tilde{e}_L$ in the ISS-NMSSM.

\begin{figure}[ht]
\centering
\includegraphics[width=0.465\textwidth]{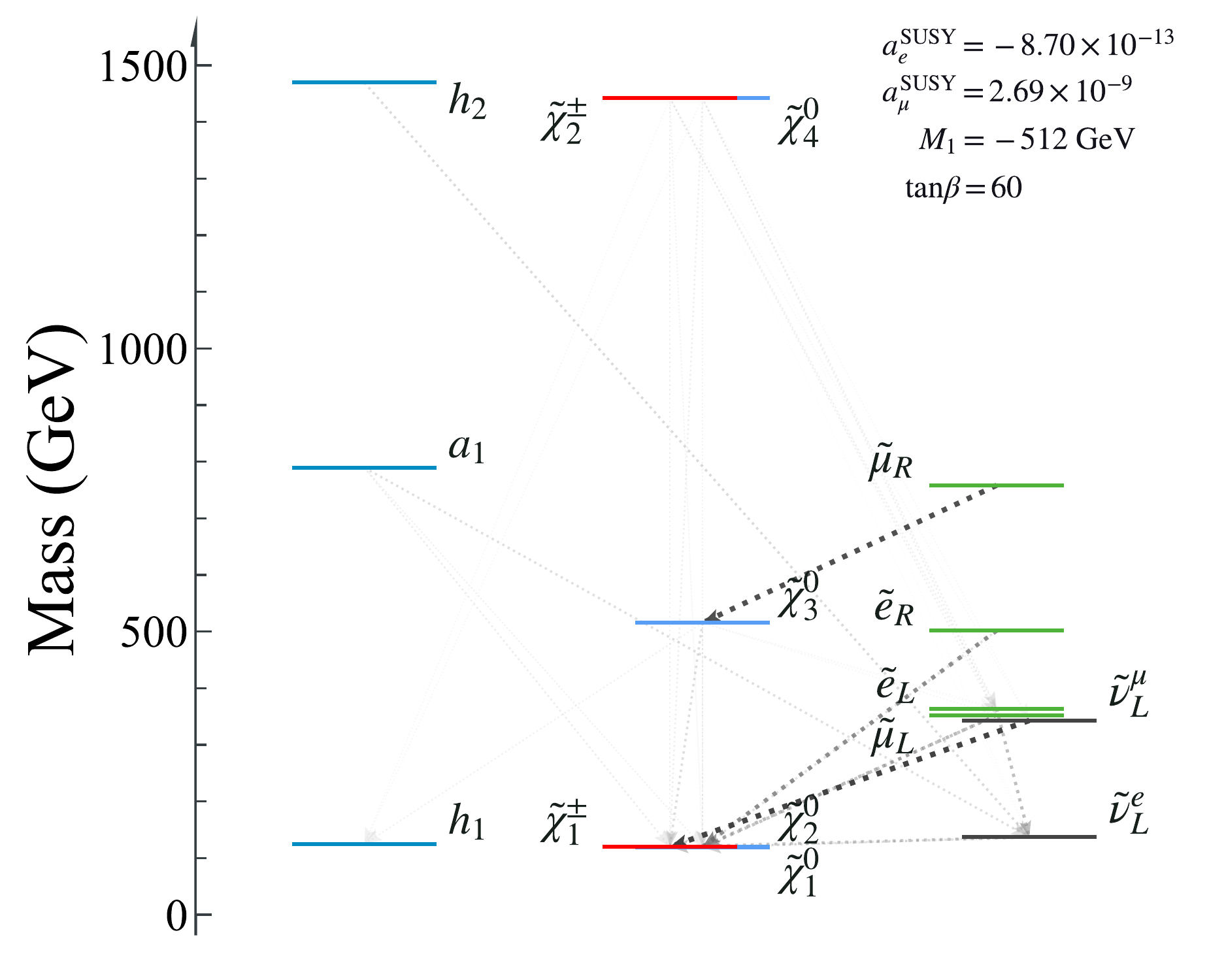}\hspace{0.5cm}
\includegraphics[width=0.465\textwidth]{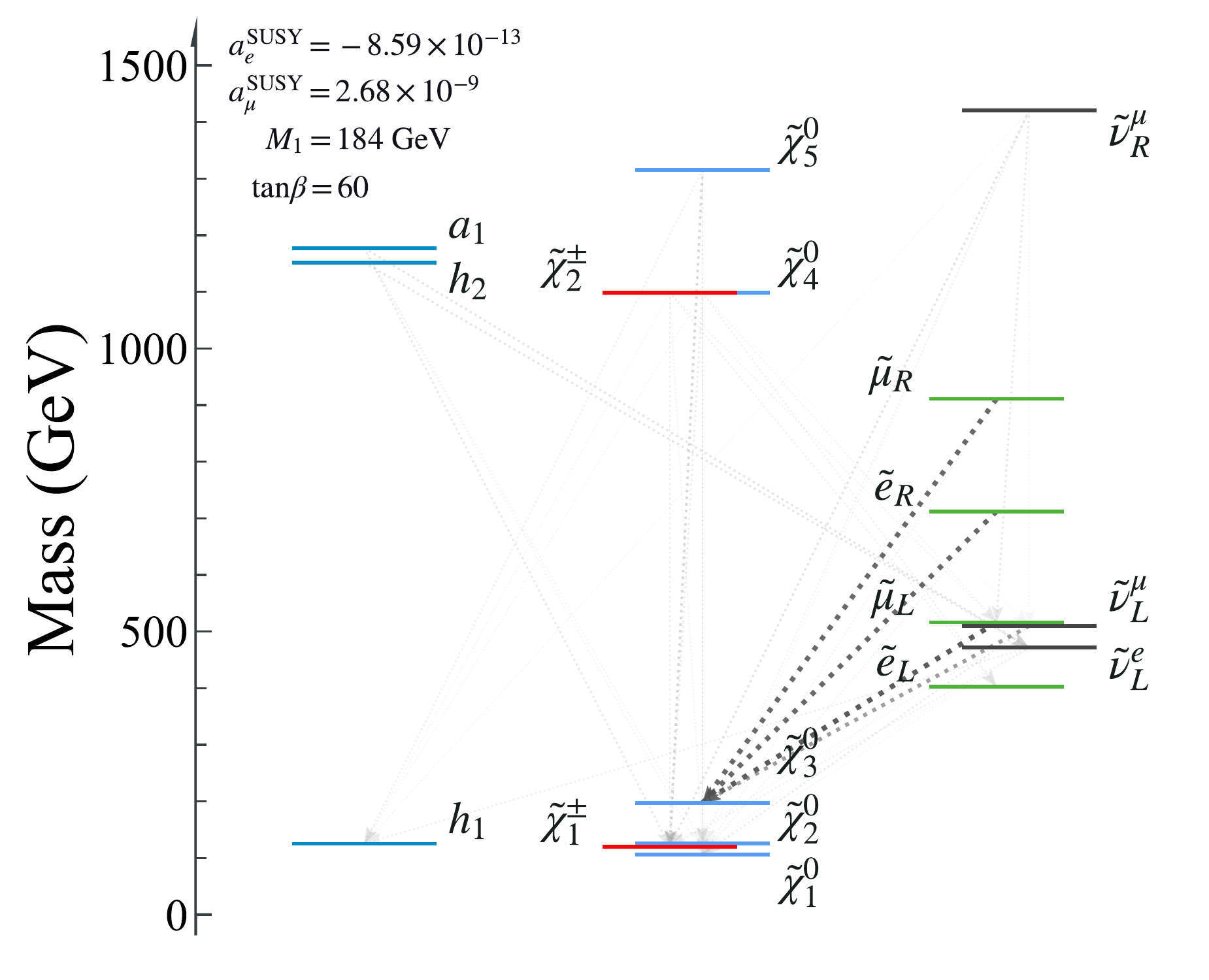}
\caption{\label{fig:typcspectr}Higgs and sparticle spectra for the typical samples in the ISS-NMSSM. Sleptons and sneutrinos are labeled by their dominant components, the decay paths are shown with branching ratios $>5\%$, and the widths of the lines are proportional to the branching ratios. }
\end{figure}

\par
\begin{figure}
	\centering
	\includegraphics[width=0.7\textwidth]{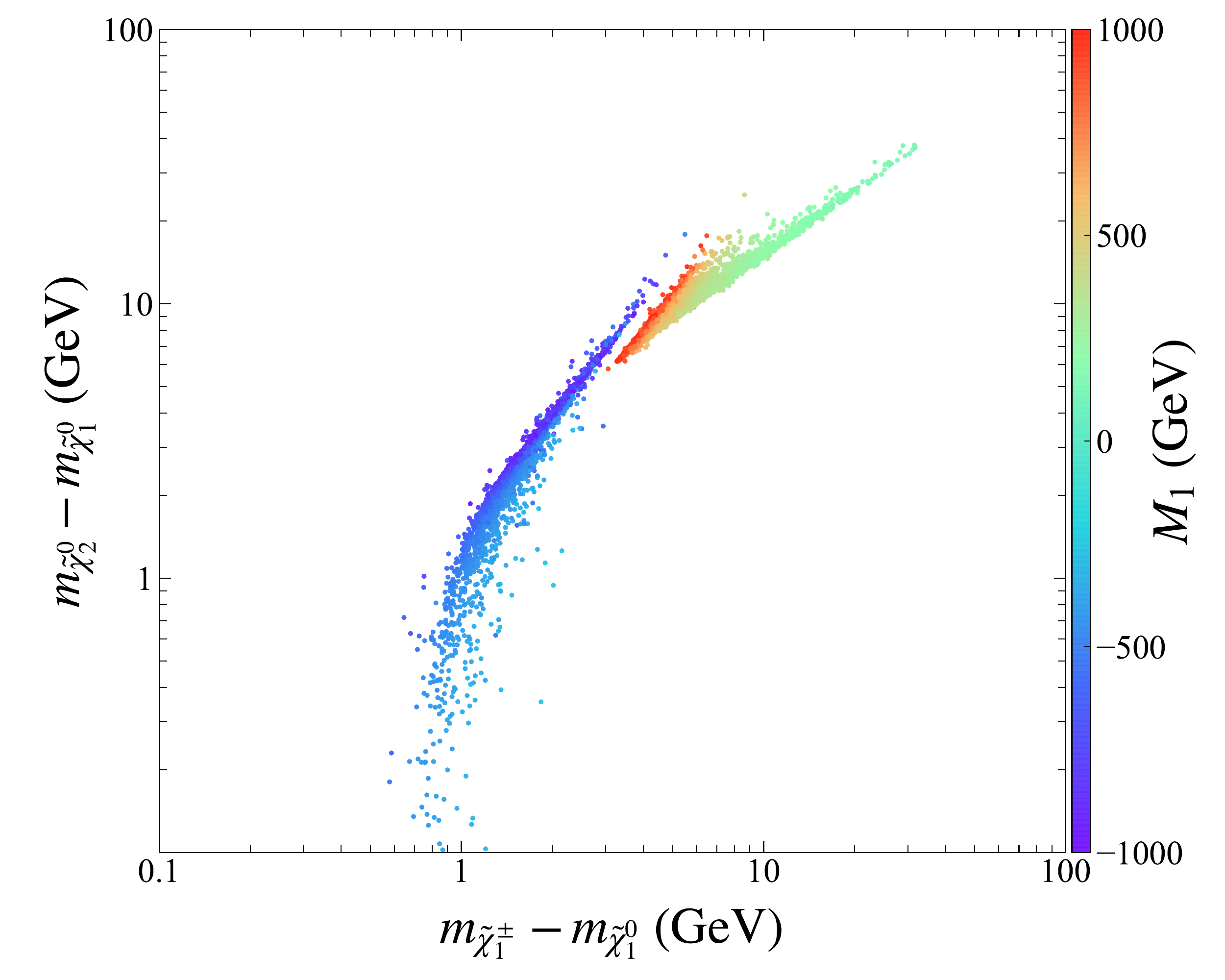}
	\caption{\label{fig:msplit}Mass splitting of Higgsino-dominated neutralinos $\tilde{\chi}_{1,2}^0$ and chargino $\tilde{\chi}_1^\pm$ for the scanned samples with the color bar indicating the value of $M_1$.}
\end{figure}

 The Higgs and sparticle spectra for two typical parameter points are shown in Fig.~\ref{fig:typcspectr}. $M_1$ may play a crucial role in the mass splitting of Higgsino-dominated neutralinos and chargino. Taking the mixing terms as a perturbation and calculating the neutralino and chargino masses to the first order in perturbation theory, the mass splitting between Higgsino-dominated electroweakinos are approximately given as follows: 

\begin{equation}\label{eq:msplit}\begin{split}
		\Delta m(\tilde{\chi}_2^0, \tilde{\chi}_1^0) &\approx \left| \frac{g_1^2 v^2 (M_1 + \mu \sin{2\beta})}{2(M_1^2 -\mu^2 )} + \frac{g_2^2 v^2(M_2 + \mu \sin{2\beta})}{2(M_2^2 - \mu^2)} + \frac{\lambda^2 v^2 (m_{\tilde{S}} - \mu\sin{2\beta})}{m_{\tilde{S}}^2 -\mu^2} \right| \\
		&\approx   \left| \frac{g_1^2 v^2 M_1}{2(M_1^2 -\mu^2 )} + \frac{g_2^2 v^2 M_2}{2(M_2^2 - \mu^2)}  \right|,\\
		\Delta m(\tilde{\chi}_1^\pm, \tilde{\chi}_1^0) & \approx \left| \frac{g_1^2 v^2 (1 + \sin{2\beta})}{2(M_1 -\mu )} + \frac{g_2^2 v^2(1  + \sin{2\beta})}{2(M_2 - \mu )} + \frac{\lambda^2 v^2 (1 - \sin{2\beta})}{m_{\tilde{S}} -\mu} \right.  \\
		&\left.+ \frac{\lambda^2 (g_1^2 + g_2^2) v^4}{2 (M_1 - \mu ) (M_{\tilde{S}} - \mu) \mu} - \frac{ g_2^2 v^2}{ 2 \mu} \left ( \frac{(m_2 \cos \beta + \mu \sin \beta)^2}{m_2^2 - \mu^2} - \cos^2 \beta \right ) \right|  \\
		&\approx \left| \frac{ g_1^2 v^2}{2 (M_1 - \mu)} +  \frac{ g_2^2 v^2}{2 (M_2 - \mu)} - \frac{g_2^2 v^2 \mu}{2(M_2^2 - \mu^2)} \right|,
\end{split}\end{equation}
where $m_{\tilde{S}} = 2\kappa v_s$ is the singlino mass. These formulae indicate that the effect of $M_1$ is negligibly small when $|M_1|$ is extremely large, while as it approaches zero from above (below), it can enhance (decrease) the splitting significantly. This characteristic is shown in Fig.~\ref{fig:msplit}, where we projected the scanned samples on the $\Delta m(\tilde{\chi}_1^\pm, \tilde{\chi}_1^0)- \Delta m(\tilde{\chi}_2^0, \tilde{\chi}_1^0)$ plane with the color bar denoting the $M_1$ value. This figure reveals that $\Delta m(\tilde{\chi}_1^\pm, \tilde{\chi}_1^0) \simeq 4~{\rm GeV}$ and $\Delta m(\tilde{\chi}_2^0, \tilde{\chi}_1^0) \simeq 10~{\rm GeV}$ when $|M_1| = 1~{\rm TeV}$. When $M_1 \simeq 200~{\rm GeV}$, the mass splittings increase to several tens of GeV, and when $M_1 \simeq -200~{\rm GeV}$, they decrease to less than $1~{\rm GeV}$. 

\section{\label{sec:dm}Dark matter phenomenology}
In the ISS-NMSSM explanation of the two leptonic anomalous magnetic moments $\Delta a_e$ and $\Delta a_\mu$, the Higgsino mass is less than $200~{\rm GeV}$ and acts as the LSP. Such a light Higgsino is not a good DM candidate due to its spin-dependent and spin-independent scattering rates with nuclei larger than the current experimental limits~\cite{Cao:2018rix, Baer:2016ucr, Athron:2017qdc, Profumo:2017ntc, Bagnaschi:2017tru, Kowalska:2018toh}. In the ISS-NMSSM, a right-handed-field- or $x$-field-dominated sneutrino can also serve as the DM candidate. The sizable neutrino Yukawa coupling $Y_\nu$ contributes significantly to the DM-nucleon scattering rate, so that $e$-type and $\mu$-type sneutrinos cannot act as DM candidates. Fortunately, $Y_\nu$ and $\lambda_N$ are chosen to be flavor diagonal, and the flavor mixing effect introduced by $\mu_X$ can be ignored. This leaves us room for taking the lightest $\tau$-type sneutrinos, $\tilde{\nu}_1^{\tau}$, and its charge conjugate sate, $\tilde{\nu}_1^{\tau \ast}$, as feasible DM candidates~\cite{Cao:2017cjf, Cao:2019aam, Cao:2019qng}.\footnote{In this study, we take $\tilde{\nu}_1^\tau$ as a complex field by setting $B_{\nu_X} = 0$. Consequently, $\tilde{\nu}_1^{\tau \ast}$ also acts as a DM candidate. The ISS-NMSSM is then a two-component DM theory~\cite{Cao:2019aam}. We add that DM physics requires non-trivial configurations in the theory's parameter space.} The candidates should be lighter than $\tilde{\chi}_1^0$ and dominated in components by the right-handed field $\tilde{\nu}_R$, the $\tilde{x}$-field, or their mixture. In addition, given the preferred parameters in the previous section, one can further restrict their properties. 
\subsection{Sneutrino dark matter}
	First, we consider the annihilation of $\tau$-type sneutrinos, whose thermally averaged cross section at the freeze-out temperature should satisfy $\langle \sigma v \rangle_{F} \sim 3 \times 10^{-26} \mathrm{cm^3/s}$ to obtain the measured abundance. As was pointed out in Ref.~\cite{Cao:2019aam}, the DM has three types of popular annihilation channels:
	\begin{itemize}
		\item $\tilde{\nu}_1^\tau  \tilde{\nu}_1^{\tau \ast} \to h_s h_s, A_s A_s$, where $h_s$ and $A_s$ denote the singlet-dominated $CP$-even and $CP$-odd Higgs bosons, respectively. Since the singlet Higgs dominated particles are much heavier than the DM, this annihilation processes are kinematically forbidden.  
		\item $\tilde{\nu}_1^\tau  \tilde{\nu}_1^{\tau \ast} \to \bar{\nu}_h \nu_h$, $\tilde{\nu}_1^\tau \tilde{\nu}_1^\tau \to \nu_h \nu_h$, and $\tilde{\nu}_1^{\tau \ast} \tilde{\nu}_1^{\tau \ast} \to \bar{\nu}_h \bar{\nu}_h$, where $\nu_h$ represents any of the heavy neutrinos. This process is mainly proceeded via $h_s$ in the $s$-channel and/or via the singlino-dominated neutralino in the $t$-channel. Evidently, the annihilation cross section is suppressed by the heavy mediator mass.  
		\item Co-annihilation with the Higgsinos. In this case, the effective annihilation rate at temperature $T$ takes the following form~\cite{Griest:1990kh}
			\begin{equation}\label{eq:annixsect}
    			\sigma_{\rm eff} = \sum_{a,b}\sigma_{ab}\frac{g_ag_b}{g^2_{\rm eff}}(1+\Delta_a)^{3/2}(1+\Delta_b)^{3/2}\times \exp[-x(\Delta_a+\Delta_b)], 
			\end{equation}
			where $\sigma_{ab}=\sigma(a b \rightarrow X Y )$, $a, b=\tilde{\nu}_1^\tau, \tilde{\nu}_1^{\tau \ast}, \tilde{\chi}_1^0, \tilde{\chi}_2^0, \tilde{\chi}_1^\pm$, $X$ and $Y$ denote any possible SM particles, $\Delta_i \equiv (m_i - m_{\tilde{\nu}_1^\tau})/m_{\tilde{\nu}_1^\tau} $ for $i=a, b$ represents the mass splitting between the initial particle $i$ and $\tilde{\nu}_1^\tau$,  $x \equiv m_{\tilde{\nu}_1^\tau}/T$, $g_i$ represents the $i$th particle's internal degrees of freedom, with $g_{\tilde{\chi}_{i}^0}=2$, $g_{\tilde{\chi}_{1}^\pm}=4$, and $g_{\tilde{\nu}_1^\tau} = g_{\tilde{\nu}_1^{\tau \ast}} = 2$, and the effective degree of freedom $g_{\rm eff}$ is
			\begin{equation}
				g_{\rm eff} \equiv   \sum_{a} g_a(1+\Delta_a)^{3/2}\exp(-x\Delta_a).
			\end{equation}
			Since the Higgsino pair annihilation cross section is much larger than $3 \times 10^{-26} {\rm cm^3}/s$ for $\mu \simeq 100~{\rm GeV}$, these formulas indicate that, even in the case of $\sigma_{\tilde{\nu}_1^\tau \tilde{\nu}_1^{\tau \ast}} \simeq 0$ and  $\sigma_{\tilde{\nu}_1^\tau \tilde{\nu}_1^{\tau}} = \sigma_{\tilde{\nu}_1^{\tau \ast} \tilde{\nu}_1^{\tau \ast}}\simeq 0$, the coannihilation can explain the abundance by adjusting appropriate $\Delta_i$s. For the samples in the Sec.~3, we verified that this is feasible by setting $Y_{\nu_\tau} \simeq 0$, $\lambda_{N_\tau} \simeq 0$ and adjusting parameters $M_{\nu_\tau}^2$ and $M_{X_\tau}^2$ to change $\Delta_i$s. We also found that the measured abundance requires $\Delta m(\tilde{\chi}_1^0, \tilde{\nu}_1^\tau) \sim 5~{\rm GeV}$ for $\Delta m(\tilde{\chi}_2^0, \tilde{\chi}_1^0) \simeq 0$ and $\Delta m(\tilde{\chi}_1^\pm, \tilde{\chi}_1^0) \simeq 1~{\rm GeV}$. With the increase in $\Delta m(\tilde{\chi}_2^0, \tilde{\chi}_1^0)$ and $\Delta m(\tilde{\chi}_1^\pm, \tilde{\chi}_1^0)$, $\Delta m(\tilde{\chi}_1^0, \tilde{\nu}_1^\tau) $ decreases monotonically to maintain the abundance.
	\end{itemize}  

Next, we consider the DM-nucleon scattering proceeded by the $t$-channel exchange of the $CP$-even Higgs bosons and $Z$ boson. The spin-dependent cross section is vanishing, and the spin-independent (SI) cross section is~\cite{Cao:2019aam}
\begin{equation}
	\sigma_{N}^{\rm SI} = \frac{1}{2}\left(\sigma_{\tilde{\nu}_1^\tau -N}^{\rm SI} + \sigma_{\tilde{\nu}_1^{\tau *}-N}^{\rm SI} \right) = \sigma_{N}^h + \sigma_{N}^Z,
\end{equation}
where $\sigma_N^h$ and $\sigma_N^Z$ with $N=p,n$ denote the Higgs-mediated and the $Z$-mediated contribution, respectively. For the preferred parameters in the last section, $\sigma_N^h$ and $\sigma_N^Z$ are approximated by
\begin{equation}\begin{split}
\sigma_N^h/{\rm cm^2} &\simeq 4.2 \times 10^{-44} \times \frac{C_{\tilde{\nu}_1^{\tau*} \tilde{\nu}_1^{\tau} \Re[H_u^0]}^2}{m_{\tilde{\nu}_1^\tau}^2}, \\
\sigma_n^Z/{\rm cm^2} &= 7.4 \times 10^{-39} \times \left ( Z^{\tau}_{11} \right )^4, \quad \quad \sigma_n^Z/{\rm cm^2}  =  4.2 \times 10^{-41} \times \left ( Z^{\tau}_{11} \right )^4,
\end{split}\end{equation}
where $C_{\tilde{\nu}_1^{\tau*} \tilde{\nu}_1^{\tau} \Re[H_u^0]}$ represents the coupling of the DM pair to the $CP$-even $H_u^0$ field and takes the following form:
\begin{equation}
C_{\tilde{\nu}_1^{\tau \ast} \tilde{\nu}_1^\tau {\rm Re}[H_u^0]} \simeq - \sqrt{2} \lambda_{N_\tau} A_{Y_{\nu_\tau}} Z_{11}^\tau Z_{12}^\tau - \lambda_{N_\tau} Y_{\nu_\tau} v_s Z_{11}^\tau Z_{13}^{\tau} - Y_{\nu_\tau}^2 v_u Z_{12}^\tau Z_{12}^{\tau}.
\end{equation}
Noting that $Z^\tau_{11}$ is proportional to $Y_{\nu_\tau}$, one can conclude that the scattering is suppressed in the case of a small $Y_{\nu_\tau}$ and $\lambda_{N_\tau}$. This case is favored by current and future DM direct detection experiments~\cite{Benabderrahmane:2019yhk}.

\subsection{Effect of dark matter embedding on sparticle signal at the LHC}
\begin{figure}[thbp]
	\centering
	\includegraphics[width=0.465\textwidth]{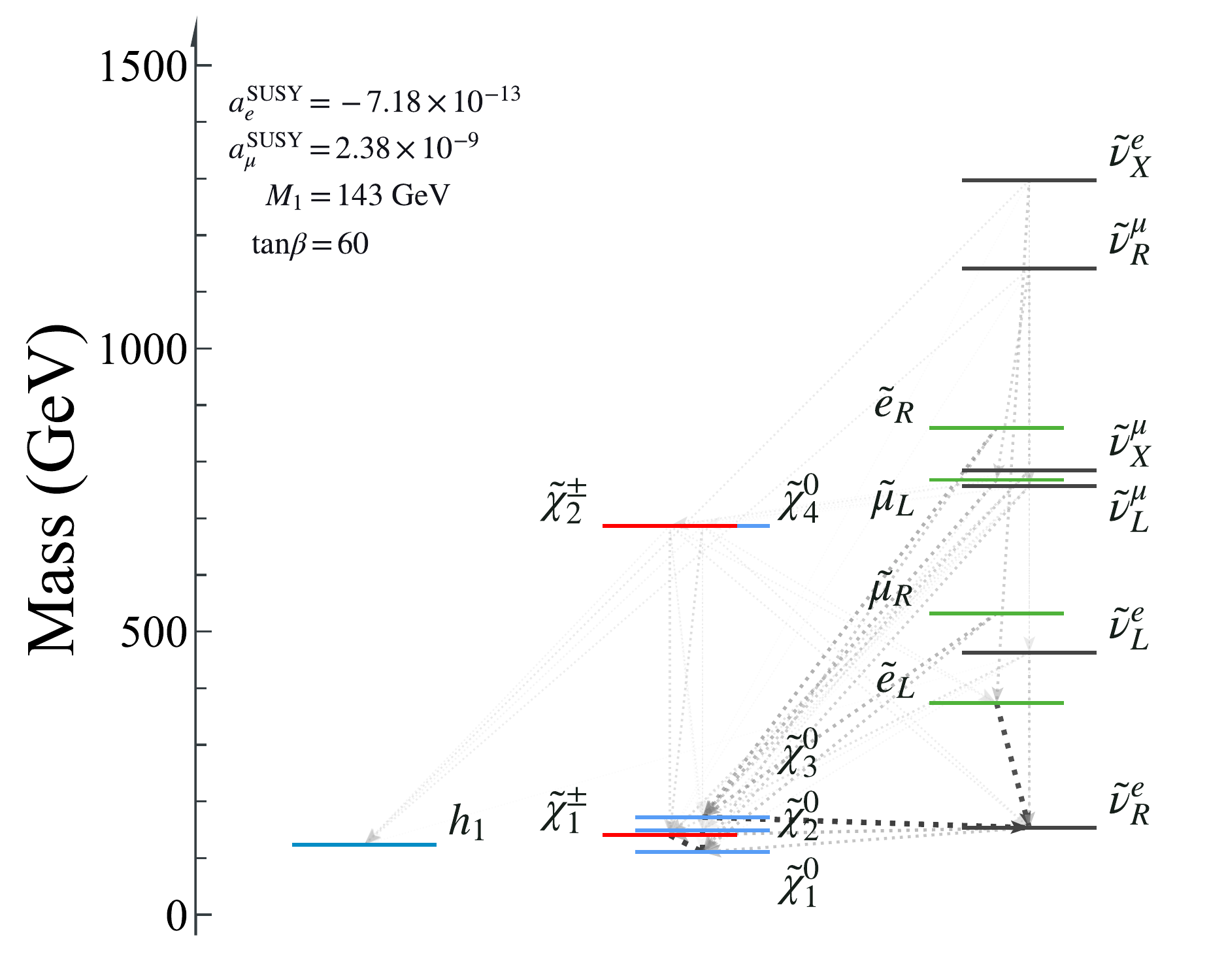}\hspace{0.5cm}
	\includegraphics[width=0.465\textwidth]{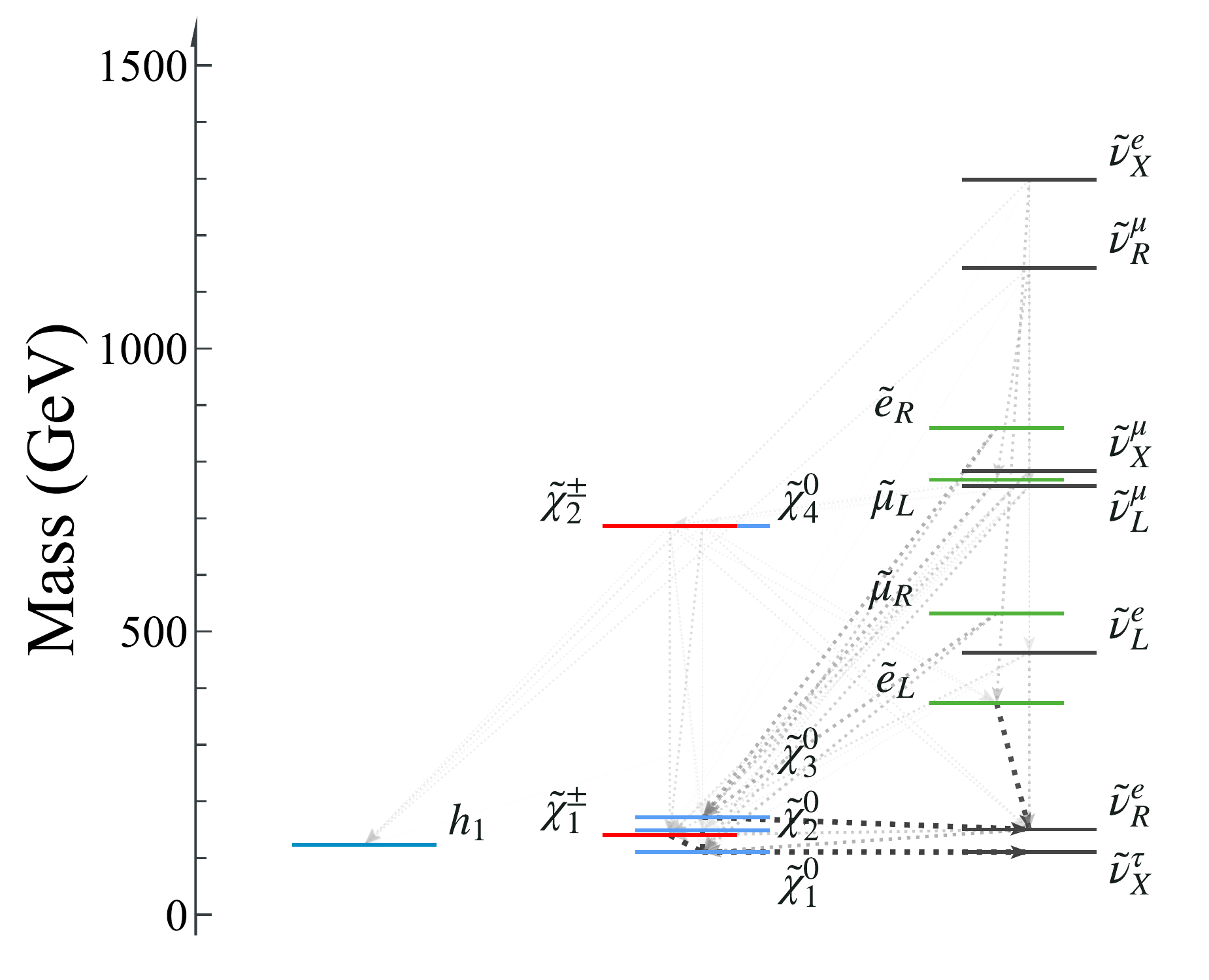}
	\caption{\label{fig:dminsert} Higgs and sparticle spectra for the benchmark sample with large Higgsino mass splitting before (left) and after (right) being inserted into  $\tau$-type sneutrino dark matter (DM). Sleptons and sneutrinos are labeled by their dominant components, the decay paths are shown with branching ratios $>5\%$, and the widths of the lines are proportional to the branching ratios.}
\end{figure}
In the ISS-NMSSM, the coupling of the DM to electroweakinos are
\begin{equation}\label{eq:coupdmnc}\begin{split}
	g_{\tilde{\nu}_1^{\tau}\bar{\tau} \tilde{\chi}_i^\pm} &=   Y_{e_\tau} U_{i2}^\ast Z^{\tau}_{11} P_L - \left( g_2 V_{i1} {Z_{11}^{\tau}}  - Y_{\nu_\tau}V_{i2} {Z_{12}^\tau} \right) P_R, \\
	g_{\tilde{\nu}_1^{\tau} \bar{\nu}_\tau \tilde{\chi}_i^0} &= \frac{1}{\sqrt{2}}(g_1N_{i1}-g_2 N_{i2}) P_{11} Z_{11}^{\tau} P_R  - (Y_{\nu_\tau} N_{i4} P_{11} Z_{12}^\tau  + \lambda_{N_\tau} N_{i5} P_{13} Z_{12}^{\tau} ) P_R \nonumber \\
&  - ( \lambda_{N_\tau} N_{i5}^\ast P_{12}^\ast Z_{13}^\tau +  Y_{\nu_\tau} N_{i4}^\ast P_{12}^\ast Z_{11}^\tau ) P_L,
\end{split}\end{equation}
where $P$ is a unitary matrix to diagonalize the neutrino mass matrix in one-generation bases ($\nu_L^\tau, \nu_R^\tau, x^{\tau}$). These expressions show that the coupling strengths are determined by $Y_{\nu_\tau}$ and $\lambda_{N_\tau}$, and they vanish when $Y_{\nu_\tau} = 0$. This characteristic has crucial applications in the phenomenology at the LHC. Concretely speaking, in the case where $Y_{\nu_\tau}$ and $\lambda_{N_\tau}$ are tiny and the DM achieves the measured abundance by the co-annihilation mechanism, the collider signal of the sparticle is roughly identical to that of the NMSSM with $\tilde{\chi}_1^0$ acting as the LSP~\cite{Cao:2019qng}.
This conclusion can be understood from the following aspects:
\begin{itemize}
\item The coannihilation mechanism is effective only when $m_{\tilde{\chi}_1^0} \simeq m_{\tilde{\nu}_1^\tau}$.
\item Since $\tilde{\chi}_1^0$ of the NLSP decays as $\tilde{\chi}_1^0 \to \tilde{\nu}_1^\tau \nu_\tau$, it appears as a missing track in the LHC detectors.
\item $\tilde{\chi}_2^0$ may decay by $\tilde{\chi}_2^0 \to \tilde{\nu}_1^\tau \nu_\tau$ and $\tilde{\chi}_2^0 \to \tilde{\chi}_1^0 Z^\ast \to \tilde{\chi}_1^0 f \bar{f}$. The two-body decay is always suppressed by the tiny coupling strength, while the phase space may suppress the three-body decay. We checked that in the case of $Y_{\nu_\tau} = \lambda_{N_\tau} = 10^{-3}$ and $m_{\tilde{\chi}_2^0} - m_{\tilde{\chi}_1^0} \lesssim 2~{\rm GeV}$, these two decay branching ratios are comparable in size, but the current LHC has no detection capability for such a small mass splitting of Higgsino-dominated triplet states~\cite{Aad:2019vnb}. In the case of  $m_{\tilde{\chi}_2^0} - m_{\tilde{\chi}_1^0} > 2~{\rm GeV}$, $\tilde{\chi}_2^0$ mainly decays by the three-body channel, so its signal is identical to that of the NMSSM with $\tilde{\chi}_1^0$ as the LSP.
\item The situation of $\tilde{\chi}_1^\pm$ is quite similar to that of $\tilde{\chi}_2^0$ except that it decays by $\tilde{\chi}_1^\pm \to \tilde{\nu}_1^\tau \tau$ and $\tilde{\chi}_1^\pm \to \tilde{\chi}_1^0 W^\ast \to \tilde{\chi}_1^0 f \bar{f}^\prime$.
\item For the other sparticles, their interactions with $\tilde{\nu}_1^\tau$ are weak, and thus, their decay chains do not change significantly.
\item Due to the constraints of DM direct detection experiments, the values of $Y_{\nu_\tau}$ and $\lambda_{N_\tau}$ tend to be relatively small. However, in order to satisfy the DM freeze-out mechanism, it is generally believe that the value of $Y_{\nu_\tau}$ and $\lambda_{N_\tau}$ should be greater than 0.001. We checked that the lifetime of NLSP $\tilde{\chi}_1^0$ is always less than $10^{15}~{\rm GeV}^{-1}$. \footnote{In some models of right-handed sneutrino as LSP~\cite{Evans:2016zau, Banerjee:2018uut}, the stau $\tilde{\tau}_1$ decays into the right-handed sneutrino via $\tilde{\tau}_1 \to W^{(*)} \tilde{\nu}_R$ driven by the tiny neutrino Yukawa coupling $Y_\nu \sin{\beta} \sim 10^{-13}$. Depending on the decay modes and mixing in $\tilde{\tau}$ and $\tilde{\nu}$ sectors, the lifetime of $\tilde{\tau}_1$ ranges from a few seconds to over ten years. However, the stau NLSP will not be a long-lived particle in our theory. One reason is that $Y_{\nu_\tau}$ should not so small to keep the LSP $\tilde{\nu}_1$ in thermal equilibrium in the early universe. Another reason is that the neutralino NLSP is usually predicted by the DM physics, and there is no requirement on the stau mass. }
\end{itemize}

Table~\ref{tab:dminsert} shows the details of a sample obtained in the previous section, including the mass spectra and decay modes of some moderately light sparticles. In Fig.~\ref{fig:dminsert}, we show the decay path of these sparticles to illustrate their properties further. This result indicates that the collider signature will not change after being embedded into the DM. Hence a natural way of a $\tau$-type sneutrino DM to achieve the measured relic density is coannihilation with Higgsino. In practice, the Yukawa parameters  $Y_{\nu_\tau}$ and $\lambda_{N_\tau}$ should be small to suppress the DM-nucleon scattering rate, the DM mass $m_{\tilde{\nu}_1^\tau}$ should be slightly lighter than the Higgsino mass, which is $m_{\tilde{\nu}_1^\tau} \approx 100~{\rm GeV}$. 
\par Since the embedded DM will increase the dimension of the scanning parameters, so we will ignore the DM constraints when studying the collider phenomenology. 
\begin{table}[th]
\centering
\makebox[\textwidth][c]{
\resizebox{1.01\textwidth}{!}{
\begin{tabular}{l c|l r|ll}
\cline{1-6}
\textbf{Parameters}          	   & \textbf{Value}    	& \textbf{Particles} 		& \textbf{Mass}  & \multicolumn{2}{c}{\textbf{Before}}   \\ \cline{1-4}
$M_1$                              & 143.2 GeV    	 	& $\tilde{\chi}_1^0$       	& 110.5 GeV & \textbf{Decays} & \textbf{Branching ratio[\%]}  \\ \cline{5-6}
$M_2$                              & 650.0 GeV    		& $\tilde{\chi}_2^0$      	& -149.4 GeV & $\tilde{\chi}_1^0$ (LSP) &  -  \\
$\mu$                              & 137.6 GeV   		& $\tilde{\chi}_3^0$      	& 172.3 GeV  &  $\tilde{\chi}_2^0 \to \tilde{\chi}_1^0 q \bar{q} / \tilde{\chi}_1^0 \ell \ell / \tilde{\chi}_1^0 \nu \nu /\tilde{\chi}_1^\pm q q\prime / \tilde{\chi}_1^\pm \ell \nu   $ & 60.67 / 9.29 / 29.44 /0.20  / 0.39 \\
$\lambda$, $\kappa$                & 0.0920, 0.6081     & $\tilde{\chi}_4^0$    	& 686.7 GeV &  $\tilde{\chi}_1^\pm \to \tilde{\chi}_1^0 q q^\prime / \tilde{\chi}_1^0 \ell \nu $ & 65.38 / 34.72   \\
$Y_{\nu_e}$, $\lambda_{N_e}$       & 0.9565, 0.7334		& $\tilde{\chi}_5^0$    	& 1807 GeV & $\tilde{\chi}_3^0 \to \tilde{\nu}_R^e \nu / \tilde{\chi}_1^{\pm} q q^\prime / \tilde{\chi}_1^{\pm} \ell \nu $ & 99.87 / 0.07 / 0.04   \\
$A_{\nu_e}$                        & 387.5 GeV   		& $\tilde{\chi}_1^\pm$  	& 140.3 GeV & $\tilde{\chi}_4^0 \to \tilde{\chi}_1^{\pm} W^\mp /\tilde{\chi}_i^0 h/ \tilde{\chi}_i^0 Z /\tilde{e}_L e /\tilde{\nu} \nu  $ & 31.91 / 14.35 / 15.31 / 15.58 / 22.56  \\
$M_{\ell_e}$                       & 332.0 GeV   		& $\tilde{\chi}_2^\pm$   	& 686.9 GeV & $\tilde{\chi}_2^\pm \to \tilde{\chi}_1^{\pm} h / \tilde{\chi}_1^\pm Z / \tilde{\chi}_i^0 W^{\pm} / \tilde{\nu} \ell / \tilde{e} \nu$ &  14.68 / 15.44 / 31.37 / 22.60 / 15.62 \\ \cline{5-6}
$M_{E_e}$                          & 883.5 GeV  		& $\tilde{e}_L$       		& 373.5 GeV   & \multicolumn{2}{c}{\textbf{After}}  \\ 
$Y_{\nu_\mu}$, $\lambda_{N_\mu}$   & 0.1701, 0.8588     & $\tilde{e}_R$      		& 859.9 GeV  & \textbf{Decays} &  \textbf{Branching ratio[\%]}  \\ \cline{5-6}
$A_{\nu_\mu}$                      & -1943 GeV 			& $\tilde{\mu}_L$        	& 767.8 GeV  & $\tilde{\chi}_1^0 \to \tilde{\nu}_X^\tau \nu$ &  100.0  \\
$M_{\ell_\mu}$                     & 746.4 GeV  		& $\tilde{\mu}_R$          	& 532.2 GeV  & $\tilde{\chi}_2^0 \to \tilde{\chi}_1^0 q \bar{q} / \tilde{\chi}_1^0 \ell \ell / \tilde{\chi}_1^0 \nu \nu /\tilde{\chi}_1^\pm q q\prime / \tilde{\chi}_1^\pm \ell \nu  $   &  57.81 / 8.85 / 32.54 / 0.20 / 0.58 \\
$M_{E_\mu}$                        & 569.2 GeV  		& $\tilde{\nu}_L^e$        	& 462.5 GeV   & $\tilde{\chi}_1^\pm \to \tilde{\chi}_1^0 q q^\prime / \tilde{\chi}_1^0 \ell \nu$  & 65.38 / 34.72   \\
$Y_{\nu_\tau}$, $\lambda_{N_\tau}$ & 0.0037, 0.0368 & $\tilde{\nu}_R^e$         & 150.5 GeV   & $\tilde{\chi}_3^0 \to \tilde{\nu}_R^e \nu / \tilde{\chi}_1^\pm q q^\prime / \tilde{\chi}_1^{\pm} \ell \nu $ & 99.91 / 0.06 / 0.03  \\
$Z_{11}^{\tau}$                    & $1.112\times 10^{-5}$     		& $\tilde{\nu}_X^e$         & 1297 GeV  & $\tilde{\chi}_4^0 \to \tilde{\chi}_1^{\pm} W^\mp /\tilde{\chi}_i^0 h/ \tilde{\chi}_i^0 Z / \tilde{e}_L e / \tilde{\nu} \nu $ & 31.89 / 14.34 / 15.30 / 15.57 / 22.60 \\
$m_{\tilde{\nu}_X^\tau}$ & 110.3 GeV	& $\tilde{\nu}_L^\mu$      	& 757.2 GeV & $\tilde{\chi}_2^\pm \to \tilde{\chi}_1^\pm h / \tilde{\chi}_1^\pm Z / \tilde{\chi}_i^0 W^\pm / \tilde{\nu} \ell / \tilde{e} \nu $ & 14.67 / 15.44 / 31.35 / 22.64 / 15.61  \\
$\Omega h^2$                       & 0.120              & $\tilde{\nu}_R^\mu$      	& 1142 GeV &  &    \\
$\sigma_{\tilde{\nu} -p}^{\rm SI}$ & $5.324\times 10^{-52}~{\rm cm}^2$ & $\tilde{\nu}_X^\mu$ &783.5 GeV &  &    \\ 
\cline{1-6}
\end{tabular}}}
\caption{\label{tab:dminsert} Input parameters, mass spectrum, and decay modes of the sample in Fig.~\ref{fig:dminsert} before and after inserting $\tau$-type sneutrino DM.}
\end{table}

\section{\label{sec:collider}Constraints from LHC sparticle searches}
Figure~\ref{fig:spmass} shows that to simultaneously explain $\Delta a_e$ and $\Delta a_\mu$, very light electroweakinos and sleptons are necessary in many samples. Generally, such light sparticles are strongly constrained by the current LHC SUSY searches. In a specific SUSY model, the mass hierarchy and decay modes of sparticles can be significantly different from the simplified model on which the experimental interpretations are based. In this section, we elaborated on the experimental searches we considered, discussed the sparticle decay modes on the ISS-NMSSM parameter space in interpreting $\Delta a_e$ and $\Delta a_\mu$, and summarized the impact of current LHC sparticle searches on the interpretation. 
\par Using the data taken at $\sqrt{s}=7,8$, and $13~{\rm TeV}$, searches for SUSY particles have been conducted for several years. First, we perform analyses at 8 and 13 TeV in \textsc{SModelS~v1.2}~\cite{Kraml:2013mwa, Ambrogi:2018ujg} to refine the samples. After this, any point that passes \textsc{SModelS} is further tested by analyses in \textsc{CheckMATE-2.0.26}~\cite{Drees:2013wra, Dercks:2016npn, Kim:2015wza}. The physical processes considered in our work are as follows:
\begin{equation}
\begin{split}
	pp \to \tilde{\chi}_i^0 \tilde{\chi}_j^\pm,& \quad i=2,3,4,5;\quad j=1,2;\\
	pp \to \tilde{\chi}_i^\pm \tilde{\chi}_j^\mp,& \quad i=1,2; \quad j=1,2;\\
	pp \to \tilde{\chi}_i^0 \tilde{\chi}_j^0,& \quad i=2,3,4,5; \quad j=2,3,4,5;\\
	pp \to \tilde{\ell}_i \tilde{\ell}_i,&\quad  i = e, \mu. 
\end{split}
\end{equation}
The cross section of $\sqrt{s}=8, 13~{\rm TeV}$ were normalized at the NLO using the \textsc{Prospino2} package~\cite{Beenakker:1996ed}. The Monte Carlo events were generated by \textsc{MadGraph\_aMC@NLO}~\cite{Alwall:2011uj, Conte:2012fm} with the \textsc{PYTHIA8} package ~\cite{Sjostrand:2014zea} for parton showering and hadronization. The event files were then input into \textsc{CheckMATE} for analysis with \textsc{Delphes}~\cite{deFavereau:2013fsa} for detector simulation. 
\par In addition to the analysis that has been implemented before, we added the following newly released LHC analyses to \textsc{CheckMATE}. 
\begin{itemize}
	\item \textbf{ATLAS search for chargino and neutralino production using recursive jigsaw reconstruction in three-lepton final states}~\cite{Aad:2019vvi}: this analysis is optimized for signals from $\tilde{\chi}_2^0 \tilde{\chi}_1^\pm$ production with on shell $WZ$ decay modes. The signal regions (SRs) are split into a low-mass region (jet veto) and the initial state radiation (ISR) region (contains at least one energetic jet) using a variety of kinematic variables, including the dilepton invariant mass $m_{\ell\ell}$, the transverse mass $m_{\rm T}$, and variables arising from the application of the emulated recursive jigsaw reconstruction technique. The smallness of the mass splittings lead to events with lower-$p_{\rm T}$ leptons or smaller $E_{\rm T}^{\rm miss}$ in the final state. Cuts in the low-mass SR are designed to reduce the $WZ$ background and the number of fake or nonprompt leptons, and cuts in the ISR region requiring large $E_{T}^{\rm miss}$ to identify events have a real $E_{\rm T}^{\rm miss}$ source. This search is sensitive to samples with relative light winos.
	\item \textbf{ATLAS search for chargino and slepton pair production in two lepton final states}~\cite{Aad:2019vnb}: this analysis targets pair production of charginos and/or sleptons decaying into final states with two electrons or muons. Signal events are required to have an exactly opposite-sign (OS) signal lepton pair with a large invariant mass $m_{\ell\ell} > 100~{\rm GeV}$ to reduce diboson and $Z+{\rm jets}$ backgrounds. SRs are separated into same-flavor and different-flavor categories with variables  $m_{\ell\ell}$, the stransverse mass $m_{\rm T2}$~\cite{Barr:2003rg}, $E_{\rm T}^{\rm miss}$ and $E_{\rm T}^{\rm miss}$ significance, and the number of non-$b$-tagged jets. The sensitivity of this analysis to the slepton mass can reach 700 GeV, and that to the chargino mass can reach about 1 TeV (420 GeV) of the decay mode $\tilde{\chi}_1^{\pm} \to \tilde{\ell} \nu / \ell \tilde{\nu} \to \ell \nu \tilde{\chi}_1^0$ ($\tilde{\chi}_1^\pm \to W^{\pm}(\to \ell \nu) \tilde{\chi}_1^0$). 
	\item \textbf{ATLAS search for electroweak production of supersymmetric particles with compressed mass spectra}~\cite{Aad:2019qnd}: this was optimized on a simplified model of mass-degenerated Higgsino triplets that assumed $\tilde{\chi}_2^0 \tilde{\chi}_1^\pm$ production followed by the decays $\tilde{\chi}_1^\pm \to W^{*}\tilde{\chi}_1^0$ and $\tilde{\chi}_2^0 \to Z^* \tilde{\chi}_1^0$. It is also sensitive to the degenerate slepton-LSP mass spectrum. The selected events have exactly two OS same-flavor leptons or one lepton plus at least one OS track, and at least one jet is required. The preselection requirements include the requirements that the invariant mass $m_{\ell\ell}$ is derived from the $J/\psi$ meson mass window, that $E_{\rm T}^{\rm miss}$ is greater than 120 GeV, and that the $p_{\rm T}$ of the leading jet is larger than 100 GeV. After applying the preselection requirements, SRs are further optimized for the specific SUSY scenario into three categories: SR-E (for electroweakino recoiling against ISR), SR-VBF [electroweakino produced through vector boson fusion (VBF)], and SR-S (sleptons recoiling against ISR). A variety of kinematic variables and the recursive jigsaw reconstruction technique are used to identify the SUSY signals. Assuming Higgsino production, this search occurs at the minimum mass of $\tilde{\chi}_2^0$ at 193 GeV at a mass splitting of 9.3 GeV. 
	\item \textbf{ATLAS search for chargino and neutralino pair production in final state with three-leptons and missing transverse momentum}~\cite{ATLAS:2021moa}: this search targets chargino-neutralino pair production decaying via $WZ$, $W^* Z^*$, or $Wh$ into three-lepton final states. This analysis uses the full LHC~run~\rom{2} dataset. The simplified model has an $\tilde{\chi}_2^0$ mass of up to 640 GeV for on shell $WZ$ decay mode with massless $\tilde{\chi}_1^0$, up to 300 GeV for the off shell $WZ$ decay mode, and up to 185 GeV for the $Wh$ decay mode with an $\tilde{\chi}_1^0$ mass below 20 GeV. 
	\item \textbf{ATLAS search for supersymmetric states with a compressed mass spectrum}~\cite{Aaboud:2017leg}: this analysis uses the OS lepton pair and large $E_{\rm T}^{\rm miss}$, searching for the electroweakino and slepton pair production with a compressed mass spectrum. Two sets of SRs are constructed separately for the production of electroweakinos and sleptons. The electroweakino SRs require the invariant mass of the lepton pair $m_{\ell \ell}$ to be less than 60 GeV, and the slepton SRs require the stransverse mass $m_{\rm T2}^{m_\chi}$ to be greater than 100 GeV, where the hypothesized mass of the LSP $m_\chi$ is equal to 100 GeV. The most sensitive location of the mass splitting is at about 5--10 GeV. The 95\% confidence level exclusion limits of the Higgsino, wino, and slepton are up to 145, 175, and 190 GeV, respectively. 
	\item \textbf{CMS combined search for charginos and neutralinos}~\cite{Sirunyan:2018ubx}: various simplified models of the SUSY are used in this combined search to interpret the results. Related to our work, the simplified model scenario interpretation of $\tilde{\chi}_2^0 \tilde{\chi}_1^{\pm}$ with decays $\tilde{\chi}_2^0 \to Z^{(*)} \tilde{\chi}_1^0 / h \tilde{\chi}_1^0$ and $\tilde{\chi}_1^{\pm} \to W^{(*)} \tilde{\chi}_1^0$ represents the most stringent constraints from the CMS to date for electroweakino pair production. Compared with the results of individual analyses, this interpretation improves the observed limit in $\tilde{\chi}_1^\pm$ to about 650 GeV for $WZ$ topology.
	\item \textbf{ATLAS search for electroweakino production in $Wh$ final states}~\cite{Aad:2019vvf}: this was optimized on a simplified model that assumed $\tilde{\chi}_2^0 \tilde{\chi}_1^\pm$ production with decay modes $\tilde{\chi}_1^\pm \to W^{\pm} \tilde{\chi}_1^0$ and $\tilde{\chi}_2^0 \to h \tilde{\chi}_1^0$. Signal events were selected with exactly one lepton, two $b$-jets requiring $100~{\rm GeV} < m_{bb} < 140~{\rm GeV}$ and $E_{\rm T}^{\rm miss} > 240~{\rm GeV}$, a transverse mass of the lepton-$E_{\rm T}^{\rm miss}$ system $m_{\rm T}$ greater than 100 GeV, and a ``contransverse mass'' $m_{\rm CT}$~\cite{Tovey:2008ui, Polesello:2009rn} greater than 180 GeV. Masses of the winos up to 740 GeV are excluded at 95\% confidence level for the massless LSP. 
	\item \textbf{CMS search in final states with two OS same-flavor leptons, jets, and missing transverse momentum}~\cite{CMS:2020bfa}: This search is sensitive to the on shell and off shell $Z$ boson from BSM processes and to direct slepton production. Search regions are split into on-$Z$ SRs, off-$Z$ SRs, and slepton SRs via various kinematic variables, including the invariant mass of the lepton pair $m_{\ell\ell}$, $M_{\rm T2}$, the scalar sum of jet $p_{\rm T}$, the missing transverse momentum $E_{\rm T}^{\rm miss}$, the number of jets $n_j$, and the number of $b$-tagging jets $n_b$. The result interpretation using the simplified model assuming direct slepton pair production with 100\% decay into dilepton final states shows that the probing limit of the slepton mass $m_{\tilde{\ell}}$ is up to 700 GeV. Certainly, this search is sensitive to the sparticles in the ISS-NMSSM interpretation of $\Delta a_\ell$.  
\end{itemize} 
Appendix~\ref{app:lhcana} shows a part of the validation table of the analyses above. We used the $R$ values obtained from \textsc{CheckMATE} to apply the LHC constraints. Here, $R\equiv \max\{S_{i}/S_{i,95}^{\rm obs}\}$ for individual analysis, in which $S_i$ represents the simulated event number of the $i_{\rm th}$ SR or bin of the analysis, and $S_{i, 95}^{\rm obs}$ is the 95\% confidence level upper limit of the event number in the corresponding SR or bin. The combination procedure of the CMS electroweakino search~\cite{Sirunyan:2018ubx} was also performed though the $\rm CL_s$ method~\cite{Read:2002hq} with \textsc{RooStats}~\cite{Schott:2012zb} using the likelihood function described previously \cite{Sirunyan:2018ubx}. 
\begin{figure}
	\centering
	\includegraphics[width=\textwidth]{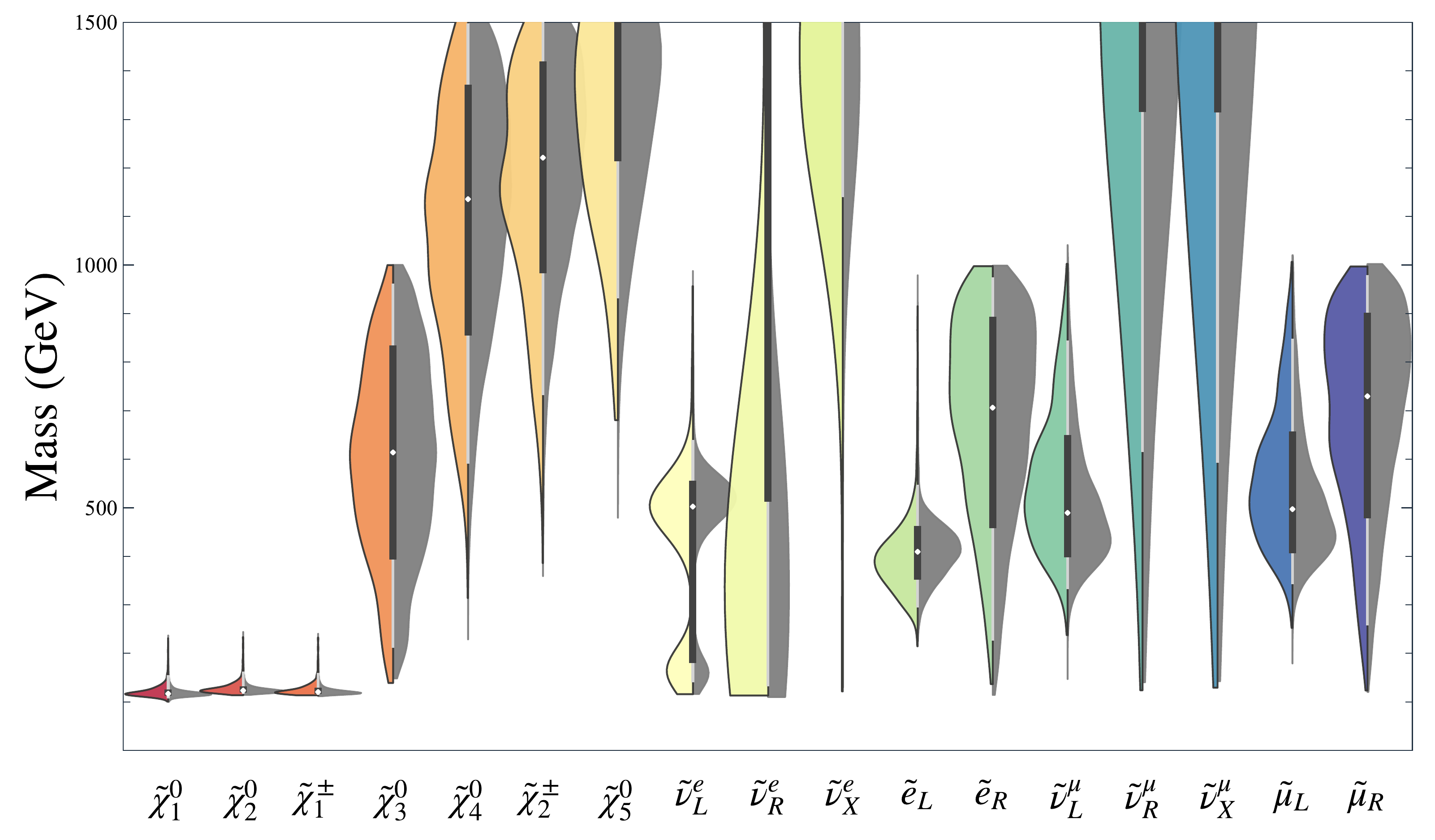}
	\caption{\label{fig:violhc}Split violin plot showing the SUSY particle mass distributions of samples based on the LHC result. The left colorful kernels are the mass distributions of samples that survive the LHC search results, and the right gray kernels indicate samples excluded by the LHC results. The medians, interquartile ranges, and 95\% confidence intervals are the same as those in Fig.~\ref{fig:spmass}. }
\end{figure}
\par
\begin{figure}[th]
	\centering
	\includegraphics[width=0.495\textwidth]{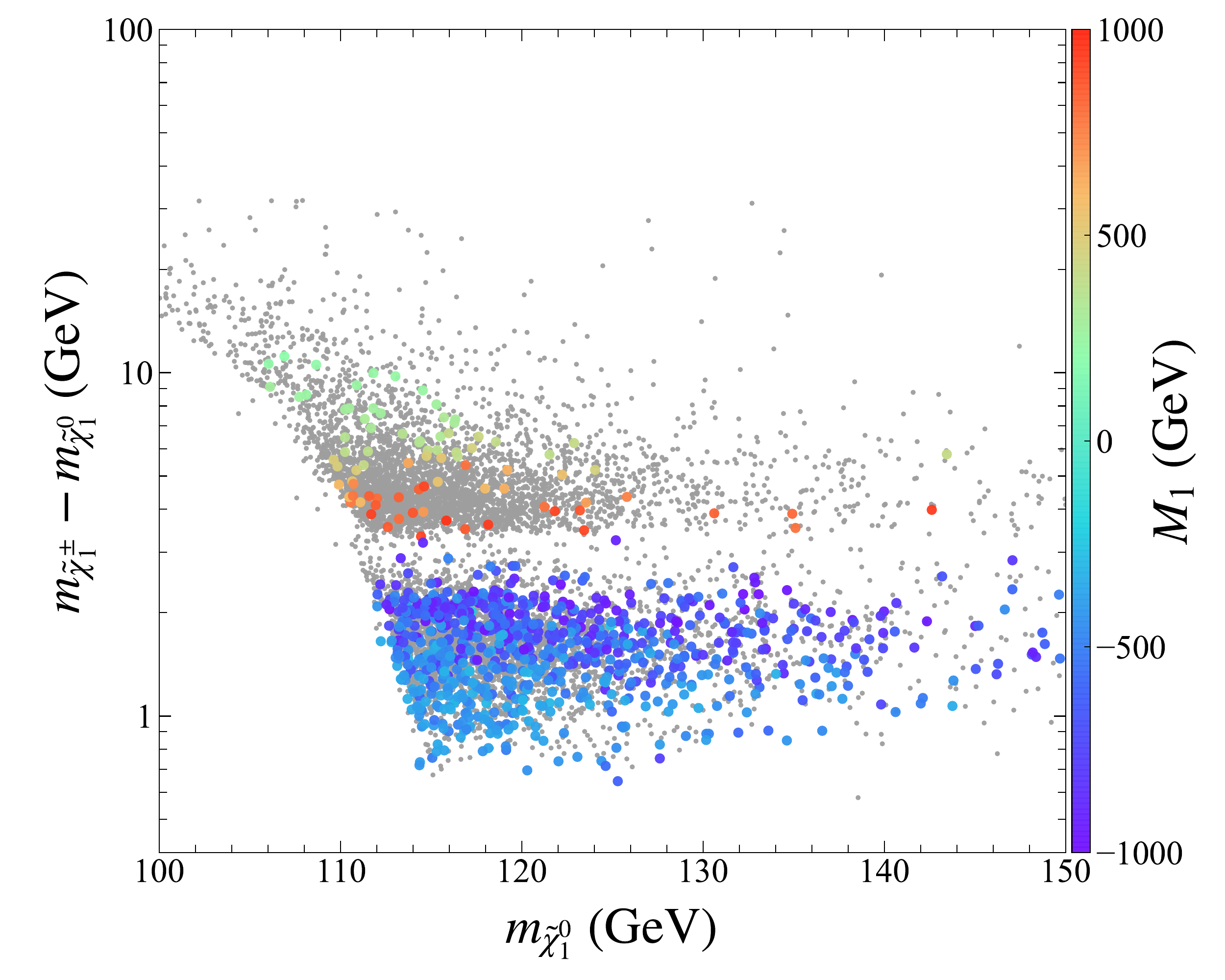}
	\includegraphics[width=0.495\textwidth]{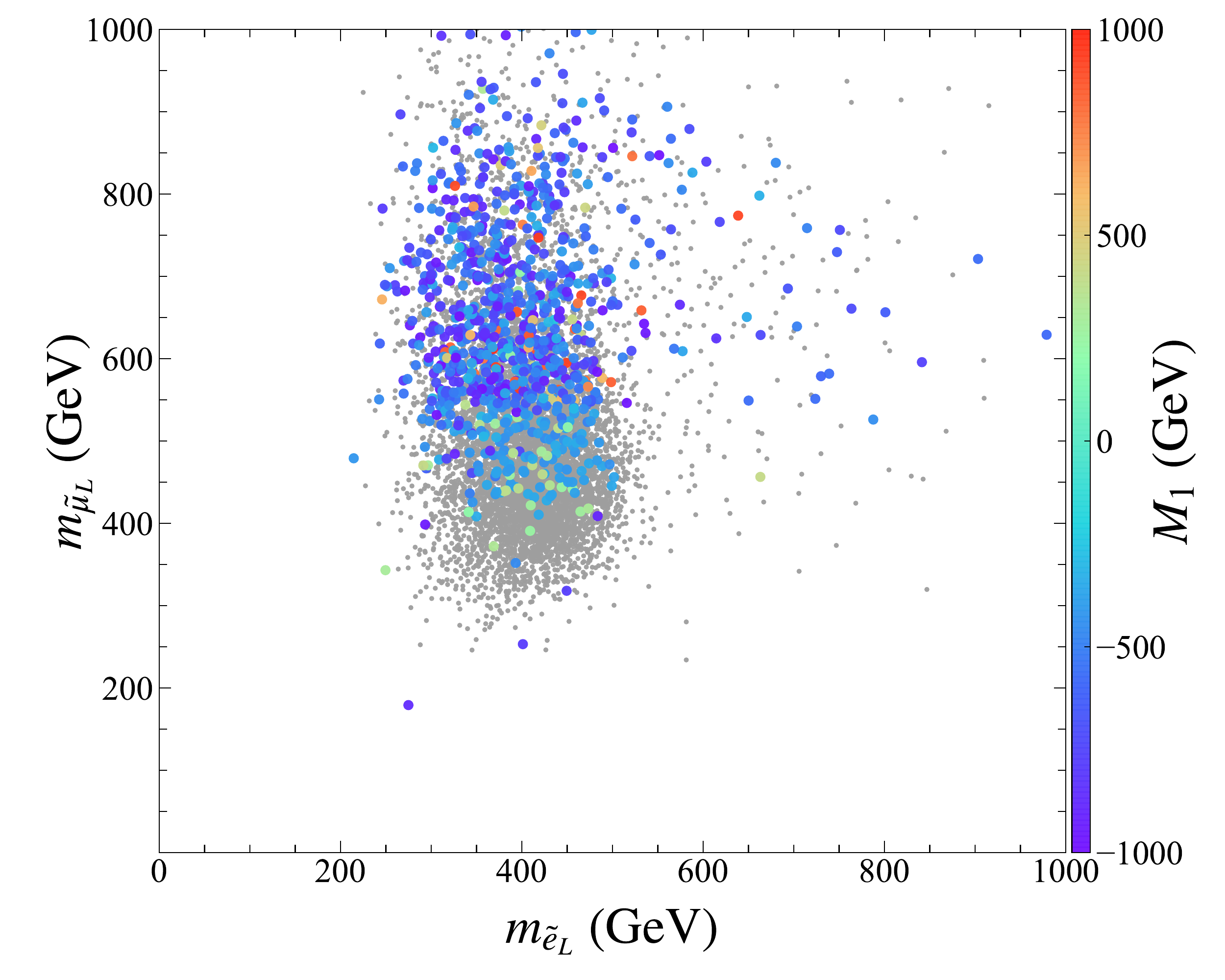}	
	\caption{\label{fig:scalhc}Samples projected on $m_{\tilde{\chi}_1^0} - \Delta m (\tilde{\chi}_1^0, \tilde{\chi}_1^\pm)$ plane and $m_{\tilde{e}_L} - m_{\tilde{\mu}_L}$ plane, where the color indicates the value of $M_1$. In each plane, the gray points indicate the samples excluded by the LHC results.} 
\end{figure}

 The impacts of LHC constraints on the samples were relatively strong, as 7029 samples were excluded, and only 960 samples survived, corresponding to a total posterior probability of 11.2\%. In Fig.~\ref{fig:violhc}, we plot the SUSY particle mass of the surviving samples and excluded samples via a split violin plot.\footnote{Similar to a violin plot, the split violin plot splits the violins in half to see the difference between two sample groups. Note that the widths of both sides of the violin are fixed, so that the ratio of the two widths does not represent the relative probability or relative number. } The LHC constraints have not significantly changed the distributions of the SUSY particle mass. In Fig.~\ref{fig:scalhc}, we plot the samples on the $m_{\tilde{\chi}_1^0} - \Delta m (\tilde{\chi}_1^0, \tilde{\chi}_1^\pm)$ plane and the $m_{\tilde{e}_L} - m_{\tilde{\mu}_L}$ plane with $M_1$ valued color.  The detection ability of the LHC was mainly affected by the decay modes of SUSY particles.  
\begin{enumerate}
	\item For the light left-handed slepton pair production processes, 
	\begin{itemize}
	\item when $\tilde{\nu}_1^e$ is very light, as shown by the samples in Fig.~\ref{fig:dminsert}, the dominant decay mode was $\tilde{e}_L \to W \tilde{\nu}_1^e$ (the sample in Fig.~\ref{fig:dminsert} for example), where $\tilde{\nu}_1^e$ contains large left-handed ingredients~\footnote{There is only one left-handed sneutrino $\tilde{\nu}_L^e$ in the MSSM or NMSSM, and its mass is slightly lighter than that of the left-handed selectron $\tilde{e}_L$}.  Such samples are difficult to detect by the current LHC experiments. In the right plane of Fig.~\ref{fig:scalhc} and Fig.~\ref{fig:violhc}, we find that the $m_{\tilde{e}_L}$ values of the surviving samples can reach $200~{\rm GeV}$, and such a light $\tilde{e}_L$ is good for explaining $\Delta a_e$. 
	\item when $|M_1|$ is lighter than $m_{\tilde{e}_L}$ or $m_{\tilde{\nu}_L}$, if $\tilde{e}_L \to W \tilde{\nu}_1^e$ is kinematically forbidden, as shown in the right panel of Fig.~\ref{fig:typcspectr}, the dominated decay mode of $\tilde{\ell}_L $ is into $\ell$ plus a bino-like neutralino $\tilde{\chi}^0$. In this case, the LHC slepton pair production searches provide the most sensitivity to the slepton mass. Figure~\ref{fig:scalhc} shows that samples with $\tilde{\mu}_L < 400~{\rm GeV}$ and small $|M_1|$ have difficulty escaping the LHC constraints. 
	\item when $|M_1|$ is very large and $\tilde{\nu}^e_1$ is very heavy, the left-handed slepton will decay into $\ell \tilde{\chi}_1^0$, $\ell \tilde{\chi}_2^0$, and $\nu  \tilde{\chi}_1^\pm$, as depicted in the left panel of Fig.~\ref{fig:typcspectr}. Because the decay products of $\tilde{\chi}_1^\pm$ are too soft, the LHC constraints for these samples are weaker than those in the light bino case. 
	\end{itemize}
	\item For light right-handed slepton pair production processes, the production cross section is about 2.7 times smaller than that of left-handed slepton. The right-handed sleptons mainly decay into $\ell\tilde{\chi}_1^0$ and $\ell \tilde{\chi}_2^0$. Only a few samples contain light $\tilde{\mu}_R$ or $\tilde{e}_R$, so the right-handed slepton had little effect on the result. 
	\item For the Higgsino-dominated electroweakino pair $\tilde{\chi}_2^0 \tilde{\chi}_1^\pm$ production process, the analysis in Ref.~\cite{Aad:2019qnd} provides a strong constraint on the samples featured by the small positive $M_1$, as shown in the left plane of Fig.~\ref{fig:scalhc}.
	\item For the wino-dominated neutralino/chargino pair production process, the explanation of the two lepton anomalies require $M_2 > 400~{\rm GeV}$, and in about more than 97\% samples, the wino-like particle mass is greater than $700~{\rm GeV}$. Therefore, in this study, because the decay modes of the winos are more complicated, the winos did contribute to the results, but they were not the main effect. 
\end{enumerate}
\begin{figure}[th]
	\centering
	\includegraphics[width=0.7\textwidth]{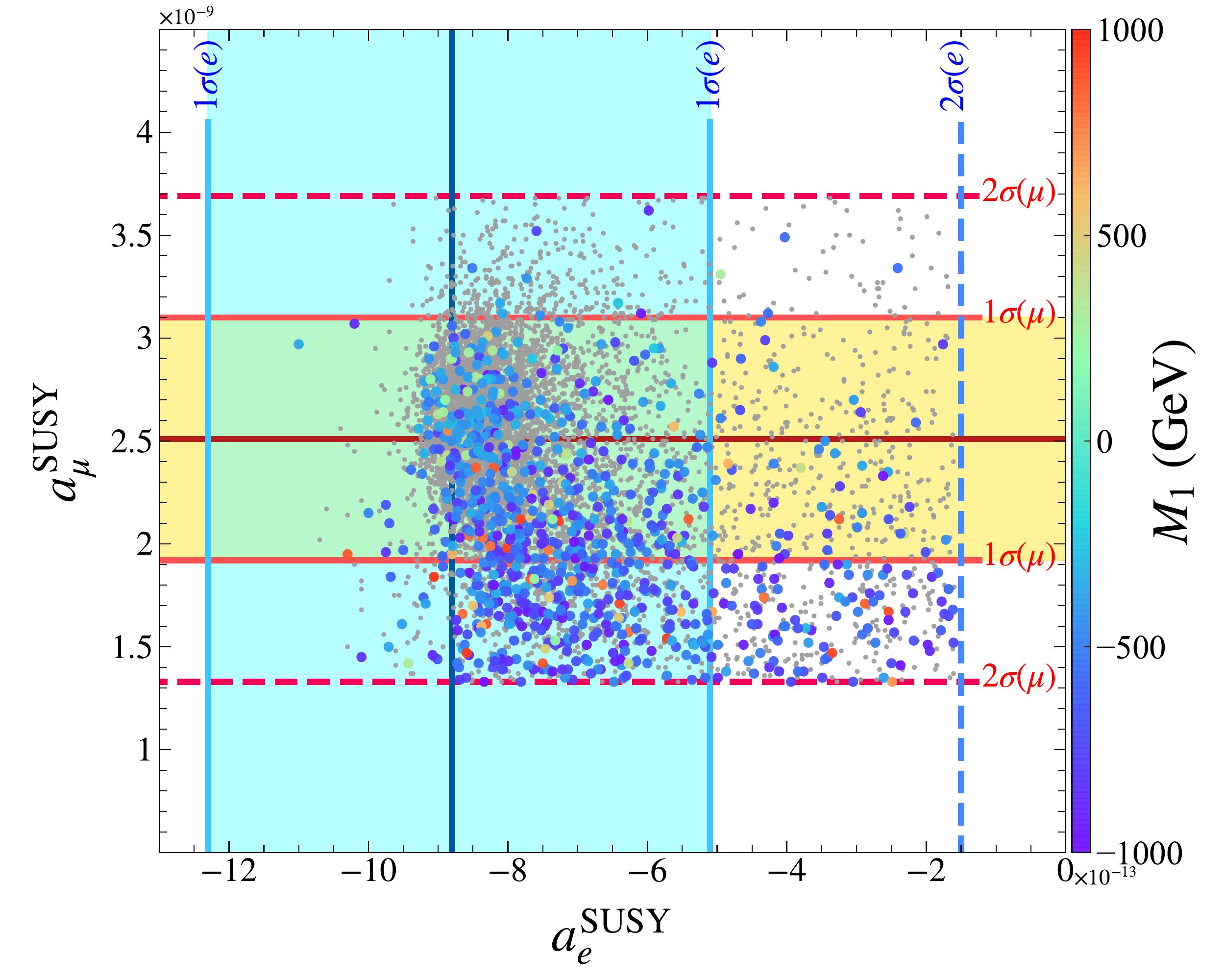}
	\caption{\label{fig:ammlhc}Similar to Fig.~\ref{fig:scalhc} but on $a_e^{\rm SUSY} - a_{\mu}^{\rm SUSY}$ plane with the $M_1$ values represented by the color.}
\end{figure}
In conclusion, except for some samples where $\tilde{e}_L$ decayed to $W\tilde{\nu}_1^e$, the samples with positive $M_1$ were strongly restricted by the current LHC experiments. The searches for the compressed electroweakino spectrum and the searches for sleptons at the LHC are complementary in detecting samples via leptonic final states. Finally, we plot the $a_e^{\rm SUSY}$ and $a_\mu^{\rm SUSY}$ values of the samples in Fig.~\ref{fig:ammlhc}. After considering the constraints from the LHC, there were still a large number of samples that could simultaneously fit $\Delta a_e$ and $\Delta a_\mu$ at the $1\sigma$ level. 
\par Before we conclude this section, we briefly comment that the signal of the sparticles may differ significantly from that of the NMSSM in the case of sizable $Y_{\nu_\tau}$ and $\lambda_{N_\tau}$~\cite{Cao:2017cjf, Cao:2019aam, Cao:2019qng}. In general, because the decay chain becomes lengthened and the signal contains at least two $\tau$ leptons, the sparticles of the ISS-NMSSM are more challenging to detect at the LHC than those of the NMSSM. We drew this conclusion by globally fitting the ISS-NMSSM with various experimental data and studying the scanned samples, similar to our previous work ~\cite{Cao:2019aam}. Therefore, we may overestimate the LHC constraints in this work, but this does not affect the main conclusion that the ISS-NMSSM can easily explain both anomalies. 
\par When this work was about to be finished, a new measurement of fine structure constant, via the rubidium atom, was reported~\cite{Morel:2020dww}:
\begin{equation}
	\alpha^{-1}({\rm Rb}) = 137.035999206(11),
\end{equation}     
which leads to a $+1.6\sigma$ discrepancy of $\Delta a_e$:
\begin{equation}\label{eq:daeRb}
	\Delta a_e({\rm Rb}) = a_e^{\rm exp} - a_{e}^{\rm SM,Rb} = (48\pm30)\times 10^{-14}. 
\end{equation}
What needs to be emphasized is that there is a $5.4\sigma$ discrepancy between two measurements $\alpha^{-1}({\rm Cs})$ and $\alpha^{-1}({\rm Rb})$~\cite{Morel:2020dww}. It is currently suspected that this difference may be caused by speckle or by a phase shift during the measurement, which requires further study. 
\par Currently, the difference of $\alpha({\rm Rb})$ and $\alpha({\rm Cs})$ are in dispute. We are more concerned about the impacts on the theory. If the future experimental study confirm the correctness of $\alpha({\rm Rb})$, that is there is no significance excess of $a_e$ within $2\sigma$ confidence level, then our previous work \cite{Cao:2019evo} shows that in ISS-NMSSM has a large parameter space to explain $\Delta a_\mu$. Especially, the HS contribution alone can explain the central value of $\Delta a_\mu$, and $\mu$ can be larger than $500~{\rm GeV}$ (see the benchmark point in Table 2 in \cite{Cao:2019evo}) correspondingly. On the other hand, if $\alpha ({\rm Cs})$ is confirmed in the future, then this work provides a supersymmetric solution without introducing leptonic flavor violation to explain $\Delta a_e$ and $\Delta a_\mu$ simultaneously.

\section{\label{sec:summary}Summary}
We proposed scenarios in the ISS-NMSSM that explain both electron and muon $g-2$ anomalies without introducing leptonic flavor violation. The results are summarized as follows:
\begin{itemize}
	\item The HS loop induced by the neutrino Yukawa coupling $Y_\nu$ plays an important role in the explanation. The advantage of this explanation is that the sign of the HS contribution to $a_\ell$ can be determined by the sign of $A_{\nu_\ell}$, which greatly reduces the correlation between $a_e$ and $a_\mu$. A larger HS contribution corresponds to a light $\mu$, which is natural for predicting $m_Z$. 
	\item The features of the sparticle spectrum preferentially had large $\tan{\beta}$, light Higgsino mass $\mu \simeq 110~{\rm GeV}$, and heavy wino $M_2 \gtrsim 600~{\rm GeV}$. Moreover, the masses of the left-handed selectron and smuon were around 400 and 500 GeV, respectively, and the singlet Higgs-dominated particles are often heavier than 1 TeV. $M_1$ affects the mass splitting of Higgsino triplets and the decay mode of the sleptons. 
	\item The mass spectrum can be introduced into a right-handed or $x$-field-dominated $\tau$-type sneutrino as a proper DM candidate, which coannihilated with Higgsinos to obtain the observed DM relic density. As indicated by the previous discussion, choosing $\lambda_{N_\tau}$ and $Y_{\nu_\tau}$ to be less than about 0.001 can avoid the change of the LHC signal caused by the introduction of the DM, and the corresponding DM direct detection cross section is much smaller than the detection ability of the current experiment.
	\item The signals of the electroweakino/slepton pairs produced at the LHC are sensitive to the parameter space explaining the two anomalies, especially for a positive $M_1$. However, due to the compressed mass spectrum, the insensitive decay mode of $\tilde{e}_L \to W \tilde{\nu}_1^e$, and the very heavy wino, the surviving samples can satisfy the current LHC constraints.
\end{itemize}
 
\section*{Acknowledgments}
Junjie Cao and Pengxuan Zhu wish to thank Jia Liu for the helpful discussion of the experimentally measuring the fine structure constant. This work is supported by the National Natural Science Foundation of China (NNSFC) under Grant No. 11575053, No. 11905044 and No. 12075076. 
\appendix
\section{\label{app:lhcana}Validations of LHC analyses}
This appendix verifies the correctness of our implementation of the necessary analyses in the package \textsc{CheckMATE}. For the sake of brevity, we only provide validation of the latest analyses. Table~\ref{tab:app-vld1908} shows the cut-flow validation of the analysis in Ref.~\cite{Aad:2019vnb} for chargino pair production channel. Table~\ref{tab:app-vld1911} shows the cut-flow validation for the slepton pair production channel of analysis~\cite{Aad:2019qnd}. All the cut-flow data were provided by experimental groups. The results indicate that our simulations were in good agreement with the analysis of the experimental groups. 

\begin{table}[ht]
	\centering
	\caption{\label{tab:app-vld1908}Cut-flow validation of ATLAS analysis~\cite{Aad:2019vnb} for mass point $m(\tilde{\chi}_1^{\pm}, \tilde{\chi}_1^0) = (300, 50)~{\rm GeV}$ in search channel of $\tilde{\chi}_1^\pm \tilde{\chi}_1^\mp$ production with $W$-boson-mediated decays. The yields in \enquote{Baseline} of \enquote{CheckMATE} are normalized to \enquote{Baseline} of \enquote{ATLAS}. \enquote{Efficiency} is defined as the ratio of the event number passing though the cut-flow to the event number of the previous event.}
	\vspace{0.3cm}
	\resizebox{0.75\textwidth}{!}{
	\begin{tabular}{l|rrrr} 
			\hline
			Process & \multicolumn{4}{c}{$pp \to \tilde{\chi}_1^{+}\tilde{\chi}_1^{-}, \tilde{\chi}_1^\pm \to W^{\pm} \tilde{\chi}_1^0$} \\ 
			Point $m(\tilde{\chi}_1^\pm, \tilde{\chi}_1^0)$ & \multicolumn{4}{c}{(300, 50) GeV} \\ 
			Generated events & \multicolumn{4}{c}{500,000} \\ \hline
			\multirow{2}{*}{Selection} & \multicolumn{2}{c}{ATLAS} & \multicolumn{2}{c}{CheckMATE}  \\ 
			& events & efficiency & events & efficiency \\ \hline
			\bf Preselection & & & & \\ 
			Baseline & 1144.0 & $\cdots$ & 1144.0 & $\cdots$  \\ 
			Trigger & 793.0 & 69.3\% & 766.0 & 66.9\%  \\ 
			OS signal leptons & 661.0 & 83.4\% & 725.1 & 94.7\%   \\ 
			$p_{\rm T}^{\ell_1,\ell_2} >$ 25 GeV & 565.0 & 85.5\% & 548.6 & 75.7\%  \\ 
			$m_{\ell\ell} > 25~{\rm GeV}$ & 559.0 & 98.9\% & 542.5 & 98.9\%  \\ 
			$n_{b-{\rm jet}} = 0$ & 526.0 & 94.1\% & 507.8 & 93.6\%  \\ \hline
			\bf SR-DF-0J & & & & \\ 
			Different flavor $\&~n_{\rm jets} = 0$ & 122.7 & 23.3\% & 134.3  & 26.4\% \\ 
			$m_{\ell\ell} > 100~{\rm GeV}$ & 94.2 & 76.8\% & 96.1 & 71.6\%  \\ 
			$E_{\rm T}^{\rm miss} > 110~{\rm GeV}$ & 46.5 & 49.4\% & 43.4 & 45.2\% \\ 
			$E_{\rm T}^{\rm miss}~{\rm significance} > 10$ & 42.2 & 90.8\% & 39.5 & 90.9\% \\ 
			$m_{\rm T2} > 100~{\rm GeV}$ & 26.4 & 62.6\% & 26.3 & 66.7\% \\ \hline 
			\bf SR-DF-1J& & & &  \\ 
			Different flavor $\&~n_{\rm jets} = 1$ & 81.9 & 15.6\% & 84.6 & 16.7\% \\ 
			$m_{\ell\ell} > 100~{\rm GeV}$ & 62.3 & 76.1\% & 57.7 & 68.2\%  \\ 
			$E_{\rm T}^{\rm miss} > 110~{\rm GeV}$ & 33.8 & 54.3\% & 30.3 & 52.5\% \\ 
			$E_{\rm T}^{\rm miss}~{\rm significance} > 10$ & 27.2 & 80.5\% & 28.2 & 93.3\% \\ 
			$m_{\rm T2} > 100~{\rm GeV}$ & 15.3 & 56.3\% & 14.7 & 52.0\% \\ \hline
			\bf SR-SF-0J & & & & \\ 
			Same flavor $\&~n_{\rm jets} = 0$ & 138.7 & 26.4\% & 111.1 & 21.9\%  \\ 
			$m_{\ell\ell} > 121.2~{\rm GeV}$ & 92.4 & 75.3\% & 74.2 & 66.8\% \\ 
			$E_{\rm T}^{\rm miss} > 110~{\rm GeV}$ & 47.1 & 50.0\% & 40.4 & 54.5\%  \\ 
			$E_{\rm T}^{\rm miss}~{\rm significance} > 10$ & 42.9 & 92.3\% & 35.9 & 88.7\% \\ 
			$m_{\rm T2} > 100~{\rm GeV}$ & 25.4 & 61.9\% & 21.8 & 60.7\% \\ \hline
			\bf SR-SF-1J & & & & \\ 
			Same flavor $\&~n_{\rm jets} = 1$ & 88.8 & 16.9\% & 82.4 & 16.2\% \\ 
			$m_{\ell\ell} > 121.2~{\rm GeV}$ & 58.9 & 66.3\% & 48.0 & 58.3\%  \\ 
			$E_{\rm T}^{\rm miss} > 110~{\rm GeV}$ & 32.6 & 55.4\% & 25.4 & 52.8\%  \\ 
			$E_{\rm T}^{\rm miss}~{\rm significance} > 10$ & 26.9 & 82.5\% & 24.9 & 98.0\% \\ 
			$m_{\rm T2} > 100~{\rm GeV}$ & 14.0 & 52.0\% & 11.2 & 45.1\%  \\ \hline
		\end{tabular}}
\end{table}

\begin{table}[ht]
	\centering
	\caption{\label{tab:app-vld1911}Similar to Table~\ref{tab:app-vld1908}, but for the cut-flow validation of ATLAS analysis~\cite{Aad:2019vvi} in the search channel of the slepton pair production with mass point $m(\tilde{\ell}, \tilde{\chi}_1^0) = (150, 140)~{\rm GeV}$.}
	\vspace{0.3cm}
	\resizebox{0.85\textwidth}{!}{
		\begin{tabular}{l|rrrr}
			\hline
			Process & \multicolumn{4}{c}{$p p \to \tilde{\ell}\tilde{\ell}, \tilde{\ell} \to \ell \tilde{\chi}_1^0$} \\ 
			Point & \multicolumn{4}{c}{$m_{\tilde{\ell}}$ = 150 GeV; $m_{\tilde{\chi}_1^{0}}$ = 140 GeV} \\ 
			Generated events & \multicolumn{4}{c}{100,000} \\ \hline
			\multirow{2}{*}{Selection} & \multicolumn{2}{c}{ATLAS} & \multicolumn{2}{c}{CheckMATE} \\ 
			& events & efficiency & events & efficiency \\ \hline
		    $E_{\rm T}^{\rm miss}$ trigger & 2355.37 & $\cdots$ & 2355.37 & $\cdots$  \\ 
			3rd lepton veto & 1014.55 & 43.1\% & 1079.07 & 45.8\%  \\ 
			$3~{\rm GeV} < m_{\ell\ell} < 3.2~{\rm GeV}$ veto & 1013.21 & 99.9\% & 1077.69 & 99.9\%    \\ 
			Lepton author 16 veto & 1009.48 & 99.6\% & 1077.69 & 100.0\%  \\ 
			min$(\Delta\phi({\rm jet}, p_{\rm T}^{\rm miss})) >$ 0.4 & 970.36 & 96.1\% & 1049.11 & 97.4\%  \\ 
			$\Delta\phi(j_1, p_{\rm T}^{\rm miss}) >$ 2.0 & 961.15 & 99.1\% & 1027.05 & 97.9\%  \\ 
			lepton truth matching & 958.99 & 99.8\% & 1027.05 & 100.0\%  \\ 
			$1~{\rm GeV}< m_{\ell \ell} < 60~{\rm GeV}$ & 827.86 & 86.3\% & 883.55 & 86.0\%  \\ 
			$\Delta R_{ee} >$ 0.3, $\Delta R_{\mu\mu} >$ 0.05, $\Delta R_{e\mu} >$ 0.2 & 826.19 & 99.8\% & 883.48 & 99.9\%  \\ 
			$p_{\rm T}^{\ell_1} >$ 5 GeV & 823.70 & 99.7\% & 880.95 & 99.7\% \\
			$n_{\rm jets}\geq 1$ & 810.59 & 98.4\% & 880.95 & 100.0\%  \\ 
			$p_{\rm T}^{j_1} >$ 100 GeV & 705.86 & 87.1\% & 702.58 & 79.8\% \\ 
			$n_{b-{\rm jets}} = 0$ & 611.05 & 86.6\% & 643.78 & 91.6\% \\ 
			$m_{\tau\tau} < 0$ or $m_{\tau\tau} > 160~{\rm GeV}$ & 533.29 & 87.3\% & 569.78 & 88.5\% \\ 
			same flavor & 532.33 & 99.8\% & 569.01 & 99.9\% \\ \hline
			{\bf SR-highMass} \\ 
			$E_{\rm T}^{\rm miss} >$ 200 GeV & 229.81 & 43.2\% & 265.83 & 46.7\%  \\ 
			$\max(0.85, 0.98 - 0.02\times m_{\rm T2}^{100}) < R_{\rm ISR} < 1.0$ & 160.30 & 69.8\% & 165.78 & 62.4\%\\ 
			$p_{\rm T}^{\ell_2} > \min(20.0, 2.5 + 2.5\times (m_{T_2}^{100} - 100))$ & 70.71 & 44.1\% & 72.51 & 43.7\% \\ 
			$m_{\rm T2}^{100} < 140~{\rm GeV}$ & 70.71 & 100.0\% & 72.51 & 100.0\% \\ 
			$m_{\rm T2}^{100} < 130~{\rm GeV}$ & 70.71 & 100.0\% & 72.51 & 100.0\%    \\ 
			$m_{\rm T2}^{100} < 120~{\rm GeV}$ & 70.71 & 100.0\% & 72.31 & 99.7\% \\ 
			$m_{\rm T2}^{100} < 110~{\rm GeV}$ & 70.71 & 100.0\% & 72.23 & 99.9\%   \\ 
			$m_{\rm T2}^{100} < 105~{\rm GeV}$ & 53.72 & 76.0\% & 57.10 & 79.1\%   \\ 
			$m_{\rm T2}^{100} < 102~{\rm GeV}$ & 20.21 & 37.6\% & 23.77 & 41.6\%   \\ 
			$m_{\rm T2}^{100} < 101~{\rm GeV}$ & 9.38 & 46.4\% & 9.90 & 41.6\%   \\ 
			$m_{\rm T2}^{100} < 100.5~{\rm GeV}$ & 4.68 & 49.9\% & 4.86 & 49.1\%\\ \hline
			{\bf SR-lowMass} \\
			$150~{\rm GeV} < E_{\rm T}^{\rm miss} < 200~{\rm GeV} $& 146.36 & 27.5\% & 167.63 & 29.5\%  \\
			$0.8 < R_{\rm ISR} < 1.0$ & 107.82 & 73.7\% & 93.17 & 55.6\%  \\
			$p_{\rm T}^{\ell_2} > \min(15.0, 7.5 + 0.75\times (m_{T_2}^{100} - 100))$ & 52.74 & 48.9\% & 42.29 & 45.4\% \\
			$m_{\rm T2}^{100} < 140~{\rm GeV}$ & 52.74 & 100.0\% & 42.29 & 100.0\% \\ 
			$m_{\rm T2}^{100} < 130~{\rm GeV}$ & 52.74 & 100.0\% & 42.29 & 100.0\%    \\ 
			$m_{\rm T2}^{100} < 120~{\rm GeV}$ & 52.74 & 100.0\% & 42.29 & 100.0\%  \\ 
			$m_{\rm T2}^{100} < 110~{\rm GeV}$ & 52.64 & 99.8\% & 41.65 & 98.5\%    \\ 
			$m_{\rm T2}^{100} < 105~{\rm GeV}$ & 38.05 & 72.3\% & 29.09 & 69.9\%    \\ 
			$m_{\rm T2}^{100} < 102~{\rm GeV}$ & 16.66 & 43.8\% & 11.24 & 38.6\%  \\ 
			$m_{\rm T2}^{100} < 101~{\rm GeV}$ & 8.70 & 52.2\% & 5.60 & 49.8\%    \\ 
			$m_{\rm T2}^{100} < 100.5~{\rm GeV}$ & 4.39 & 50.5\% & 2.29 & 40.9\%    \\ \hline
		\end{tabular}}
\end{table}
\bibliography{references}
\bibliographystyle{CitationStyle}

\end{document}